\documentclass[aps,prd,showpacs,amssymb,amsmath,amsfonts,superscriptaddress,
twocolumn,floatfix,preprintnumbers,altaffilletter]{revtex4}
\usepackage{graphicx}
\usepackage{hyperref}
\usepackage{amssymb}
\usepackage{amsmath}
\usepackage{longtable}
\usepackage{rotating}
\usepackage{color}
\usepackage{epsfig}
\usepackage{epsf}
\usepackage{fancyhdr}

\def\H{\textrm{\mbox{\tiny{H}}}}
\def\th{\textrm{\mbox{\tiny{th}}}}
\def\Tobs{T_{\textrm{\mbox{\tiny{obs}}}}}
\def\Tcoh{T_{\textrm{\mbox{\tiny{coh}}}}}

\def\max{\textrm{\mbox{\tiny{max}}}}

\def\iSubSupInd{i}
\def\srchTemplateInd{{}}

\newcommand{\ee}[1]{\!\times\!10^{#1}}

\def\bea{\begin{eqnarray}}
\def\eea{\end{eqnarray}}
\def\beq{\begin{equation}}
\def\eeq{\end{equation}}
\def\vec#1{\mathbf{#1}}
\def\be{\begin{equation}}
\def\ee{\end{equation}}

\newcommand{\ba}{\begin{eqnarray}}
\newcommand{\ea}{\end{eqnarray}}

\newcommand{\bml}{\begin{mathletters}}
\newcommand{\eml}{\end{mathletters}}

\def \S {{\cal S}}

\def\sci#1#2{#1\times10^{#2}}

\def\fhat{\hat f}
\def\hpluszero{A_{+}}
\def\hcrosszero{A_{\times}}
\def\fhatzero{\hat{f}_0}
\def\fdot{\dot{f}}

\begin{document}
\pagestyle{fancy}
\rhead[]{}
\lhead[]{}
\title{ 
 All-sky search for periodic gravitational waves in LIGO S4 data
}

\affiliation{Albert-Einstein-Institut, Max-Planck-Institut f\"ur Gravitationsphysik, D-14476 Golm, Germany}
\affiliation{Albert-Einstein-Institut, Max-Planck-Institut f\"ur Gravitationsphysik, D-30167 Hannover, Germany}
\affiliation{Andrews University, Berrien Springs, MI 49104 USA}
\affiliation{Australian National University, Canberra, 0200, Australia}
\affiliation{California Institute of Technology, Pasadena, CA  91125, USA}
\affiliation{Caltech-CaRT, Pasadena, CA  91125, USA}
\affiliation{Cardiff University, Cardiff, CF24 3AA, United Kingdom}
\affiliation{Carleton College, Northfield, MN  55057, USA}
\affiliation{Charles Sturt University, Wagga Wagga, NSW 2678, Australia}
\affiliation{Columbia University, New York, NY  10027, USA}
\affiliation{Embry-Riddle Aeronautical University, Prescott, AZ   86301 USA}
\affiliation{Hobart and William Smith Colleges, Geneva, NY  14456, USA}
\affiliation{Inter-University Centre for Astronomy  and Astrophysics, Pune - 411007, India}
\affiliation{LIGO - California Institute of Technology, Pasadena, CA  91125, USA}
\affiliation{LIGO Hanford Observatory, Richland, WA  99352, USA}
\affiliation{LIGO Livingston Observatory, Livingston, LA  70754, USA}
\affiliation{LIGO - Massachusetts Institute of Technology, Cambridge, MA 02139, USA}
\affiliation{Louisiana State University, Baton Rouge, LA  70803, USA}
\affiliation{Louisiana Tech University, Ruston, LA  71272, USA}
\affiliation{Loyola University, New Orleans, LA 70118, USA}
\affiliation{Moscow State University, Moscow, 119992, Russia}
\affiliation{NASA/Goddard Space Flight Center, Greenbelt, MD  20771, USA}
\affiliation{National Astronomical Observatory of Japan, Tokyo  181-8588, Japan}
\affiliation{Northwestern University, Evanston, IL  60208, USA}
\affiliation{Rochester Institute of Technology, Rochester, NY 14623, USA}
\affiliation{Rutherford Appleton Laboratory, Chilton, Didcot, Oxon OX11 0QX United Kingdom}
\affiliation{San Jose State University, San Jose, CA 95192, USA}
\affiliation{Southeastern Louisiana University, Hammond, LA  70402, USA}
\affiliation{Southern University and A\&M College, Baton Rouge, LA  70813, USA}
\affiliation{Stanford University, Stanford, CA  94305, USA}
\affiliation{Syracuse University, Syracuse, NY  13244, USA}
\affiliation{The Pennsylvania State University, University Park, PA  16802, USA}
\affiliation{The University of Texas at Brownsville and Texas Southmost College, Brownsville, TX  78520, USA}
\affiliation{Trinity University, San Antonio, TX  78212, USA}
\affiliation{Universitat de les Illes Balears, E-07122 Palma de Mallorca, Spain}
\affiliation{Universit\"at Hannover, D-30167 Hannover, Germany}
\affiliation{University of Adelaide, Adelaide, SA 5005, Australia}
\affiliation{University of Birmingham, Birmingham, B15 2TT, United Kingdom}
\affiliation{University of Florida, Gainesville, FL  32611, USA}
\affiliation{University of Glasgow, Glasgow, G12 8QQ, United Kingdom}
\affiliation{University of Maryland, College Park, MD 20742 USA}
\affiliation{University of Michigan, Ann Arbor, MI  48109, USA}
\affiliation{University of Oregon, Eugene, OR  97403, USA}
\affiliation{University of Rochester, Rochester, NY  14627, USA}
\affiliation{University of Salerno, 84084 Fisciano (Salerno), Italy}
\affiliation{University of Sannio at Benevento, I-82100 Benevento, Italy}
\affiliation{University of Southampton, Southampton, SO17 1BJ, United Kingdom}
\affiliation{University of Strathclyde, Glasgow, G1 1XQ, United Kingdom}
\affiliation{University of Washington, Seattle, WA, 98195}
\affiliation{University of Western Australia, Crawley, WA 6009, Australia}
\affiliation{University of Wisconsin-Milwaukee, Milwaukee, WI  53201, USA}
\affiliation{Washington State University, Pullman, WA 99164, USA}
\author{B.~Abbott}\affiliation{LIGO - California Institute of Technology, Pasadena, CA  91125, USA}
\author{R.~Abbott}\affiliation{LIGO - California Institute of Technology, Pasadena, CA  91125, USA}
\author{R.~Adhikari}\affiliation{LIGO - California Institute of Technology, Pasadena, CA  91125, USA}
\author{J.~Agresti}\affiliation{LIGO - California Institute of Technology, Pasadena, CA  91125, USA}
\author{P.~Ajith}\affiliation{Albert-Einstein-Institut, Max-Planck-Institut f\"ur Gravitationsphysik, D-30167 Hannover, Germany}
\author{B.~Allen}\affiliation{Albert-Einstein-Institut, Max-Planck-Institut f\"ur Gravitationsphysik, D-30167 Hannover, Germany}\affiliation{University of Wisconsin-Milwaukee, Milwaukee, WI  53201, USA}
\author{R.~Amin}\affiliation{Louisiana State University, Baton Rouge, LA  70803, USA}
\author{S.~B.~Anderson}\affiliation{LIGO - California Institute of Technology, Pasadena, CA  91125, USA}
\author{W.~G.~Anderson}\affiliation{University of Wisconsin-Milwaukee, Milwaukee, WI  53201, USA}
\author{M.~Arain}\affiliation{University of Florida, Gainesville, FL  32611, USA}
% ...
\collaboration{The LIGO Scientific Collaboration, http://www.ligo.org}
\noaffiliation

\date{\today}

\begin{abstract}
  We report on an all-sky search with the LIGO detectors for periodic gravitational waves in the frequency
  range $50\,$--$\,1000$~Hz and with the frequency's time derivative in the range 
  $-\sci{1}{-8}~\mathrm{Hz}~\mathrm{s}^{-1}$ to zero.
  Data from the fourth LIGO science run (S4) have been used in this search.
  Three different semi-coherent methods of transforming and summing strain power from
  Short Fourier Transforms (SFTs) of the calibrated data have been used.
  The first, known as ``StackSlide'', averages normalized power from each SFT.
  A ``weighted Hough'' scheme is also developed and used, and which also allows for a
  multi-interferometer search. The third method, known as ``PowerFlux'', 
  is a variant of the StackSlide method in which the power is weighted before summing.
  In both the weighted Hough and PowerFlux methods, the weights are chosen according
  to the noise and detector antenna-pattern to maximize the signal-to-noise ratio.
  The respective advantages and disadvantages of these methods are discussed. 
  Observing no evidence of periodic gravitational radiation, we report upper limits;
  we interpret these as limits on this radiation from isolated rotating neutron stars.
  The best population-based upper limit with $95\%$ confidence on the gravitational-wave
  strain amplitude, found for simulated sources distributed isotropically across the sky
  and with isotropically distributed spin-axes, 
  is $4.28 \times 10^{-24}$ (near 140 Hz). Strict upper limits are also
  obtained for small patches on the sky for best-case and worst-case inclinations
  of the spin axes.
\end{abstract}
\preprint{LIGO-P060010-06-Z}
\maketitle
%
\section{Introduction}
\label{sec:introduction}

We report on a search with the LIGO (Laser Interferometer
Gravitational-wave Observatory) detectors \cite{ligo1, ligo2} for
periodic gravitational waves in the frequency range $50\,$--$\,1000$~Hz
and with the frequency's time derivative in the range 
$-\sci{1}{-8}~\mathrm{Hz}~\mathrm{s}^{-1}$ to zero.
The search is carried out over the entire sky using 
data from the fourth LIGO science run (S4). 
Isolated rotating neutron stars in our galaxy are the prime target.

Using data from earlier science runs, the LIGO Scientific Collaboration (LSC) has previously
reported on searches for periodic gravitational radiation, using a long-period coherent
method to target known pulsars \cite{S1PulsarPaper,S2TDPaper,S3S4TDPaper}, using a
short-period coherent method to target Scorpius X-1 in selected bands and search the entire sky 
in the $160.0\,$--$\,728.8$~Hz band \cite{S2FstatPaper}, and using a long-period
semi-coherent method to search the entire sky in the $200\,$--$\,400$~Hz band \cite{S2HoughPaper}. 
Einstein@Home, a distributed home computing effort running under the BOINC architecture \cite{BOINC},
has also been searching the entire sky using a coherent first stage, followed by a simple coincidence
stage \cite{S3EatH}. In comparison, this paper: 1) examines more sensitive data;
2) searches over a larger range in frequency and its derivative; and 3) uses three
alternative semi-coherent methods for summing measured strain powers to detect excess
power from a continuous gravitational-wave signal.

The first purpose of this paper is to present results from our search for periodic
gravitational waves in the S4 data. Over the LIGO frequency band of sensitivity,
the S4 all-sky upper limits presented here are approximately  an order of magnitude better than published
previously from earlier science runs~\cite{S2FstatPaper,S2HoughPaper}. 
After following up on outliers in the data, we find that no 
candidates survive, and thus report upper limits. These are interpreted
as limits on radiation from rotating neutron stars, which can be expressed as functions of
the star's ellipticity and distance, allowing for an astrophysical interpretation.
The best population-based upper limit with $95\%$ confidence on the gravitational-wave
strain amplitude, found for simulated sources distributed isotropically across the sky
and with isotropically distributed spin-axes,
is $4.28 \times 10^{-24}$ (near 140 Hz). Strict upper limits are also obtained for
small patches on the sky for best-case and worst-case inclinations of the spin axes.

The second purpose of this paper, along with the previous coherent~\cite{S2FstatPaper}
and semi-coherent~\cite{S2HoughPaper} papers, is to lay the foundation for the methods that
will be used in future searches. It is well known that the search for periodic gravitational waves is
computationally bound; to obtain optimal results will require a hierarchical approach that uses coherent
and semi-coherent stages \cite{hough04,pss01,BC00,cgk}. A fifth science run (S5), which started in November 2005,
is generating data at initial LIGO's design sensitivity. We plan to search this data using the
best methods possible, based on what is learned from this and previous analyses.

In the three methods considered here, one searches for cumulative excess power from a 
hypothetical periodic gravitational wave signal by examining successive spectral estimates
based on Short Fourier Transforms (SFTs) of the calibrated detector strain data channel, taking into account
the Doppler modulations of detected frequency due to the Earth's rotational and orbital motion
with respect to the Solar System Barycenter (SSB), and the time derivative of the
frequency intrinsic to the source. The simplest method presented, known
as ``StackSlide'' \cite{BCCS,BC00,cgk,StackSlideTechNote}, averages
normalized power from each SFT. In the Hough method reported previously \cite{S2HoughPaper,hough04},
referred to here as ``standard Hough'', the sum is of binary zeroes or ones,
where an SFT contributes unity if the power exceeds a normalized power threshold. 
In this paper a ``weighted Hough'' scheme, henceforth also referred to as ``Hough'', has been developed
and is similar to that described in Ref.~\cite{Palomba2005}. This scheme also allows for a multi-interferometer search.
The third method, known as ``PowerFlux'' \cite{PowerFluxTechNote}, 
is a variant of the StackSlide method in which the power is weighted before summing.
In both the weighted Hough and PowerFlux methods, the weights are chosen according
to the noise and detector antenna pattern to maximize the signal-to-noise ratio.
 
The Hough method is computationally faster and more robust against large transient power artifacts,
but is slightly less sensitive than StackSlide for stationary data \cite{S2HoughPaper,StackSlideTechNote}. 
The PowerFlux method is found in most frequency ranges to have better detection efficiency than
the StackSlide and Hough methods, the exceptions occurring in bands with large non-stationary
artifacts, for which the Hough method proves more robust. However, the StackSlide and Hough methods
can be made more sensitive by starting with the maximum likelihood statistic
(known as the $\cal F$-statistic \cite{jks,hough04,S2FstatPaper}) rather than SFT power as
the input data, though this improvement comes with increased computational cost.
The trade-offs among the methods means that each could play a role in our future searches.

In brief, this paper makes several important contributions. It sets the best all-sky upper limits
on periodic gravitational waves to date, and shows that these limits are becoming
astrophysically interesting. It also introduces methods that are crucial to the
development of our future searches.
  
This paper is organized as follows:
Section~\ref{sec:detectordata} briefly describes the LIGO interferometers,
focusing on improvements made for the S4 data run, and 
discusses the sensitivity and relevant detector artifacts. 
Section~\ref{sec:waveforms} precisely defines the waveforms
we seek and the associated assumptions we have made. Section~\ref{sec:analysismethodoverview}
gives a detailed description of the three analysis
methods used and summarizes their similarities and
differences, while Section \ref{sec:analysismethoddetails}
gives the details of their implementations and the pipelines used.
Section~\ref{sec:valandhwinj} discusses the validation of the software
and, as an end-to-end test, shows the detection of simulated pulsar
signals injected into the data stream at the hardware level. 
Section~\ref{sec:results} describes the search results, and
Section~\ref{sec:comparisonresults} compares the results from the three respective
methods. Section~\ref{sec:summary} concludes
with a summary of the results, their astrophysical implications, and future plans.

\section{The LIGO Detector Network and the S4 Science Run}
\label{sec:detectordata}

The LIGO detector network consists of a 4-km interferometer in
Livingston Louisiana (called L1) and two interferometers in Hanford
Washington, one 4-km and another 2-km (H1 and H2, respectively). 

\begin{figure}
  \begin{center}
  \includegraphics[height=6.0cm]{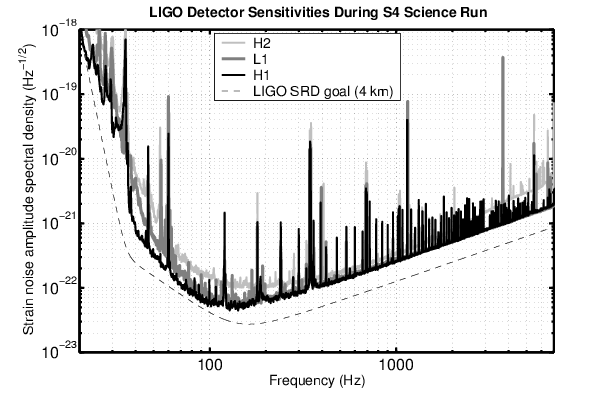}
  \caption{Median amplitude strain noise spectral densities 
            from the three LIGO interferometers during the S4 run,
            along with the Initial LIGO design sensitivity goal.}
  \label{fig:S4sensitivity} 
  \end{center}
\end{figure}

The data analyzed in this paper were produced during LIGO's 29.5-day
fourth science run (S4)~\cite{S4detpaper}. This run started at noon 
Central Standard Time (CST) on February 22 and ended
at midnight CST on March 23, 2005. During the run, all three LIGO
detectors had displacement spectral amplitudes near 
$\sci{2.5}{-19}~{\rm m}~{\rm Hz}^{-1/2}$ in their most sensitive frequency band near 150 Hz.
In units of gravitational-wave strain amplitude, the sensitivity of H2
is roughly a factor of two worse than that of H1 and L1 over
much of the search band.
The typical strain sensitivities in
this run were within a factor of two of the design goals.
Figure~\ref{fig:S4sensitivity} shows representative strain spectral
noise densities for the three interferometers during the run.
As discussed in Section~\ref{sec:analysismethoddetails} below, however,
non-stationarity of the noise was significant.

Changes to the interferometers before the S4 run included the following improvements~\cite{S4detpaper}:
\begin{itemize}
\item Installation of active seismic isolation of support structures at Livingston
to cope with high anthropogenic ground motion in the 1-3 Hz band.
\item Thermal compensation with a CO$_2$ laser of mirrors subject
to thermal lensing from the primary laser beam to a greater or lesser
degree than expected.
\item Replacement of a synthesized radio frequency oscillator for phase modulation
with a crystal oscillator before S4 began (H1)
and mid-way through the S4 run (L1), reducing 
noise substantially above 1000 Hz and eliminating a comb of $\sim 37$~Hz lines.
(The crystal oscillator replacement for H2 occurred after the S4 run.)
\item Lower-noise mirror-actuation electronics (H1, H2, \&\ L1).
\item Higher-bandwidth laser frequency stabilization (H1, H2, \&\ L1) and intensity stabilization (H1 \&\ L1).
\item Installation of radiation pressure actuation of mirrors for calibration validation (H1).
\item Commissioning of complete alignment control system for the L1 interferometer (already
implemented for H1 \&\ H2 in S3 run).
\item Refurbishment of lasers and installation of photodiodes and electronics 
to permit interferometer operation with increased laser power (H1, H2, \&\ L1).
\item Mitigation of electromagnetic interference (H1, H2, \&\ L1) and acoustic interference (L1).
\end{itemize}

The data were acquired and digitized at a rate of 16384 Hz.
Data acquisition was periodically interrupted by disturbances
such as seismic transients,
reducing the net running time of the interferometers.
The resulting duty factors for the interferometers
were 81\% for H1 and H2, and 74\% for L1.
While the H1 and H2 duty factors were somewhat higher than those in
previous science runs, the L1 duty factor was dramatically
higher than the $\simeq$40\% typical of the past, thanks to
the increased stability from the installation of the active 
seismic isolation system at Livingston.

\section{Signal Waveforms}
\label{sec:waveforms}

The general form of a gravitational-wave signal is described in terms of
two orthogonal transverse polarizations defined as ``$+$'' with waveform $h_+(t)$ 
and ``$\times$'' with waveform $h_\times(t)$.
The calibrated response seen by an interferometric gravitational-wave detector is then \cite{jks}
\begin{equation}
h(t) = F_+(t,\alpha,\delta,\psi)h_+(t) + F_\times(t,\alpha,\delta,\psi)h_\times(t) \label{eq:detoutput},
\end{equation}
where $t$ is time in the detector frame, $\alpha$ is the source right ascension, 
$\delta$ is the source declination, $\psi$ is the polarization
angle of the wave, and $F_{+,\times}$ are the detector antenna pattern
functions for the two orthogonal polarizations. For
periodic (nearly pure sinusoidal) gravitational waves, which in general are elliptically polarized,
the individual components $h_{+,\times}$ have the form
\begin{eqnarray}
h_+(t) \quad & = & \hpluszero \cos\Phi(t) , \label{eq:sinusoidCosPhi} \\
h_\times(t) \quad & = & \hcrosszero \sin\Phi(t) , \label{eq:sinusoidSinPhi}
\end{eqnarray} 
where $\hpluszero$ and $\hcrosszero$ are the amplitudes
of the two polarizations, and $\Phi(t)$ is the phase of the signal at the detector.
(One can also define
the initial phase of the signal, $\Phi_0$, but in this paper it can be
taken to be an unknown and irrelevant constant).  

For an isolated quadrupolar gravitational-wave emitter,
characterized by a rotating triaxial ellipsoid mass distribution,
the amplitudes $\hpluszero$ and $\hcrosszero$ are related 
to the inclination angle of the source, $\iota$, and the
wave amplitude, $h_0$, by:
\begin{eqnarray}
\hpluszero &=& \frac{1}{2}h_0 \left(1+ \cos^2\iota\right),\\
\hcrosszero &=& h_0 \cos\iota, 
\end{eqnarray}
where $\iota$ is the angle of its spin axis with respect to the
line of sight between source and detector. 
For such a star, the gravitational-wave frequency, $f$, is twice the rotation
frequency, $\nu$, and the amplitude
$h_0$ is given by
\begin{equation} \label{eq:h0} 
h_0 = \frac{16\pi^2G}{c^4}\frac{I\epsilon \nu^2}{d}\,. 
\end{equation}
Here $d$ is the distance to the star, $I$ is the
principal moment of inertia with respect to its spin axis,
and 
$\epsilon$ is the equatorial ellipticity of the star \cite{jks}.
Assuming that all of the frequency's derivative, $\dot{f}$, is due to emission of
gravitational radiation and that $I$ takes the canonical value
$10^{38}~{\rm kg}{\rm m}^2$, we can relate $\epsilon$ to $f$ and $\dot{f}$
and use Eq.~(\ref{eq:h0}) to obtain
\begin{equation}
\label{hsd}
h_\mathrm{sd} = 4.54\times10^{-24} \left( \mbox{1 kpc} \over d \right)
\left( \mbox{250 yr} \over -f/(4\dot{f}) \right)^{\frac{1}{2}} ,
\end{equation}
by eliminating $\epsilon$, or
\begin{equation}
\label{esd}
\epsilon_\mathrm{sd} = 7.63\times10^{-5} \left( -\dot{f} \over
10^{-10}~\mathrm{Hz}~\mathrm{s}^{-1} \right)^{\frac{1}{2}} \left( \mbox{100 Hz} \over f
\right)^{\frac{5}{2}} ,
\end{equation}
by eliminating $d$.
These are referred to, respectively, as the \textit{spin-down limits} on
strain and ellipticity.
(See Eqs.~(8), (9), and (19) of \cite{S2FstatPaper} for more details of the
derivation.)

Note that the methods used in this paper are sensitive to periodic signals 
from any type of isolated gravitational-wave source (e.g., freely precessing
or oscillating neutron stars as well as triaxial ones), though
we present upper limits in terms of $h_0$ and $\epsilon$.
Because we use semi-coherent methods, only the
instantaneous signal frequency in the detector
reference frame, $2\pi f(t) = d\Phi(t)/dt$, needs to be calculated.
In the detector reference
frame this can, to a very good approximation, be related
to the instantaneous SSB-frame frequency $\fhat(t)$ by~\cite{S2HoughPaper}
\begin{equation}\label{eq:freqatdetector}
f(t) - \hat{f}(t) = \hat{f}(t)\frac{ {\bf v} (t)\cdot\bf{\hat{n}}}{c}  ,
\end{equation}
where ${\bf v}(t)$ is the detector's velocity with respect to the SSB frame,
and $\bf{\hat{n}}$ is the unit-vector corresponding to the sky-location of
the source.  In this analysis, we search for $\fhat(t)$ signals well
described by a nominal frequency $\fhatzero$ at the start of the S4
run $t_0$ and a constant first time derivative $\fdot$, such that
\begin{equation} 
\hat{f}(t) = \hat f_0 + \fdot\left(t - t_0\right).
\end{equation}
These equations ignore corrections to the time
interval $t - t_0$ at the detector compared with that at the SSB and
relativistic corrections. These corrections are negligible for
the one month semi-coherent searches described here, though
the LSC Algorithm Library (LAL) code \cite{LAL} used by our searches
does provide routines that make all the corrections needed to provide
a timing accuracy of 3 $\mu s$. (The LAL code also can calculate $f(t)$
for signals arriving from periodic sources in binary systems. Including
unknown orbital parameters in the search, however, would greatly
increase the computational cost or require new methods beyond the scope
of this article.)

\section{Overview of the Methods}
\label{sec:analysismethodoverview}

\subsection{Similarities and Differences}
\label{subsec:simsanddiffs}

The three different analysis methods presented here have many features in
common, but also have important differences, both major and minor. 
In this Section we give a brief overview of the methods.

\subsubsection{The parameter space}
\label{subsubsec:theparamspace}

\begin{figure}
  \begin{center}
  \includegraphics[height=3.0cm]{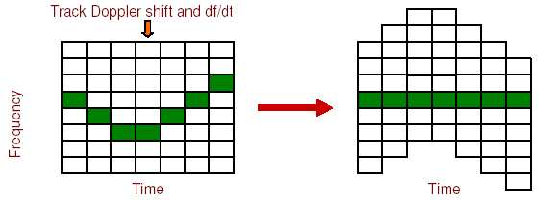}
  \caption{An illustration of the discrete frequency bins of the Short Fourier Transform (SFTs) of
  the data are shown vertically, with the discrete start times of the SFTs shown horizontally.
  The dark pixels represent a signal in the data. Its frequency changes with time due to
  Doppler shifts and intrinsic evolution of the source. By sliding the frequency bins, the power from a source
  can be lined up and summed after appropriate weighting or transformation. 
  This is, in essence, the starting point for all of the semi-coherent search methods
  presented here, though the actual implementations differ significantly.}
  \label{fig:StackingAndSlidingGraphic}
  \end{center}
\end{figure}

All three methods are based on summing measures of strain power from
many SFTs that have been created
from 30-minute intervals of calibrated strain data. Each method
also corrects explicitly for sky-position dependent
Doppler modulations of the apparent source
frequency due to the Earth's rotation and its orbital motion around
the SSB, and the frequency's time derivative, intrinsic to
the source (see Fig.~\ref{fig:StackingAndSlidingGraphic}). 
This requires a search in a four-dimensional parameter space;
a template in the space refers to a set of values: 
${\mathbf \lambda} = \{ \fhatzero, \dot{f}, \alpha, \delta \}$.
The third method, PowerFlux, also searches explicitly
over polarization angle, so
that ${\mathbf \lambda} = \{ \fhatzero, \dot{f}, \alpha, \delta, \psi \}$.

All three methods search for initial frequency $\fhatzero$ in the
range $50\,$--$\,1000$~Hz with a uniform grid spacing equal to the size of an SFT frequency bin,
\begin{equation} \label{eq:deltaf}
\delta f = {1 \over \Tcoh} = 5.556 \times 10^{-4} \, {\rm Hz} \,.
\end{equation}
where $\Tcoh$ is the time-baseline of each SFT.
The range of $\fhatzero$ is determined by the noise curves of the
interferometers, likely detectable source frequencies \cite{palomba},
and limitations due to the increasing computational cost at
high frequencies. 

The range of $\dot{f}$ values searched is $[-\sci{1}{-8}$,~$\,0]~\mathrm{Hz}~\mathrm{s}^{-1}$
for the StackSlide and PowerFlux methods and 
$[-\sci{2.2}{-9}$,~$\,0]~\mathrm{Hz}~\mathrm{s}^{-1}$ for
the Hough method. The ranges of $\dot{f}$ are
determined by the computational cost, as well as by the low
probability of finding an object with $|\dot{f}|$ higher than the values
searched---in other words, the ranges of $\dot{f}$ are narrow enough to
complete the search in a reasonable amount of time, yet wide enough to
include likely signals.  All known isolated pulsars spin down more slowly
than the two values of $|\dot{f}|_\max$ used here, and as seen in the results
section, the ellipticity required for higher $|\dot{f}|$ is improbably high
for a source losing rotational energy primarily via gravitational radiation at low
frequencies.
A small number of isolated pulsars in globular clusters exhibit slight spin-up, 
believed to arise from acceleration in the Earth's direction;  such spin-up values have
magnitudes small enough to be detectable with the zero-spin-down templates
used in these searches, given a strong enough signal.
The parameter ranges correspond to a minimum spin-down timescale
$f/|4\dot{f}|$ (the gravitational-wave spin-down age) of 40 years for a source emitting at 50~Hz 
and 800 years for a source at 1000~Hz.
Since for known pulsars \cite{ATNF} this characteristic timescale is at
least hundreds of years for frequencies on the low end of our range and
tens of millions of years for frequencies on the high end, we see again
that the ranges of $|\dot{f}|$ are wide enough to include sources from
this population.

As discussed in our previous
reports \cite{S2HoughPaper,S2FstatPaper}, the number of sky points
that must be searched grows quadratically with
the frequency $\fhatzero$, ranging here from about five thousand at
50 Hz to about two million at 1000 Hz.  All three methods use
nearly isotropic grids which cover the entire sky. The PowerFlux search
also divides the sky into regions according to susceptibility to stationary
instrumental line artifacts. Sky grid and spin-down spacings and other details are provided
below.

\subsubsection{Upper limits}
\label{subsubsec:themethodandul}

While the parameter space searched is similar for the three methods,
there are important differences in the way upper limits are
set. StackSlide and Hough both
set population-based frequentist limits on $h_0$ by carrying out
Monte Carlo simulations of a random population of pulsar sources
distributed uniformly over the sky and with isotropically distributed spin-axes.
PowerFlux sets strict frequentist limits on circular and linear polarization
amplitudes $h_0^{\rm Circ-limit}$ and  $h_0^{\rm Lin-limit}$, which correspond
to limits on most and least favorable pulsar inclinations, respectively.
The limits are placed separately on tiny patches of the sky, with the
highest strain upper limits presented here. In this context ``strict''
means that,  regardless of its polarization angle $\psi$
or inclination angle $\iota$, regardless of its sky location
(within fiducial regions discussed below), and regardless
of its frequency value and spin-down within the frequency and spin-down step
sizes of the search template, an isolated pulsar of
true strain amplitude $h_0 =2h_0^{\rm Lin-limit}$,
would have yielded 
a higher measured amplitude than what we measure, in at least 95\% 
of independent observations.
The circular polarization limits $h_0^{\rm Circ-limit}$ apply only to the most
favorable inclinations ($\iota\approx0$, $\pi$), regardless of sky location
and regardless of frequency and spin-down, as above.

Due to these different upper limit setting methods, sharp instrumental
lines are also handled differently. StackSlide and Hough carry out
removal of known instrumental lines of varying widths in individual
SFTs. The measured powers in those bins are replaced with random
noise generated to mimic the noise observed in neighboring bins.
This line cleaning technique can lead to a true signal being missed
because its apparent frequency may coincide with an instrumental line
for a large number of SFTs. However, population-averaged upper limits
are determined self-consistently to include loss of detection efficiency
due to line removal, by using Monte Carlo simulations.

Since its limits are intended to be strict, that is, valid for any
source inclination and for any source location within its
fiducial area, PowerFlux must handle
instrumental lines differently. Single-bin lines are flagged during
data preparation so that when searching for a particular source
an individual SFT bin power is ignored when it coincides with
the source's apparent frequency. If more than 80\% of otherwise
eligible bins are excluded for this reason, no attempt is made to
set a limit on strain power from that source. In practice, however, 
the 80\% cutoff is not used because we have found
that all such sources lie in certain unfavorable regions of the sky, which
we call ``skybands'' and which we exclude when setting upper limits.
These skybands depend on source frequency and 
its derivative, as described in Sec.~\ref{subsubsec:skybanding}.

\subsubsection{Data Preparation}
\label{subsubsec:dataprep}

Other differences among the methods 
concern the data windowing and filtering used in
computing Fourier transforms and concern the noise estimation.
StackSlide and Hough apply high pass filters to the data above 
$40 {\rm Hz}$, in addition to the filter used to produce the
calibrated data stream, and 
use Tukey windowing. PowerFlux applies no additional filtering and uses Hann
windowing with 50\% overlap between adjacent SFT's. StackSlide
and Hough use median-based noise floor tracking~\cite{mohanty02b,mohanty02a,badri}. 
In contrast, Powerflux uses a time-frequency decomposition. Both of
these noise estimation methods are described in Sec.~\ref{sec:analysismethoddetails}.

The raw, uncalibrated data channels containing the strain measurements from the three 
interferometers are converted to a calibrated ``$h(t)$'' data stream, following the
procedure described in~\cite{hoftpaper}, using calibration reference functions
described in~\cite{S4CalibrationNote}. 
SFTs are generated directly from the calibrated
data stream, using 30-minute intervals of data for which the interferometer is
operating in what is known as science-mode. The choice of 30 minutes is a tradeoff between
intrinsic sensitivity, which increases with SFT length, and robustness against
frequency drift during the SFT interval due to the Earth's motion,
source spin-down, and non-stationarity of the data \cite{S2HoughPaper}.
The requirement that each SFT contain contiguous data
at nominal sensitivity introduces duty factor loss from edge effects, 
especially for the Livingston interferometer ($\simeq$20\%) which had
typically shorter contiguous-data stretches.
In the end, the StackSlide and Hough searches used 1004 SFTs from H1 and 899
from L1, the two interferometers with the best broadband sensitivty. 
For PowerFlux, the corresponding numbers of overlapped SFTs were 1925 and 1628. 
The Hough search also used 1063 H2 SFTs.
In each case, modest requirements were placed on data quality to avoid short periods with
known electronic saturations, unmonitored calibration strengths, and the periods
immediately preceding loss of optical cavity resonance.

\subsection{Definitions And Notation}
\label{subsec:basicdefsandnotation}

Let $N$ be the number of SFTs, $\Tcoh$ the
time-baseline of each SFT, and $M$ the number of uniformly spaced data
points in the time domain from which the SFT is constructed.  If the
time series is denoted by $x_j$ ($j=0,1,2\ldots M-1$), then our convention
for the discrete Fourier transform is 
\begin{equation} \label{eq:DFT}
\tilde{x}_k = \Delta t \sum_{j=0}^{M-1}x_j e^{-2\pi {\mathrm i} jk/M} \, ,
\end{equation}
where $k=0,1,2\ldots (M-1)$, and $\Delta t = \Tcoh/M$.   
For $0\leq k \leq M/2$, the frequency index $k$ corresponds 
to a physical frequency of $f_k= k/\Tcoh$. 

In each method, the ``power'' (in units of spectral density)
associated with frequency bin $k$ and
SFT $i$ is taken to be 
\begin{equation} \label{eq:defpower}
P_k^{\iSubSupInd}=\frac{2|\tilde{x}_k^{\iSubSupInd}|^2}{\Tcoh}.
\end{equation}
It proves convenient to define
a normalized power by
\begin{equation} \label{eq:normpower}
\rho_k^{\iSubSupInd} = \frac{P_k^{\iSubSupInd}}{S_k^{\iSubSupInd}} \,.
\end{equation}
The quantity $S_k^{\iSubSupInd}$ is the single-sided power spectral density of the
detector noise at frequency $f_k$, the estimation of which is described below.
Furthermore, a threshold, $\rho_{\th}$, can be used to
define a \emph{binary count} by \cite{hough04}:
\begin{equation} \label{eq:1}
  n_k^{\iSubSupInd} = \left\{ 
  \begin{array}{ccc} 
    1  & \textrm{if} & \rho_k^{\iSubSupInd} \geq \rho_{\th}  \\
    0  & \textrm{if} & \rho_k^{\iSubSupInd} < \rho_{\th}  
  \end{array}  \right.\,.
\end{equation}

\begin{table}
\begin{tabular}{cc}\hline
Quantity                              & Description           \\
\hline \hline
$P^\srchTemplateInd_{\iSubSupInd}$ & Power for SFT $i$ \& template ${\mathbf \lambda}$ \\
$\rho^\srchTemplateInd_{\iSubSupInd}$ & Normalized power for SFT $i$ \& template ${\mathbf \lambda}$ \\
$n^\srchTemplateInd_{\iSubSupInd}$ & Binary count for SFT $i$ \& template ${\mathbf \lambda}$ \\
$S^\srchTemplateInd_{\iSubSupInd}$ & Power spect.\ noise density for SFT $i$ \& template ${\mathbf \lambda}$ \\
$F_+^{\iSubSupInd}$ & $F_+$ at midpoint of SFT $i$ for template ${\mathbf \lambda}$ \\
$F_\times^{\iSubSupInd}$ & $F_\times$ at midpoint of SFT $i$ for template ${\mathbf \lambda}$ \\
\hline
\end{tabular}
\caption{Summary of notation used.}
\label{tab:BasicDefsAndNotation}
\end{table}

When searching for a signal using template ${\mathbf \lambda}$ the detector antenna pattern
and frequency of the signal are found at the midpoint time of the data used to 
generate each SFT. Frequency dependent quantities are then evaluated at a frequency index
$k$ corresponding to the bin nearest this frequency. 
To simplify the equations in the rest
of this paper we drop the frequency index $k$
and use the notation given in Table~\ref{tab:BasicDefsAndNotation} to
define various quantities for SFT $i$ and template ${\mathbf \lambda}$.

\subsection{Basic StackSlide, Hough, and PowerFlux Formalism}
\label{subsec:methodsummaryeqns}

We call the detection statistics used in this search the ``StackSlide Power'', $P$, 
the ``Hough Number Count'', $n$, and the ``PowerFlux Signal Estimator'', $R$.  
The basic definitions of these quantities are given below.

Here the simple StackSlide method described in
\cite{StackSlideTechNote} is used; the ``StackSlide Power''
for a given template is defined as
\begin{equation} \label{eq:stackslidepower}
P = {1 \over N} \sum_{i=0}^{N-1} \rho^\srchTemplateInd_{\iSubSupInd} \,,
\end{equation}
This normalization results in values of $P$ with a mean value of unity and, for Gaussian
noise, a standard deviation of $1/\sqrt{N}$.  
Details about the value and statistics of $P$ in the presence and absence of a signal
are given in Appendix~\ref{sec:stackslidepowerandstats} and \cite{StackSlideTechNote}.

In the Hough search, instead of summing the normalized power, the
final statistic used in this paper is a weighted sum of the binary counts,
giving the ``Hough Number Count'':
\begin{equation}
  \label{eq:2}
  n = \sum_{i=0}^{N-1} w^{\srchTemplateInd}_{\iSubSupInd} n^{\srchTemplateInd}_{\iSubSupInd}\,.
\end{equation}
where the Hough weights are defined as
\begin{equation} \label{eq:wipropto}
  w^{\srchTemplateInd}_{\iSubSupInd} \propto  \frac{1}{S^{\srchTemplateInd}_{\iSubSupInd}}\left\{
    \left(F_{+}^{\iSubSupInd}\right)^2 +
    \left(F_{\times}^{\iSubSupInd}\right)^2\right\},
\end{equation}
and the weight normalization is chosen according to
\begin{equation}
  \label{eq:3}
  \sum_{i=0}^{N-1} w^{\srchTemplateInd}_{\iSubSupInd} = N\,.
\end{equation}
With this choice of normalization the Hough Number Count $n$ lies within the range $[0,N]$.
Thus, we take a binary count $n^{\srchTemplateInd}_{\iSubSupInd}$ to
have greater weight if the SFT $i$ has a lower noise floor and if, in the time-interval
corresponding to this SFT, the beam pattern functions are larger for a
particular point in the sky.  Note that the sensitivity of the search is governed by 
the ratios of the different weights, not by  the choice of overall scale. In the next
section we show that these weights maximize the sensitivity, averaged over the
orientation of the source. This choice of $w^{\srchTemplateInd}_{\iSubSupInd}$ was originally
derived in \cite{Palomba2005} using a different argument and is similar to that
used in the PowerFlux circular polarization projection described next.
More about the Hough method is given in \cite{S2HoughPaper,hough04}.

The PowerFlux method takes advantage of the fact that less weight should be given 
to times of greater noise variance or smaller detector antenna response to a signal.
Noting that power estimated from the data divided by 
the antenna pattern increases the variance of the data at times of
small detector response, the problem reduces to finding weights that minimize
the variance, or in other words that maximize the signal-to-noise
ratio.  The resulting PowerFlux detection statistic is  \cite{PowerFluxTechNote},
\begin{equation} \label{eq:PFRk}
R = {2 \over \Tcoh}
{
\sum_{i=0}^{N-1} W^{\srchTemplateInd}_{\iSubSupInd}
 P^\srchTemplateInd_{\iSubSupInd} /  (F_{\psi}^{\iSubSupInd})^2
\over
\sum_{i=0}^{N-1} W^{\srchTemplateInd}_{\iSubSupInd}
} ,
\end{equation}
where the PowerFlux weights are defined as
\begin{equation} \label{eq:PFWeights}
W^\srchTemplateInd_{\iSubSupInd} = [(F_{\psi}^{\iSubSupInd})^2]^2/S^2_{\iSubSupInd} ,
\end{equation}
and where 
\begin{equation} \label{PFFpsi}
  (F_{\psi}^{\iSubSupInd})^2 = \left\{ 
  \begin{array}{cc} 
    (F_{+}^{\iSubSupInd})^2  & \textrm{linear polarization} \\
    (F_{+}^{\iSubSupInd})^2 + (F_{\times}^{\iSubSupInd})^2  & \textrm{circular polarization}
  \end{array}  \right.\,.
\end{equation}
As noted previously, the PowerFlux method searches
using four linear polarization projections and one circular polarization
projection.
For the linear polarization projections, note that  $(F_{+}^{\iSubSupInd})^2$ is 
evaluated at the angle $\psi$, which is the same
as $(F_{\times}^{\iSubSupInd})^2$ evaluated at the angle $\psi - \pi/4$; for
circular polarization, the value of 
$(F_{+}^{\iSubSupInd})^2 + (F_{\times}^{\iSubSupInd})^2$ is independent of $\psi$.
Finally note that the factor of $2/\Tcoh$ in Eq.~(\ref{eq:PFRk}) makes $R$ dimensionless
and is chosen to make it directly related to an estimate of the squared amplitude of the
signal for the given polarization. Thus $R$ is also called in this paper
the ``PowerFlux Signal Estimator''. (See \cite{PowerFluxTechNote} and Appendix~\ref{sec:polarization}
for further discussion.)

We have shown in Eqs.~(\ref{eq:stackslidepower})-(\ref{PFFpsi}) how to compute the
detection statistic (or signal estimator) for a given template.
The next section gives the details of the implementation and pipelines used, 
where these quantities are calculated for a set of templates ${\mathbf \lambda}$
and analyzed.

\section{Implementations and Pipelines}
\label{sec:analysismethoddetails}


\subsection{Running Median Noise Estimation}
\label{subsec:runningmedian}

The implementations of the StackSlide and Hough methods described below use a ``running
median'' to estimate the mean power and, from this estimate, the power spectral density
of the noise, for every frequency bin of every SFT. PowerFlux uses a different noise
decomposition method described in its implementation section below.

Note that for Gaussian noise, the single-sided power spectral density can be estimated using
\begin{equation} \label{eq:Sn}
S_k^{\iSubSupInd} \cong 
\frac{2 \langle |\tilde{x}_k^{\iSubSupInd}|^2\rangle}{\Tcoh}
\end{equation}
where the angle brackets represent an ensemble average.  The estimation of
$S_k^{\iSubSupInd}$ must guard against any biases introduced by the presence of a
possible signal and also against narrow spectral disturbances.  For
this reason the mean,
$\langle|\tilde{x}^{\iSubSupInd}_k|^2\rangle$, is estimated via the median.  We
assume that the noise is stationary within a single SFT, but allow for
non-stationarities across different SFTs.  In every SFT we calculate
the ``running median'' of $|\tilde{x}_k^{\iSubSupInd}|^2$ for every $101$ frequency
bins centered on the $k^{\mathrm{th}}$ bin, and then estimate $\langle
|\tilde{x}_k^{\iSubSupInd}|^2\rangle$~\cite{mohanty02b,mohanty02a,badri} by 
dividing by the expected ratio of the median to the mean.  
 
Note, however, that in the StackSlide search, after the estimated mean
power is used to compute $S_k^{\iSubSupInd}$ in the denominator
of Eq.~(\ref{eq:normpower}) these terms are summed in
Eq.~(\ref{eq:stackslidepower}), while the Hough search applies
a cutoff to obtain binary counts in Eq.~(\ref{eq:1}) before summing.  
This results in the use of a different correction to get the mean in
the StackSlide search from that used in the Hough search. For a running
median using 101 frequency bins, the effective ratio of the median to
mean used in the StackSlide search was $0.691162$
(which was chosen to normalize the data so that the mean value of the
StackSlide Power equals one) compared
with the expected ratio for an exponential distribution of $0.698073$
used in the Hough search (which
is explained in Appendix~A of \cite{S2HoughPaper}).
It is important to realize that the results reported here are
valid independent of the factor used, since any overall constant
scaling of the data does not affect the selection of outliers or
the reported upper limits, which are based on Monte Carlo injections
subjected to the same normalization.

\subsection{The StackSlide Implementation}
\label{subsec:stackslide}

\subsubsection{Algorithm and parameter space}
\label{subsubsec:stackslideimpl}

The StackSlide method uses power averaging to gain sensitivity by decreasing the variance of the
noise \cite{BCCS,BC00,cgk,StackSlideTechNote}. Brady and Creighton \cite{BC00} first described this approach
in the context of gravitational-wave detection as a part of a hierarchical search for periodic sources. 
Their method consists of averaging the power from a demodulated time series, but as an approximation
did not include the beam pattern response of the detector. In Ref.~\cite{StackSlideTechNote}, a simple
implementation is described that averages the normalized power given in Eq.~(\ref{eq:normpower}). Its
extension to averaging the maximum likelihood statistic (known as the $\cal F$-statistic) which
does include the beam pattern response is mentioned in Ref.~\cite{StackSlideTechNote} (see also
\cite{jks,hough04,S2FstatPaper}), and further extensions of the StackSlide method are given
in \cite{cgk}.

As noted above, the simple StackSlide method given in
\cite{StackSlideTechNote} is used here and the detection 
statistic, called the ``StackSlide Power'', is defined by Eq.~(\ref{eq:stackslidepower}).
The normalization is chosen so that the mean value of
$P$ is equal to $1$ and its standard deviation is $1/\sqrt{N}$ for Gaussian noise alone.
For simplicity, the StackSlide Power signal-to-noise ratio (in general the value of 
$P$ minus its mean value and then divided by the standard deviation of $P$)
will be defined in this paper as $(P - 1)\sqrt{N}$, even for non-Gaussian noise.

\begin{figure}
  \begin{center}
  \includegraphics[height=6.5cm]{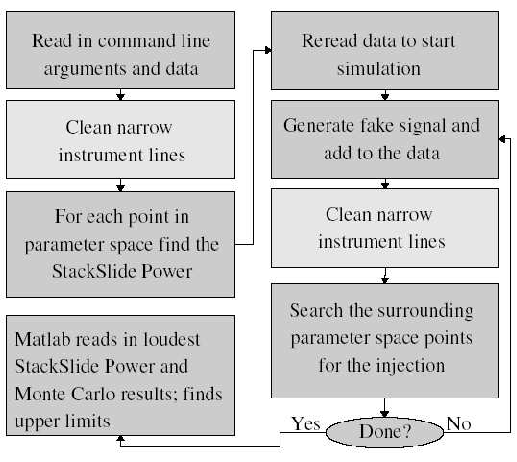}
  \caption{Flow chart for the pipeline used to find the upper limits presented in this paper
   using the StackSlide method.}
  \label{fig:StackSlideFlowChart}
  \end{center}
\end{figure}

The StackSlide code, which implements the method described above,
is part of the C-based LSC Algorithms Library Applications (LALapps) stored in the lscsoft CVS
repository \cite{LAL}. The code is run 
in a pipeline with options set to produce the results from a search and from
Monte Carlo simulations. Parallel jobs are run on computer clusters within the LSC, in the
Condor environment \cite{condor}, and the final post processing steps are performed using Matlab
\cite{matlab}. The specific StackSlide pipeline used to find the upper limits presented in this
paper is shown in Fig.~\ref{fig:StackSlideFlowChart}. The first three boxes on the left side
of the pipeline can also be used to output candidates for follow-up searches.

A separate search was run for each successive $0.25$~Hz band within $50-1000$~Hz.
The spacing in frequency used is given by Eq.~(\ref{eq:deltaf}).
The spacing in $\dot{f}$ was chosen as that which changes the frequency by one SFT frequency bin during the observation
time $T_{\rm obs}$, i.e., so that $\dot{f} T_{\rm obs} = \delta f$. For
simplicity $\Tobs = 2.778 \times 10^{6}$ seconds $\simeq 32.15$ days was chosen, which is greater
than or equal to $\Tobs$ for each interferometer. Thus, the $\dot{f}$ part of the parameter
space was over-covered by choosing 
\begin{equation} \label{eq:deltafdot}
|\delta \dot{f}| = {\delta f \over \Tobs} = {1 \over \Tcoh \Tobs} =  2 \times 10^{-10} \, 
\mathrm{Hz}~\mathrm{s}^{-1} \,.
\end{equation}
Values of $\dot{f}$ in the range
$[-1 \times 10^{-8}~\mathrm{Hz}~\mathrm{s}^{-1}, 0~\mathrm{Hz}~\mathrm{s}^{-1}]$
were searched.
This range corresponds to a search over $51$ values of $\dot{f}$, which is
the same as PowerFlux used in its low-frequency search
(discussed in Section.~\ref{subsec:powerflux}).  

The sky grid used is similar to that used for the all-sky search in \cite{S2FstatPaper}, but
with a spacing between sky-grid points appropriate for the StackSlide search.
This grid is isotropic on the celestial sphere, with an angular spacing between points chosen
for the $50$-$225$ Hz band, such that the maximum change in Doppler shift from one sky grid point
to the next would shift the frequency by half a bin. This is given by
\begin{equation} \label{eq:deltatheta}
\delta \theta_0 = {0.5\, c\, \delta f \over \hat{f} (v \,{\rm sin}\theta)_{\rm max}}
= 9.3 \times 10^{-3} \, {\rm rad} \left ( {300 {\rm Hz} \over \hat{f}} \right ) \,,
\end{equation}
where $v$ is the magnitude of the velocity $\mathbf{v}$ of the detector
in the SSB frame, and $\theta$ is the angle between $\mathbf{v}$ and
the unit-vector $\mathbf{\hat{n}}$ giving the sky-position of the source.  
Equations~(\ref{eq:deltafdot}) and (\ref{eq:deltatheta}) are the same as Eqs.~(19) and (22) 
in \cite{S2HoughPaper}, which represent conservative choices that over-cover
the parameter space. Thus, the parameter space used
here corresponds to that in Ref.~\cite{S2HoughPaper}, adjusted
to the S4 observation time, and with the exception that a stereographic projection of the
sky is not used. Rather an isotropic sky grid is used like the one used
in \cite{S2FstatPaper}.

One difficulty is that the computational cost of the search increases quadratically with
frequency, due to the increasing number of points on the sky grid. To reduce the computational
time, the sky grid spacing given in Eq.~(\ref{eq:deltatheta}) was increased by a factor
of $5$ above $225$~Hz. This represents a savings of a factor of $25$ in computational cost.
It was shown through a series of simulations, comparing the upper limits in various frequency
bands with and without the factor of 5 increase in grid spacing, that this changes the upper
limits on average by less than than $0.3\%$, with a standard deviation of $2\%$.
Thus, this factor of $5$ increase was used to allow the searches in the $225-1000$~Hz band
to complete in a reasonable amount of time.

It is not surprising that the sky grid spacing can be increased, 
for at least three reasons.
First, the value for $\delta \theta_0 $ given
in Eq.~(\ref{eq:deltatheta}) applies to only a small annular region on the sky, and is
smaller than the average change. Second,
only the net change in Doppler shift during the observation time is important, which is less than the
maximum Doppler shift due to the Earth's orbital motion during a one month run. (If the Doppler shift were
constant during the entire observation time, one would not need to search sky positions even if
the Doppler shift varied across the sky. A source frequency would be shifted by a constant amount
during the observation, and would be detected, albeit in a frequency bin different from that at the SSB.)
Third, because of correlations on the sky, one can detect a signal
with negligible loss of SNR much farther from its sky location than the spacing above suggests. 

\subsubsection{Line cleaning}
\label{subsubsec:stackslidelinecleaning}

\begin{figure}
  \begin{center}
  \includegraphics[height=6.5cm]{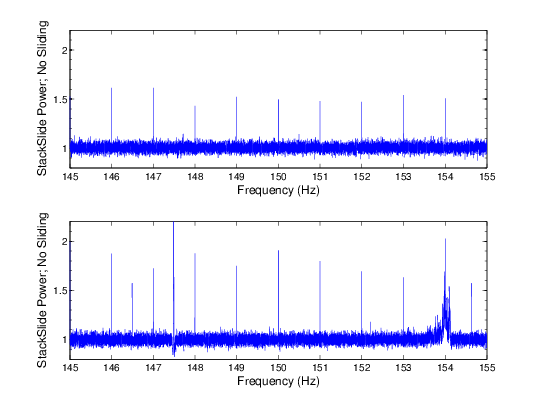}
  \caption{The StackSlide Power for the $145-155$~Hz band with no sliding.
  Harmonics of 1 Hz instrumental lines are clearly seen in H1 (top) and L1 (bottom).
  These lines are removed from the data by the StackSlide and Hough searches
  using the method described in the text, while PowerFlux search tracks these lines
  and avoids them when setting upper limits.}
  \label{fig:S4H1StackSlidePowerOneHzLines145to155Hz}
  \end{center}
\end{figure}

\begin{figure}
  \begin{center}
  \includegraphics[height=6.5cm]{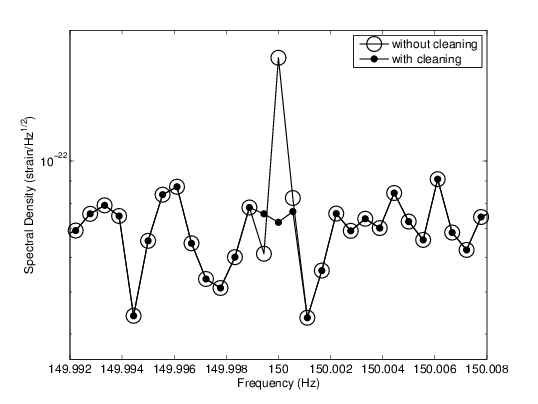}
  \caption{The L1 amplitude spectral density in a narrow frequency band estimated from 10 SFTs before and after
  the line cleaning used by the StackSlide pipeline. In the band shown, the $150$~Hz bin, and one
  bin either side of this bin have been replaced with estimates of the noise based on neighboring
  bins.}
  \label{fig:S4L1StackSlideWithoutAndWithCleaning150Hz}
  \end{center}
\end{figure}

\begin{table}
\begin{tabular}{ccccccc}\hline
IFO & $f_{\rm start}$ & $f_{\rm step}$ & Num. & $\Delta f_{\rm left}$ & $\Delta f_{\rm right}$ & Description \\
    & Hz              & Hz             &      & Hz                    & Hz                     &             \\
\hline \hline
H1  & 46.7            & ---            & 1    & 0.0                   & 0.0                    & Cal. Line   \\
H1  & 393.1           & ---            & 1    & 0.0                   & 0.0                    & Cal. Line   \\
H1  & 973.3           & ---            & 1    & 0.0                   & 0.0                    & Cal. Line   \\
H1  & 1144.3          & ---            & 1    & 0.0                   & 0.0                    & Cal. Line   \\
H1  & 0.0             & 1.0            & 1500 & 0.0006                & 0.0006                 & 1 Hz Comb   \\
\hline
L1  & 54.7            & ---            & 1    & 0.0                   & 0.0                    & Cal. Line   \\
L1  & 396.7           & ---            & 1    & 0.0                   & 0.0                    & Cal. Line   \\
L1  & 1151.5          & ---            & 1    & 0.0                   & 0.0                    & Cal. Line   \\
L1  & 0.0             & 1.0            & 1500 & 0.0006                & 0.0006                 & 1 Hz Comb   \\
\hline
\end{tabular}
\caption{Instrumental lines cleaned during the StackSlide search. The frequencies cleaned are found
by starting with that given in the first column, and then taking steps in frequency
given in the second column, repeating this the number of times shown in the third column; the fourth and
fifth columns show how many additional Hz are cleaned to the immediate left and right of each line. }
\label{tab:StackSlideCleanedLines}
\end{table}

Coherent instrumental lines exist in the data which can mimic a continuous gravitational-wave
signal for parameter space points that correspond to little Doppler modulation.
Very narrow instrumental lines are removed (``cleaned'') from the data.
In the StackSlide search, a line is considered ``narrow'' if 
its full width is less than $5\%$ of the $0.25$~Hz band, or less than $0.0125$~Hz. 
The line must also have been identified {\it a priori} as a known instrument artifact.
Known lines with less than this width were cleaned
by replacing the contents of bins corresponding to lines with random values generated by 
using the running median to find the mean power using
101 bins from either side of the lines. This method is also used to estimate the noise,
as described in Section~\ref{subsec:runningmedian}. 

It was found when characterizing the data that a comb
of narrow $1$~Hz harmonics existed in the H1 and L1 data, as
shown in Fig.~\ref{fig:S4H1StackSlidePowerOneHzLines145to155Hz}. 
Table~\ref{tab:StackSlideCleanedLines} shows the lines cleaned during
the StackSlide search. As the table shows, only this comb of narrow $1$~Hz harmonics and 
injected lines used for calibration were removed.
As an example of the cleaning process, Fig.~\ref{fig:S4L1StackSlideWithoutAndWithCleaning150Hz} shows
the amplitude spectral density estimated from $10$ SFTs before and after line cleaning, for the
band with the $1$~Hz line at $150$~Hz.

\begin{table}
\begin{tabular}{cc}\hline
Excluded Bands                     & Description           \\
  Hz                               &                       \\
\hline \hline
$[57, 63)$                         & Power lines           \\
$[n60-1,n60 + 1)$ $n = 2$ to $16$  & Power line harmonics  \\
$[340, 350)$                       & Violin modes          \\
$[685, 690)$                       & Violin mode harmonics \\
$[693, 696)$                       & Violin mode harmonics \\  
\hline
\end{tabular}
\caption{Frequency bands excluded from the StackSlide search.}
\label{tab:StackSlideExcludedBands}
\end{table}

The cleaning of very narrow lines has a negligible effect on the efficiency to detect
signals. Very broad lines, on the other hand, cannot be handled in this way.
Bands with very broad lines were searched without any line cleaning. There were also
a number of highly disturbed bands, dominated either by the harmonics of $60$~Hz power
lines or by the violin modes of the suspended optics, that were excluded from the
StackSlide results. (Violin modes refer to resonant excitations of the steel wires that
support the interferometer mirrors.) These are shown in Table~\ref{tab:StackSlideExcludedBands}.
While these bands can be covered by adjusting the parameters used to find outliers and set upper
limits, we will wait for future runs to do this.

\subsubsection{Upper limits method}
\label{subsubsec:stackslideulmethod}

After the lines are cleaned, the powers in the SFTs are normalized and the parameter
space searched, with each template producing a value of the StackSlide Power, defined
in Eq.~(\ref{eq:stackslidepower}).  For this paper, 
only the ``loudest'' StackSlide Power is kept, resulting in a value $P_{\rm max}$
for each $0.25$~Hz band, and these are used to set upper limits on the gravitational-wave
amplitude, $h_0$. (The loudest coincident outliers are also identified, but none survive
as candidates after follow-up studies described in Sec.~\ref{subsubsec:stackslidelpsandoutliers}.)
The upper limits are found by a series of Monte Carlo simulations,
in which signals are injected in software with
a fixed value for $h_0$, but with otherwise randomly chosen parameters, and
the parameter space points that surround the injection are searched. The number of times the loudest
StackSlide Power found during the Monte Carlo simulations is greater than or equal to $P_{\rm max}$
is recorded, and this is repeated for a series of $h_0$ values. The $95\%$ confidence upper limit
is defined to be the value of $h_0$ that results in a detected StackSlide Power greater than
or equal to $P_{\rm max}$ $95\%$ of the time. As shown in Fig.~\ref{fig:StackSlideFlowChart}, the
line cleaning described above is done after each injection is added to
the input data, which folds any loss of detection efficiency due to line cleaning
into the upper limits self-consistently.

\begin{figure}
  \begin{center}
  \includegraphics[height=6.5cm]{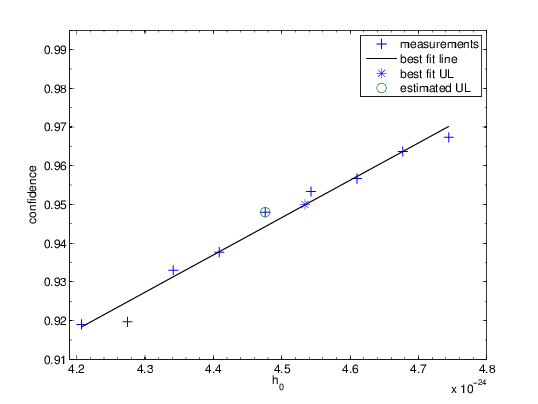}
  \caption{Measure confidence vs. $h_0$ for an example band ($140-140.25$~Hz in H1).
  A best-fit straight line is used to find the value of $h_0$ corresponding
  to $95\%$ confidence and to estimate the uncertainties in the results (see text).}
  \label{fig:S4H1Exampleh0vsconf}
  \end{center}
\end{figure}

Figure~\ref{fig:S4H1Exampleh0vsconf} shows the
measured confidence versus $h_0$ for an example frequency band. 
The upper limit finding process involves first making an
initial guess of its value, then refining this guess using a single set of injections
to find an estimate of the upper limit, and finally using this estimate to run
several sets of injections to find the final value of the upper limit.  
These steps are now described in detail.

To start the upper limit finding process, first an initial guess, $h_0^{\rm guess}$, is
used as the gravitational-wave amplitude. 
The initial guess need not be near the sought-after upper limit, just 
sufficiently large, as explained below.
A single set of $n$ injections is done (specifically $n=3000$
was used) with random sky positions and isotropically distributed spin axes,
but all with amplitude $h_0^{\rm guess}$. The output list of StackSlide Powers
from this set of injections is sorted in ascending order and the $0.05n$'th (specifically
for $n=3000$ the $150$th) smallest value of the StackSlide Power is found, which we call $P_{0.05}$,  
Note that the goal is to find the value of $h_0$ that makes $P_{0.05} = P_{\rm max}$, so
that $95\%$ of the output powers are greater than the maximum power found during the search.
This is what we call the $95\%$ confidence upper limit. Of course, in general $P_{0.05}$ will not
equal $P_{\rm max}$ unless our first guess was very lucky. However,
as per the discussion concerning Eq.~(\ref{eq:eststackslidepwr}),
$P - 1$ is proportional to $h_0^2$ (i.e, removing the mean value due to noise leaves on average
the power due to the presence of a signal). Thus, an estimate of the $95\%$ $h_0$ confidence
upper limits is given by the following rescaling of $h_0^{\rm guess}$,
\begin{equation} \label{eq:StackSlideEstUL}
h_0^{\rm est} = { \sqrt{P_{\rm max} - 1} \over \sqrt{P_{0.05} - 1} } h_0^{\rm guess} \,.
\end{equation}
Thus an estimated upper limit, $h_0^{\rm est}$, is found from
a single set of injections with amplitude $h_0^{\rm guess}$; the only requirement is
that $h_0^{\rm guess}$ is chosen loud enough to make $P_{0.05} > 1$.

It is found that using Eq.~(\ref{eq:StackSlideEstUL})
results in a estimate of the upper limit that is typically within $10\%$ of the 
final value. For example, the estimated upper limit found in this way is indicated by the circled
point in Fig.~\ref{fig:S4H1Exampleh0vsconf}. The value of $h_0^{\rm est}$ then
becomes the first value for $h_0$ in a series of Monte Carlo simulations, each with $3000$
injections, which use this value and $8$ neighboring values, measuring the confidence
each time. The Matlab~\cite{matlab} polyfit and polyval functions are then
used to find the best-fit straight line to determine the value of $h_0$
corresponding to $95\%$ confidence and to estimate the uncertainties in the results. This is the final
step of the pipeline shown in Fig.~\ref{fig:StackSlideFlowChart}.

\subsection{The Hough Transform Implementation}
\label{subsec:hough}

\subsubsection{Description of Algorithm}

The Hough transform is a general method for pattern recognition, invented
originally to analyze bubble chamber pictures
from CERN \cite{hough1,hough2}; it has found many applications in the
analysis of digital images \cite{ik}.  This method has already been
used to analyze data from the second science run (S2) of the LIGO
detectors \cite{S2HoughPaper} and a detailed description can be found
in \cite{hough04}.  Here we present only a brief description,
emphasizing the differences between the previous S2 search and the
S4 search described here.

The Hough search uses a weighted sum of the binary counts as its final statistic, 
as given by Eqs.~(\ref{eq:1}) and (\ref{eq:3}).
In the standard Hough search as presented in \cite{hough04,S2HoughPaper},
the weights are all set to unity.  The weighted Hough
transform was originally discussed in \cite{Palomba2005}.  The
software for performing the Hough transform has been adapted to use
arbitrary weights without any significant loss in computational
efficiency. Furthermore, the robustness of the Hough transform method
in the presence of strong transient disturbances is not compromised by
using weights because each SFT contributes at most $w_i$ (which is of
order unity) to the final number count.

The following statements can be proven using the methods of
\cite{hough04}.  The mean number count in the absence of a signal is
$\bar{n} = Np$, where $N$ is the number of SFTs and
$p$ is the probability that the normalized power, of a given
frequency bin and SFT defined by 
Eq.~(\ref{eq:normpower}), exceeds a threshold $\rho_\th$, i.e., $p$ is the
probability that a frequency bin is selected in the absence of a
signal.
For unity weighting, the standard
deviation is simply $\sigma = \sqrt{Np(1-p)}$.  However, with
more general weighting, it can be shown that $\sigma$
is given by
\begin{equation}
  \label{eq:4}
  \sigma = \sqrt{||\vec{w}||^2p(1-p)}\,,
\end{equation}
where $||\vec w||^2 = \sum_{i=0}^{N-1}w_i^2$.
A threshold $n_{\th}$ on the number count
corresponding to a false alarm rate $\alpha_\H$ is given by
\begin{equation}
  \label{eq:nth}
  n_\th = Np + \sqrt{2||\vec
    w||^2p(1-p)}\,\textrm{erfc}^{-1}(2\alpha_\H) \,.
\end{equation}
 Therefore $n_{\th}$  depends on the weights of the corresponding template
 $\lambda$. In this case, the natural detection statistic is not the 
 \lq\lq Hough Number
 Count" $n$, but the \emph{significance} of a number count, defined by
\begin{equation}
  \label{eq:5}
  s = \frac{n - \bar{n}}{\sigma}\,,
\end{equation}
where $\bar{n}$ and $\sigma$ are the expected mean
and standard deviation for pure noise. 
Values of $s$ can be compared directly across different templates
characterized by differing weight distributions.

  The threshold
$\rho_{\th}$ (c.f. Eq.~\ref{eq:1}) is selected to give the minimum false dismissal probability
$\beta_\H$ for a given false alarm rate.  In \cite{S2HoughPaper} it was
shown that the optimal choice for
$\rho_{\th}$ is $1.6$ which correspond to a peak selection
probability $p = e^{-\rho_{\th}} \approx 0.2$.  It can be shown that
the optimal choice is unchanged by the weights and hence
$\rho_{\th} = 1.6$ is used once more~\cite{badrisintes}.

Consider a population of sources located at a given point in the sky,
but having uniformly distributed spin axis directions. For a template that is perfectly matched in
frequency, spin-down, and sky-position, and given the optimal peak
selection threshold, it can be shown \cite{badrisintes}
that the weakest signal that can
cross the threshold $n_\th$ with a false dismissal probability $\beta_\H$
has an amplitude
\begin{equation}
  \label{eq:hough_h0}
  h_0 = 3.38\> \S^{1/2} \left( \frac{||\vec{w}||}{\vec{w}\cdot
      \vec{X}}\right)^{1/2}\sqrt{\frac{1}{\Tcoh}} ,
\end{equation}
where 
\begin{eqnarray}
  \mathcal{S}&=&\textrm{erfc}^{-1}(2\alpha_\H)+
  \textrm{erfc}^{-1}(2\beta_\H)\,,  \label{eq:sdef} \\
  X_i &=& \frac{1}{S^\srchTemplateInd_{\iSubSupInd}}\left\{ \left(F_{+}^{\iSubSupInd}\right)^2 +
    \left(F_{\times}^{\iSubSupInd}\right)^2\right\} \,. \label{eq:optimalweights}
\end{eqnarray}
As before, $F_{+}^{\iSubSupInd}$ and $F_{\times}^{\iSubSupInd}$ are the values
of the beam pattern functions at the mid-point of the $i^{th}$ SFT.
To derive \eqref{eq:hough_h0} we have assumed that the number of SFTs
$N$ is sufficiently large and that the signal is weak \cite{hough04}.  

From \eqref{eq:hough_h0} it is clear that the scaling of the weights
does not matter; $w_i\rightarrow k w_i$ leaves $h_0$ unchanged for any
constant $k$.  More importantly, it is also clear that the sensitivity
is best, i.e.\ $h_0$ is minimum, when $\vec{w}\cdot\vec{X}$ is maximum:
\begin{equation}
  w_i \propto X_i  \,.
\end{equation}
This result is equivalent to Eq.~\eqref{eq:wipropto}.  

In addition to improving sensitivity in single-interferometer analysis,
the weighted Hough method allows automatic optimal combination of Hough counts from 
multiple interferometers of differing senstivities.

\begin{figure}
  \includegraphics[width=7cm]{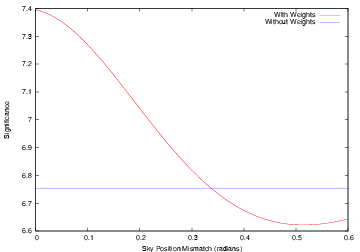}
  \caption{The improvement in the significance as a function
    of the mismatch in the sky-position.  A signal is injected in fake
    noise at $\alpha=\delta=0$ and the weights are calculated at
    $\alpha=\delta=\delta\theta$.  The curve is the observed
    significance as a function of $\delta\theta$ while the horizontal line
    is the observed significance when no weights are used.  See main
    text for more details.  }\label{fig:houghmismatch}
\end{figure}

Ideally, to obtain the maximum increase in sensitivity, we should
calculate the weights for each sky-location separately.  In practice,
we break up the sky into smaller patches and calculate one weight for
each sky-patch center. The gain from using the
weights will be reduced if the sky patches are too large.  From
equation (\ref{eq:optimalweights}), it is clear that the dependence of
the weights on the sky-position is only through the beam pattern
functions.  Therefore, the sky patch size is determined by the typical
angular scale over which $F_+$ and $F_\times$ vary; thus for a
spherical detector using the beam pattern weights would not gain us
any sensitivity.  For the LIGO interferometers, we have investigated
this issue with Monte-Carlo simulations using random Gaussian noise.
Signals are injected in this noise
corresponding to the H1 interferometer at a sky-location
$(\alpha_0,\delta_0)$, while the weights are calculated at a
mismatched sky-position $(\alpha_0+\delta\theta,
\delta_0+\delta\theta)$.    The significance values are
compared with the significance when no weights are used.  An example
of such a study is shown in Fig.~\ref{fig:houghmismatch}.  Here, we
have injected a signal at $\alpha = \delta = 0$, $\cos\iota = 0.5$,
zero spin-down, $\Phi_0 = \psi = 0$, and a signal to noise ratio
corresponding approximately to a $6$-$\sigma$ level without weights.
The figure shows a gain of $\sim 10\%$ at $\delta\theta=0$, decreasing
to zero at $\delta\theta \approx 0.3\,$rad.  We get
qualitatively similar results for other sky-locations,
independent of frequency and other parameters.  There is an additional
gain due to the non-stationarity of the noise itself, which depends, however,
on the quality of the data.  In practice, we have chosen to
break the sky up into 92 rectangular patches in which the average
sky patch size is about $0.4\,$rad wide, corresponding to a maximum sky position
mismatch of $\delta\theta=0.2\,$rad in Fig.~\ref{fig:houghmismatch}.

\subsubsection{The Hough Pipeline}

The Hough analysis pipeline for the search and for setting upper limits
follows roughly the same scheme as in \cite{S2HoughPaper}.
In this section we present a short description of
the pipeline, mostly emphasizing the differences from \cite{S2HoughPaper} and
from the StackSlide and PowerFlux searches.  
As discussed in the previous subsection, the key differences from
the S2 analysis \cite{S2HoughPaper} are (i) using the beam-pattern and
noise weights, and (ii) using SFTs from multiple interferometers.  

The total frequency range analyzed is 50-1000 Hz,
with a resolution $\delta f = 1/\Tcoh$ as in \eqref{eq:deltaf}.  The
resolution in $\dot{f}$ is $\sci{2.2}{-10}~\mathrm{Hz}~\mathrm{s}^{-1}$ given in
\eqref{eq:deltafdot}, and the reference time for defining the spin-down
is the start-time of the observation.  However, unlike StackSlide and
PowerFlux, the Hough search is carried out over only 11 values of $\dot{f}$,
including zero, in the range [$\sci{-2.2}{-9}~\mathrm{Hz}~\mathrm{s}^{-1}$,
 $0~\mathrm{Hz}~\mathrm{s}^{-1}$].  
This choice is driven by the technical design of the current implementation,
 which uses look-up-tables and partial Hough maps as in \cite{S2HoughPaper}.
 This implementation of the Hough algorithm is efficient when analyzing all resolvable points
 in $\dot{f}$, as given in \eqref{eq:deltafdot}, but this approach is incompatible
 with the larger $\dot f$ step sizes used in the other search methods, which permit
 those searches to search a larger $\dot f$ range for comparable computational cost.
  
 The sky resolution  is similar to that used by the StackSlide method for
$f < 225~{\rm Hz}$ as given by \eqref{eq:deltatheta}.  At frequencies
higher than this, the StackSlide sky-resolution is 5 times coarser, thus the Hough search
is analyzing about 25 more templates at a given frequency and spin-down value. 
In each of the 92 sky patches, by means of the stereographic projection, the
sky patch is mapped to a two dimensional plane with a uniform grid of that resolution
$\delta\theta_0$. Sky Patches slightly overlap to avoid gaps among them
(see \cite{S2HoughPaper} for further details).

\begin{figure}
  \begin{center}
  \includegraphics[height=6.5cm]{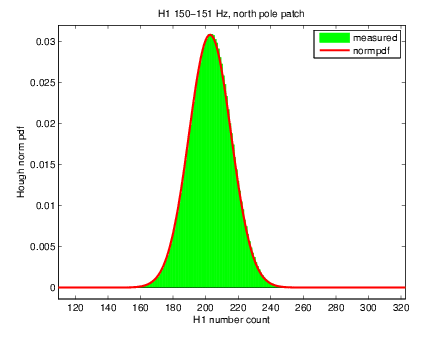}
  \includegraphics[height=6.5cm]{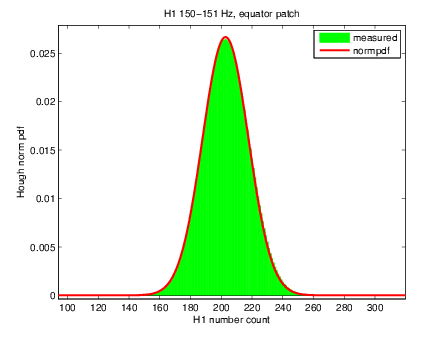}
  \caption{Two example histograms of the normalized Hough number count compared to a 
  Gaussian
  distribution for the H1 detector in the frequency band 150-151 Hz. 
  The upper figure
  corresponds to a  a patch located at the north pole for the case in which 
  the weights are used.
  The number of templates analyzed in this 1Hz band is of $11 \times 10^6$, 
  the number of SFTs 1004, the corresponding mean $\bar{n} = 202.7$ and $\sigma=12.94$
  is obtained from the weights.
  The lower figure corresponds to a patch at the equator using the same data.
  In this case the  number of templates analyzed in this 1Hz band is of $10.5 \times 10^6$, 
  and its corresponding $\sigma = 14.96$.}
  \label{fig:H1histo150}
  \end{center}
\end{figure}

Figure~\ref{fig:H1histo150} shows
examples of histograms of the number counts in two particular sky patches
for the  H1 detector in the 150-151 Hz band. In all the bands free of
instrumental disturbances, the Hough number count distributions follows the
expected theoretical distribution, which can be approximated by a Gaussian
distribution. Since the number of SFTs for H1 is 1004, the corresponding 
mean $\bar{n} = 202.7$ and the standard deviation is given by Eq.~(\ref{eq:4}).
The standard deviation is computed from the weights $\vec{w}$ and 
varies among different
sky patches because of varying antenna pattern functions.

\begin{table}
\begin{tabular}{ccccccc}\hline
IFO & $f_{\rm start}$ & $f_{\rm step}$ & $n$  & $\Delta f_{\rm left}$ & $\Delta f_{\rm right}$ & Description \\
    & Hz              & Hz             &      & Hz          & Hz                     &             \\
\hline \hline
H1  & 392.365    & ---   &   1     &    0.01     &        0.01   &  Cal. SideBand\\
H1  & 393.835    & ---   &   1     &    0.01     &        0.01   &  Cal. SideBand\\ 
\hline
H2  & 54.1           & ---            & 1    & 0.0                   & 0.0   & Cal. Line   \\
H2  & 407.3          & ---            & 1    & 0.0                   & 0.0   & Cal. Line   \\
H2  & 1159.7         & ---            & 1    & 0.0                   & 0.0   & Cal. Line   \\
H2  & 110.934        & 36.9787        & 4    & 0.02                  & 0.02  & 37 Hz Oscillator  \\
\hline
L1  & 154.6328       &8.1386          & 110  & 0.01                  & 0.01  & 8.14 Hz Comb \\
L1  & 0.0            & 36.8725        &  50  & 0.02                  & 0.02  & 37 Hz Oscillator (*) \\
\hline
\end{tabular}
\caption{Instrumental lines cleaned during the Hough search that were not 
listed in Table~\ref{tab:StackSlideCleanedLines} (see text).  
(*) These lines were removed 
only in the multi-interferometer search.}
\label{tab:HoughCleanedLines}
\end{table}

The upper limits on $h_0$ are derived from the \emph{loudest event}, 
registered over the entire sky and spin-down range 
in each $0.25\,$Hz band, not from the highest number count. As for the
StackSlide method, 
we use a frequentist method, where upper limits refer to a
hypothetical population of isolated spinning neutron stars which are uniformly
distributed in the sky and have a spin-down rate $\dot f$ uniformly distributed in
the range [$\sci{-2.2}{-9}~\mathrm{Hz}~\mathrm{s}^{-1}$,
 $0~\mathrm{Hz}~\mathrm{s}^{-1}$]. We also assume uniform distributions for the
parameters $\cos\iota \in [-1,1]$, $\psi\in [0,2\pi]$, and $\Phi_0 \in
[0,2\pi]$.  The strategy for calculating the 95$\%$ upper limits is 
roughly the same scheme as in \cite{S2HoughPaper}, except for the 
treatment of 
narrow instrumental lines.

Known spectral disturbances are removed from the
SFTs in the same way as for the StackSlide search.  The known
spectral lines are, of course, also consistently removed after each
signal injection when performing the Monte-Carlo simulations to
obtain the upper limits.   

The narrow instrumental lines ``cleaned'' from the SFT data are the same
ones cleaned during the StackSlide search shown in 
Table~\ref{tab:StackSlideCleanedLines}, together with ones listed in
Table~\ref{tab:HoughCleanedLines}. The additional lines
listed in Table~\ref{tab:HoughCleanedLines} are cleaned to prevent
large artifacts in one instrument from increasing the false alarm rate
of the Hough multi-interferometer search.
Note that the L1 36.8725~Hz comb was eliminated 
mid-way through the S4 run by replacing a synthesized radio frequency oscillator for phase
modulation with a crystal oscillator, and these lines were not removed in the 
Hough L1 single-interferometer analysis.

No frequency bands have been excluded from the Hough search, although the upper limits
reported on the bands shown in Table \ref{tab:StackSlideExcludedBands},
that are dominated by 60 Hz power line harmonics or violin modes of the suspended optics, did not always 
give satisfactory convergence to an upper limit.
In a few of these very noisy bands, upper limits were set by extrapolation, instead of interpolation,
of the Monte-Carlo simulations. Therefore the results reported on those bands have larger error bars.
No parameter tuning was performed on these disturbed bands to improve the 
upper limits.

\subsection{The Powerflux Implementation}
\label{subsec:powerflux}

\begin{figure}
  \begin{center}
  \includegraphics[height=10.0cm]{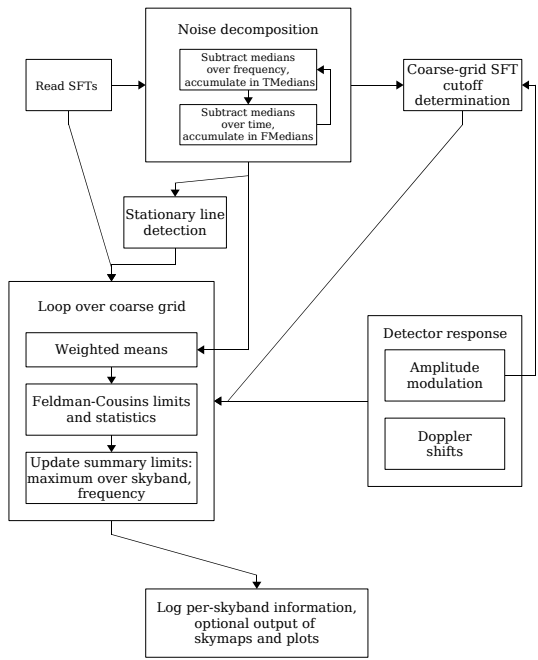}
  \caption{Flow chart for the pipeline used to find the upper limits presented in this paper
   using the PowerFlux method.}
  \label{fig:powerfluxflowchart} 
  \end{center}
\end{figure}

The PowerFlux method is a variant on the StackSlide method in which
the contributions from each SFT are weighted by the inverse square of
the average spectral power density in each band and weighted according
to the antenna pattern sensitivity of the interferometer for each point
searched on the sky. This weighting scheme has two advantages: 1) variance
on the signal strength estimator is minimized, improving signal-to-noise ratio;
and 2) the estimator is itself a direct measure of source strain power, allowing
direct parameter estimation and dramatically reducing dependence on Monte Carlo
simulations. Details of software usage and algorithms can be found in
a technical document~\cite{PowerFluxTechNote}. Figure~\ref{fig:powerfluxflowchart}
shows a flow chart of the algorithm, discussed in detail below. 

\subsubsection{Noise decomposition}
\label{subsubsec:noisedecomposition}
Noise estimation is carried out through a time/frequency noise decomposition
procedure in which the dominant variations are factorized within each nominal 0.25 Hz band
as a product of a spectral variation and a time variation across the data run.
Specifically, for each 0.25 Hz band, a matrix of logarithms of power measurements across the 
$0.56$~mHz SFT bins and across the SFT's of the run is created. Two vectors, denoted
TMedians and FMedians, are initially set to zero and then iteratively updated according
to the following algorithm:
\begin{enumerate}
\item For each SFT (row in matrix), the median value (logarithm of power) is computed
and then added to the corresponding element of TMedians while subtracted from each
matrix element in that row.
\item For each frequency bin (column in matrix), the median value is computed and
then added to the corresponding FMedians element, while subtracted from each
matrix element in that column.
\item The procedure repeats from step 1 until all medians computed in steps 1 and 2
are zero (or negligible).
\end{enumerate}
The above algorithm typically converges quickly. The size of the frequency band treated
increases with central frequency, as neighboring bins are included to allow for 
maximum and minimum Doppler shifts to be searched in the next step.

\begin{figure}
  \begin{center}
  \includegraphics[height=6.5cm]{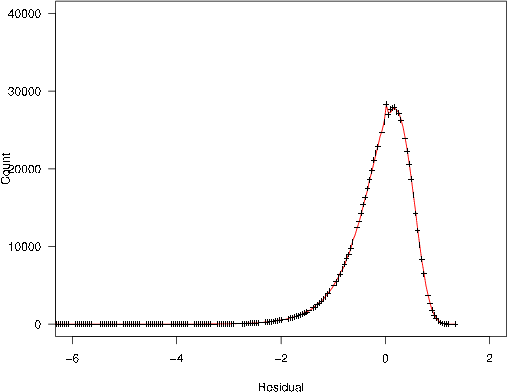}
  \caption{Typical residual logarithmic power following noise decomposition
           for a sample 0.25-Hz band of H1 data (crosses) near 575 Hz in a band containing an injected
pulsar. The residual is defined as the difference between a measured
power for a given frequency bin in a given 30-minute period and the value
predicted by the FMedians and TMedians vectors. The smooth curve is for
a simulation in Gaussian noise.}
  \label{fig:sampleresidual} 
  \end{center}
\end{figure}

For stationary, Gaussian noise and for noise that follows the above assumptions of
underlying factorized frequency and time dependence, the expected distribution of residual 
matrix values can be found from simulation. 
Figure~\ref{fig:sampleresidual} shows a sample expected residual power distribution
following noise decomposition for simulated stationary, Gaussian data, along with
a sample residual power distribution
from the S4 data (0.25-Hz band of H1 near 575 Hz, in this case) following noise decomposition.
The agreement in shape between these two distributions is very good and is typical
of the S4 data, despite sometimes large variations in the corresponding
TMedians and FMedians vectors, and despite, in this case, the presence
of a moderately strong simulated pulsar signal (Pulsar2 in Table~\ref{tab:ParametersHWInjections}). 

The residuals are examined for outliers. If the largest residual value is found
to lie above a threshold of 1.5, that corresponding 0.25 Hz band is flagged as
containing a ``wandering line'' because a strong but drifting instrumental line
can lead to such outliers. The value 1.5 is determined empirically from Gaussian
simulations.
An extremely strong pulsar could also be flagged in this
way, and indeed the strongest injected pulsars are labelled as wandering lines.
Hence in the search, the wandering lines are followed up, but no upper limits are
quoted here for the affected bands.

\subsubsection{Line flagging}
\label{subsubsec:lineflagging}

Sharp instrumental lines can prevent accurate noise estimation for 
pulsars that have detected frequencies in the same $0.56$~mHz bin as
the line. In addition, strong lines tend to degrade achievable sensitivity
by adding excess apparent power in an affected search. In early LIGO science runs,
including the S4 run, there have been sharp instrumental lines at multiples
of 1 Hz or 0.25 Hz, arising from artifacts in the data acquisition electronics.

To mitigate the most severe of these effects, the PowerFlux algorithm performs
a simple line detection and flagging algorithm. For each 0.25 Hz band, the 
detected summed powers are ranked and an estimated Gaussian sigma computed from
the difference in the 50\% and 94\% quantiles. Any bins with power greater than
5.0 $\sigma$ are marked for ignoring in subsequent processing. Specifically, when carrying
out a search for a pulsar of a nominal true frequency, its contribution to the
signal estimator is ignored when the detected frequency would lie in the same 
$0.56$~mHz bin as a detected line. As discussed below, for certain frequencies,
spin-downs and points in the sky, the fraction of time a putative pulsar has
a detected frequency in a bin containing an instrumental line can be quite large,
requiring care. The deliberate ignoring of contributing bins affected by sharp
instrumental lines does not lead to a bias in resulting limits, but it does
degrade sensitivity, from loss of data. In any 0.25 Hz band, no more than five
bins may be flagged as lines. Any band with more than five line candidates is
examined manually.

\subsubsection{Signal estimator}
\label{subsubsec:signalestimator}

Once the noise decomposition is complete, with estimates of the spectral noise
density for each SFT, the PowerFlux algorithm computes a weighted sum of the
strain powers, where the weighting takes into account the underlying time
and spectral variation contained in TMedians and FMedians and the antenna pattern
sensitivity for an assumed sky location and incident wave polarization.
Specifically, for an assumed polarization angle $\psi$ and sky location, 
the following quantity is defined 
for each bin $k$ of each SFT $i$:
\begin{equation}
Q_{\iSubSupInd} = \frac{P_{\iSubSupInd}}{ (F_\psi^{\iSubSupInd})^2},
\end{equation}
where $F_\psi^i$ is the $\psi$-dependent antenna pattern for the sky location,
defined in Eq.~(\ref{PFFpsi}). (See also Appendix~\ref{sec:polarization}.)

As in Sec.~\ref{subsec:basicdefsandnotation}, to simplify the notation we define
$Q^\srchTemplateInd_{\iSubSupInd} = P^\srchTemplateInd_{\iSubSupInd}
/ (F_\psi^{\iSubSupInd})^2$ as the value of $Q_{\iSubSupInd}$ for SFT $i$ 
and a given template ${\mathbf \lambda}$.

For each individual SFT bin power measurement $P_{\iSubSupInd}$, one expects an underlying
exponential distribution, with a standard deviation equal to the mean, a statement
that holds too for $Q_{\iSubSupInd}$. To minimize the variance of a signal estimator
based on a sum of these powers, each contribution is weighted by the inverse of
the expected variance of the contribution. Specifically, we compute the following
signal estimator:
\begin{eqnarray}
  R & = & {2 \over \Tcoh} \left(\sum_i\frac{1}{ (\bar{Q}^\srchTemplateInd_{\iSubSupInd})^{2} }\right)^{-1}
  \sum_i \frac{ Q^\srchTemplateInd_{\iSubSupInd} }{ (\bar{Q}^\srchTemplateInd_{\iSubSupInd})^2 },  \\
  & = & {2 \over \Tcoh} \left(\sum_i\frac{ [(F_\psi^{\iSubSupInd})^2]^2 }{ (\bar{P}^\srchTemplateInd_{\iSubSupInd})^2 }\right)^{-1}
  \sum_i \frac{(F_\psi^{\iSubSupInd})^2 P^\srchTemplateInd_{\iSubSupInd} }{ (\bar{P}^\srchTemplateInd_{\iSubSupInd})^2} ,
\end{eqnarray}
where $\bar{P}_\iSubSupInd$ and $\bar{Q}_\iSubSupInd$ are the expected uncorrected and
antenna-corrected powers of SFT $i$ averaged over frequency.  Since the antenna factor
is constant in this average, $\bar{Q}^\srchTemplateInd_{\iSubSupInd}
= \bar{P}^\srchTemplateInd_{\iSubSupInd}/(F_\psi^{\iSubSupInd})^2$.  Furthermore,
$\bar{P}^\srchTemplateInd_{\iSubSupInd}$ is a estimate of the power spectral
density of the noise. The replacement 
$\bar{P}^\srchTemplateInd_{\iSubSupInd} \cong S^\srchTemplateInd_{\iSubSupInd}$ gives
Eq.~(\ref{eq:PFRk}).

Note that for an SFT $i$ with low antenna pattern sensitivity $|F_\psi^{\iSubSupInd}|$, the
signal estimator receives a small contribution.  Similarly, SFT's $i$ for which
ambient noise is high receive small contributions. Because computational time
in the search grows linearly with the number of SFT's and because of large time variations
in noise, 
it proves efficient to ignore 
SFT's with sky-dependent and polarization-dependent effective noise higher
than a cutoff value. The cutoff procedure saves significant computing time,
with negligible effect on search performance. 

Specifically, the cutoff is computed as follows. Let $\sigma_j$ be the
{\it ordered} estimated standard deviations in noise, taken to be
the ordered means of 
$\bar Q_{\iSubSupInd}= {1\over k_{\rm max}}\Sigma_k \bar{Q}^{\iSubSupInd}_k$,
where $k_{\rm max}$ is the number of frequency bins used in the search template.
Define $j_{\rm opt}$ to be the index $j_{\rm max}$ for which
the quantity ${1\over j_{\rm max}}\sqrt{\Sigma_{j=1}^{j_{\rm max}}\sigma_j^2}$ is
minimized. Only SFT's for which $\sigma_j<2\sigma_{j_{\rm opt}}$ are used
for signal estimation. In words, $j_{\rm opt}$ defines the last SFT that
improves rather than degrades signal estimator variance in an unweighted mean.
For the weighted mean used here, the effective noise contributions are allowed
to be as high as twice the value found for $j_{\rm opt}$. The choice of
$2\sigma_{j_{\rm opt}}$ is determined empirically.

The PowerFlux search sets strict, frequentist, all-sky 95\% confidence-level upper limits
on the flux of gravitational radiation bathing the Earth. To be conservative
in the strict limits, numerical corrections to the signal estimator are applied:
1) a factor of $1/\cos(\pi/8)=1.082$ for maximum linear polarization
mismatch, based on twice the maximum half-angle of mismatch 
(see Appendix~\ref{sec:polarization}) and
2) a factor of $1.22$ for bin-centered signal power loss due to Hann
windowing (applied during SFT generation); and
3) a factor of $1.19$ for drift of detected signal frequency across
the width of the $0.56$~mHz bins used in the SFT's. 
Note that the use of rectangular
windowing would eliminate the need for correction 2) above, but would require a larger
correction of $1.57$ for 3) 

\begin{figure}
  \begin{center}
  \includegraphics[height=8.5cm, angle=270]{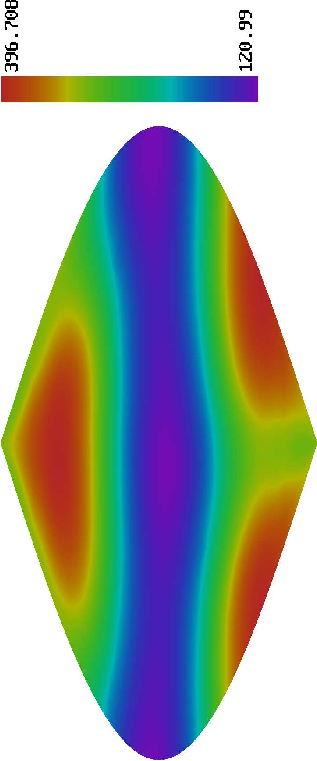}
  \caption{Sky map of run-summed PowerFlux weights for a 0.25-Hz band near
575 Hz for one choice of linear polarization
in the S4 H1 data. The normalization corresponds roughly to the effective
number of median-noise SFT's contributing to the sum.}
  \label{fig:weightedskymap} 
  \end{center}
\end{figure}

Antenna pattern and noise weighting in the PowerFlux
method allows weaker sources to be detected in certain regions of the sky, where 
run-averaged antenna patterns discriminate in declination and diurnal noise 
variations discriminate in right ascension.
Figure~\ref{fig:weightedskymap} illustrates
the resulting variation in effective noise across the sky for a 0.25-Hz H1 band near 575 Hz for
the circular polarization projection. By separately
examining SNR, one may hope to detect a signal in a sensitive region of the sky with 
a strain significantly lower than suggested by the strict worst-case all-sky frequentist limits
presented here, as discussed below in section~\ref{subsec:powerfluxmcvalidation}.
Searches are carried out for four linear polarizations, ranging over polarization angle from $\psi=0$ to 
$\psi={3\over8}\pi$ in steps of $\pi/8$ and for (unique) circular polarization.

A useful computational savings comes from defining two different sky resolutions.
A ``coarse'' sky gridding is used for setting the cutoff value defined above,
while fine grid points are used for both frequency and amplitude demodulation. 
A typical ratio of number of coarse grid points to number of fine grid points used for
Doppler corrections is 25. 

\subsubsection{Sky banding}
\label{subsubsec:skybanding}

Stationary and near-stationary instrumental spectral lines can be mistaken
for a periodic source of gravitational radiation if the nominal source parameters
are consistent with small variation in detected frequency during the time of
observation. The variation in the frequency at the detector can be found
by taking the time derivative of Eq.~(\ref{eq:freqatdetector}), which gives,
\begin{equation}\label{eq:dfdtatdetector}
{df \over dt} = \left (1 + \frac{ {\bf v} (t)\cdot\bf{\hat{n}}}{c} \right ) \dot{f}
+ \hat{f}(t)\frac{ {\bf a} (t)\cdot\bf{\hat{n}}}{c} .
\end{equation}
The detector's acceleration, ${\bf a}$ in this equation is dominated by the
Earth's orbital acceleration ${\bf a}_{\rm Earth}$,
since the diurnal part of the detector's acceleration is small and
approximately averages to zero during the observation.
Thus, it should be emphasized that a single instrumental line can mimic
sources with a range of slightly {\it different} frequencies and assumed different
positions in the sky that lie in an annular band. For a source $\dot{f}$ assumed to be
zero, the center of the band is defined by a circle 90 degrees away from the direction of
the average acceleration of the Earth during the run
where $\bar{{\bf a}}_{\rm Earth} \cdot\bf{\hat{n}} = 0$,
{\it i.e.,} toward the average
direction of the Sun during the run. For source spin-downs different from zero, there
can be a cancellation between assumed spin-down (or spinup) that is largely cancelled
by the Earth's average acceleration, leading to a shift of the annular region of
apparent Doppler stationarity toward (away from) the Sun.

A figure of merit found to be useful for discriminating regions of ``good'' sky
from ``bad'' sky (apparent detected frequency is highly stationary) is the
``$S$ parameter'':
\begin{equation}\label{eq:sparam}
S \quad = \quad \fdot + [({\mathbf \Omega} \times {\bf v}_{\rm Earth}/c)\cdot {\bf \hat{n}})]\hat{f}_0 ,
\end{equation}
where ${\bf \Omega}$ is the Earth's angular velocity vector about the solar system barycenter.
The term ${\bf \Omega} \times {\bf v}_{\rm Earth}$ is a measure of the
Earth's average acceleration during the run, where ${\bf v}_{\rm Earth}$ is taken to be 
the noise-weighted velocity of the H1 detector during the run.
Regions of sky with small $|S|$ for a given $\hat f$ and $\fdot$
have stationary detected frequency. As discussed below in section~\ref{subsec:powerfluxmcvalidation},
such regions are not only prone to high false-alarm rates, but the line flagging procedure
described in section~\ref{subsubsec:lineflagging} leads to systematically underestimated
signal strength and invalid upper limits. Hence limits are presented here for only sources
with $|S|$ greater than a threshold value denoted $S_{\rm large}$. The minimum acceptable value 
chosen for $S_{large}$ is found from software signal injections to be 
$\sci{1.1}{-9}~\mathrm{Hz}~\mathrm{s}^{-1}$ 
for the 1-month S4 run and can be understood to be
\begin{equation}\label{eq:slarge}
S_{\rm large} \quad = \quad {N_{\rm occupied\>\>bins}\over T_{obs}\cdot T_{coh}},
\end{equation}
where $N_{\rm occupied\>\>bins}\sim5$ is the minimum total number of $0.56$~mHz detection bins occupied by the 
source during the data run for reliable detection. In practice, we use 
still larger values for the H1 interferometer 
($S_{large}=\sci{1.85}{-9}~\mathrm{Hz}~\mathrm{s}^{-1}$)
and L1 interferometer ($S_{large}=\sci{3.08}{-9}~\mathrm{Hz}~\mathrm{s}^{-1}$) during the S4 run for the
limits presented here because of a pervasive and strong comb of precise 1-Hz lines in both
interferometers. These lines,
caused by a GPS-second synchronized electronic disturbance and worse in L1 than in H1, 
lead to high false-alarm rates 
from that data for lower values of $S_{large}$. For the frequency and spin-down ranges 
searched in this analysis, the average fractions of sky lost to the skyband veto are
15\%\ for H1 and 26\%\ for L1. 

\begin{figure}
  \begin{center}
  \includegraphics[height=9.0cm]{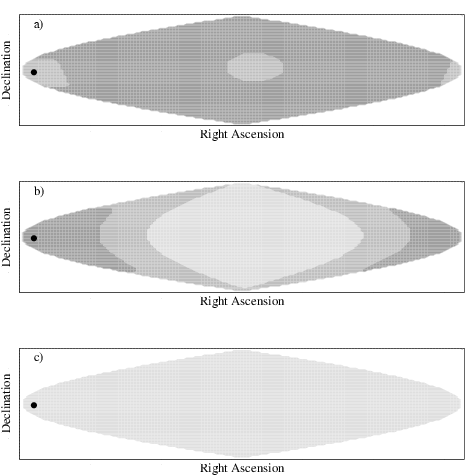}
  \caption{S4 sky band regions (good - light gray, bad for L1 - medium or dark gray, bad for H1 \&\ L1 - dark gray) 
           for a source frequency $\hat f$ = 100 Hz and three different assumed spin-down
           choices: a) zero; b) $-3\times10^{-9}$ Hz s$^{-1}$; and c) $-1\times10^{-8}$ Hz s$^{-1}$. 
           The black circle indicates the average position of the Sun during the data run.}
  \label{fig:skyband100hz} 
  \end{center}
\end{figure}

\begin{figure}
  \begin{center}
  \includegraphics[height=9.0cm]{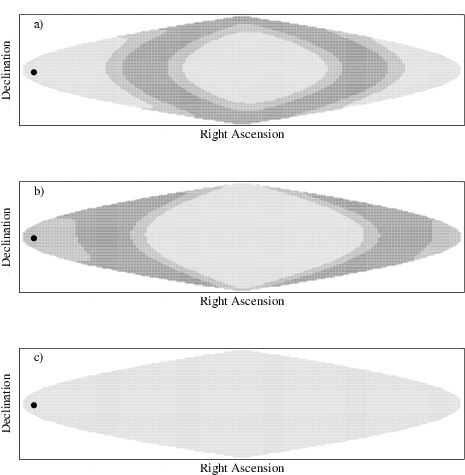}
  \caption{S4 sky band regions (good - light gray, bad for L1 - medium or dark gray, bad for H1 \&\ L1 - dark gray) 
           for a source frequency $\hat f$ = 300 Hz and three different assumed spin-down
           choices: a) zero; b) $-3\times10^{-9}$ Hz s$^{-1}$; and c) $-1\times10^{-8}$ Hz s$^{-1}$. 
           The black circle indicates the average position of the Sun during the data run.}
  \label{fig:skyband300hz} 
  \end{center}
\end{figure}

\begin{figure}
  \begin{center}
  \includegraphics[height=9.0cm]{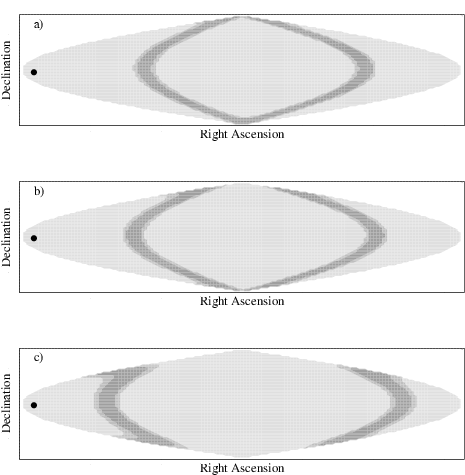}
  \caption{S4 sky band regions (good - light gray, bad for L1 - medium or dark gray, bad for H1 \&\ L1 - dark gray) 
           for a source frequency $\hat f$ = 1000 Hz and three different assumed spin-down
           choices: a) zero; b) $-3\times10^{-9}$ Hz s$^{-1}$; and c) $-1\times10^{-8}$ Hz s$^{-1}$. 
           The black circle indicates the average position of the Sun during the data run.}
  \label{fig:skyband1000hz} 
  \end{center}
\end{figure}

Figures~\ref{fig:skyband100hz}-\ref{fig:skyband1000hz} illustrate the variation in
the fraction of sky marked as ``bad'' as assumed source frequency and spin-down are
varied. Generally, at low frequencies, large sky regions are affected, but only for
low spin-down magnitude, while at high frequencies, small sky regions are affected,
but the effects are appreciable to larger spin-down magnitude. It should be noted that
the annular regions of the sky affected depend upon the start time and duration of a
data run. The longer the data run, the smaller is the region of sky for which Doppler
stationarity is small. Future LIGO data runs of longer duration should have only small
regions near the ecliptic poles for which stationary instrumental lines prove
troublesome.

\subsubsection{Grid-point upper limit determination}
\label{subsubsec:powerfluxupperlimitsetting}

An intermediate step in the PowerFlux analysis is the setting of upper limits
on signal strength for each sky-point for each $0.56$~mHz bin. The limits presented
here for each interferometer are the highest of these intermediate limits for
each 0.25-Hz band over the entire ``good'' sky. The intermediate limits are 
set under the assumption of Gaussian residuals in noise.
In brief, for each
$0.56$~mHz bin and sky-point, a Feldman-Cousins~\cite{FeldmanCousins}  95\% 
confidence-level is set
for an assumed normal distribution with a standard deviation determined robustly
from quantiles of the entire 0.25 Hz band. The Feldman-Cousins approach provides the virtues of
a well behaved upper limit even when background noise fluctuates well below its
expectation value and
of smooth transition between 1-sided and 2-sided limits,
but in practice the highest upper limit for any 0.25 Hz band is invariably 
the highest measured power plus 1.96 times the estimated standard deviation on
the background power for that bin, corresponding to a conventional 
{\it a priori} 1-sided 97.5\% upper CL.
A Kolmogorov-Smirnov (KS) statistic is computed to check the actual power 
against a Gaussian distribution for each
0.25 Hz band. Those bands that fail the KS test value of 0.07 ($>$ 5$\sigma$ deviation
for the S4 data sample) 
are flagged as ``Non-Gaussian'',
and no upper limits on pulsars are quoted here for those bands, although a full
search is carried out. Bands subject to violin modes and 
harmonics of the 60 Hz power mains tend to fail the KS test because of
sharp spectral slope (and sometimes because non-stationarity of sharp features
leads to poor noise factorization).

\begin{figure}
  \begin{center}
  \includegraphics[height=6.5cm]{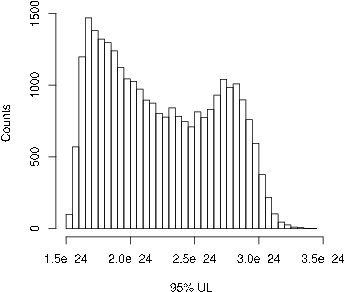}
  \caption{Histogram of Feldman Cousins 95\% confidence-level upper limits in a 0.25-Hz band near
  149 Hz in S4 H1 data. Each entry corresponds to the highest upper limit in the band
  for a single sky location.}
  \label{fig:powerfluxbackground} 
  \end{center}
\end{figure}

Figure~\ref{fig:powerfluxbackground} provides an example of derived upper limits
from one narrow band. The figure
shows the distribution of PowerFlux strain upper limits on linear polarization
amplitude $h_0^{\rm Lin}$ for a sample 0.25 Hz band of S4 H1 data near 149 Hz. The highest upper limit found
is $\sci{3.35}{-24}$ (corresponding to a worst-case pulsar upper limit on $h_0$ of 
$\sci{6.70}{-24}$). The bimodal distribution arises from different regions of 
the sky with intrinsically different antenna pattern sensitivities. The peak at 
$\sci{2.8}{-24}$ corresponds to points near the celestial equator where the run-averaged
antenna pattern sensitivity is worst.

\section{Hardware Injections and Validation}
\label{sec:valandhwinj}

\begin{table*}
\begin{tabular}{ccccccccc} \hline
Name  & $f_0$ (Hz)  & $df/dt$ ($\mathrm{Hz}~\mathrm{s}^{-1}$) & $\alpha$ (radians) & $\delta$ (radians) & $\psi$ (radians) 
& $\hpluszero$ & $\hcrosszero$ \\
\hline \hline
Pulsar0 & $265.57693318$ & $-4.15\times 10^{-12}$ & $1.248816734$ & $-0.981180225$ & $0.770087086$ 
& $4.0250\times 10^{-25}$ & $3.9212\times 10^{-25}$  \\
Pulsar1 & $849.07086108$ & $-3.00\times 10^{-10}$ & $0.652645832$ & $-0.514042406$ & $0.35603553$  
& $2.5762\times 10^{-24}$ & $1.9667\times 10^{-24}$  \\
Pulsar2 & $575.16356732$  & $-1.37\times 10^{-13}$ & $3.75692884$  & $0.060108958$  & $-0.221788475$
& $7.4832\times 10^{-24}$ & $-7.4628\times 10^{-24}$ \\
Pulsar3 & $108.85715940$ & $-1.46\times 10^{-17}$ & $3.113188712$ & $-0.583578803$ & $0.444280306$ 
& $1.6383\times 10^{-23}$ & $-2.6260\times 10^{-24}$ \\
Pulsar4 & $1402.11049084$ & $-2.54\times 10^{-08}$ & $4.886706854$ & $-0.217583646$ & $-0.647939117$ 
& $2.4564\times 10^{-22}$ & $1.2652\times 10^{-22}$  \\
Pulsar5 & $52.80832436$ & $-4.03\times 10^{-18}$ & $5.281831296$ & $-1.463269033$ & $-0.363953188$ 
& $5.8898\times 10^{-24}$ & $4.4908\times 10^{-24}$  \\
Pulsar6 & $148.44006451 $ & $-6.73\times 10^{-09}$ & $6.261385269$ & $-1.14184021$  & $0.470984879$ 
& $1.4172\times 10^{-24}$ & $-4.2565\times 10^{-25}$ \\
Pulsar7 & $1220.93315655$ & $-1.12\times 10^{-09}$ & $3.899512716$ & $-0.356930834$ & $0.512322887$ 
& $1.0372\times 10^{-23}$ & $9.9818\times 10^{-24}$  \\
Pulsar8 & $193.94977254$ & $-8.65\times 10^{-09}$ & $6.132905166$ & $-0.583263151$ & $0.170470927$ 
& $1.5963\times 10^{-23}$ & $2.3466\times 10^{-24}$  \\
Pulsar9 & $763.847316499$ & $-1.45\times 10^{-17}$ & $3.471208243$ & $1.321032538$  & $-0.008560279$
& $5.6235\times 10^{-24}$ & $-5.0340\times 10^{-24}$ \\
\hline
Pulsar10 & $501.23896714$ & $-7.03\times 10^{-16}$ & $3.113188712$ & $-0.583578803$  & $0.444280306$
& $6.5532\times 10^{-23}$ & $-1.0504\times 10^{-24}$ \\
Pulsar11 & $376.070129771$ & $-4.2620\times 10^{-15}$ & $6.132905166$ & $-0.583263151$  & $0.170470927$
& $2.6213\times 10^{-22}$ & $-4.2016\times 10^{-23}$ \\
\hline
\end{tabular}
\caption{Nominal (intended) parameters for hardware injected signals, known as Pulsar0 to Pulsar11,
for GPS reference time $= 793130413$~s (start of S4 run) at the SSB. These parameters are defined
in section~\ref{sec:waveforms}. As discussed in the text, imperfect calibration knowledge at the
time of injections led to slightly different actual injected strain amplitudes among the three
LIGO interferometers. The last two pulsars listed are binary system injections with additional
orbital parameters not shown, which were injected during only the last day of the S4 run.}
\label{tab:ParametersHWInjections}
\end{table*}

All three methods discussed in this paper have undergone extensive internal testing and review. 
Besides individual unit tests of the software, hardware injections provided
an end-to-end validation of the entire pipelines. The next subsections discuss the hardware
injections, the validations of the three methods and their pipelines. 
The detection of the hardware injections also shows in dramatic fashion that we can detect
the extremely tiny signals that the detectors were designed to find.

\subsection{Hardware injections}
\label{hwinj}

During a 15-day period in the S4 run, ten artificial isolated pulsar signals  
were injected into all three LIGO interferometers at a variety of
frequencies and time derivatives of the frequency, sky locations, and
strengths. Two additional artificial binary pulsar signals were injected
for approximately one day. These hardware injections were implemented 
by modulating the interferometer mirror positions via signals sent to
voice actuation coils surrounding magnets glued to the mirror edges.
The injections provided an end-to-end validation of the search
pipelines. Table~\ref{tab:ParametersHWInjections}
summarizes the nominal parameters used in the isolated-pulsar injections; the parameters are defined in
section~\ref{sec:waveforms}. 

Imperfect calibration knowledge at the
time of these injections led to slightly different actual strain amplitude injections
among the three LIGO interferometers. For the H1 and L1 comparisons between expected and detected signal
strengths for these injections described in section~\ref{subsec:StackSlideValidation},
corrections must be applied for the differences from nominal amplitudes.
The corrections are the ratios of the actuation function derived from final
calibration to the actuation function assumed in the preliminary calibration
used during the injections. For H1 this ratio was independent of the
injection frequency and equal to 1.12. For L1, this ratio varied slightly
with frequency, with a ratio of 1.11 for all injected pulsars except 
Pulsar1 (1.15) and Pulsar9 (1.18).

\begin{table*}
\begin{tabular}{cccccccccccc}\hline
        &          &          &    H1    &          &            & $\quad$ &          &          &    L1    &          &            \\
        & Observed & Injected & Observed & Injected & Percent    & $\quad$ & Observed & Injected & Observed & Injected & Percent    \\
Pulsar  & SNR      & SNR      & $\sqrt{P}$ & $\sqrt{P}$ & Difference & $\quad$ & SNR  & SNR      & $\sqrt{P}$ & $\sqrt{P}$ & Difference \\
\hline \hline
Pulsar0 & 0.27     & 0.23     & 1.006    & 1.005    & 0.1\%    & $\quad$ & 0.15     & 0.13     & 1.003    & 1.003    & 0.1\%    \\
Pulsar1 & 1.62     & 0.80     & 1.035    & 1.017    & 1.7\%    & $\quad$ & 0.27     & 0.69     & 1.006    & 1.016    & $-$1.0\%   \\
Pulsar2 & 8.92     & 8.67     & 1.179    & 1.175    & 0.4\%    & $\quad$ & 8.20     & 9.34     & 1.180    & 1.203    & $-$1.9\%   \\
Pulsar3 & 199.78   & 174.72   & 3.124    & 2.943    & 6.2\%    & $\quad$ & 89.89    & 104.76   & 2.304    & 2.454    & $-$6.1\%   \\
Pulsar4 & 2081.64  & 1872.24  & 9.607    & 9.116    & 5.4\%    & $\quad$ & 1279.12  & 1425.14  & 7.895    & 8.326    & $-$5.2\%   \\
Pulsar5 & 0.05     & 1.30     & 1.001    & 1.028    & $-$2.6\%   & $\quad$ & 1.02     & 0.44     & 1.024    & 1.010    & 1.4\%    \\
Pulsar6 & 0.17     & 2.94     & 1.004    & 1.063    & $-$5.5\%   & $\quad$ & 2.90     & 1.36     & 1.067    & 1.032    & 3.4\%    \\
Pulsar7 & 6.25     & 5.50     & 1.129    & 1.114    & 1.3\%    & $\quad$ & 6.07     & 5.11     & 1.136    & 1.116    & 1.8\%    \\
Pulsar8 & 98.12    & 96.21    & 2.303    & 2.285    & 0.8\%    & $\quad$ & 92.77    & 103.45   & 2.334    & 2.441    & $-$4.4\%   \\
Pulsar9 & 6.68     & 6.59     & 1.137    & 1.135    & 0.2\%    & $\quad$ & 2.61     & 3.69     & 1.061    & 1.085    & $-$2.2\%   \\
\hline
\end{tabular}
\caption{Results of StackSlide analyses of the ten hardware injected continuous gravitational-wave signals from isolated neutron stars.}
\label{tab:StackSlideHWInj}
\end{table*}

\subsection{StackSlide Validation}
\label{subsec:StackSlideValidation}

Besides individual unit tests and review of each component of the StackSlide code,
we have shown that simulated signals are detected with the expected
StackSlide Power,
including the hardware injections listed in Table~\ref{tab:ParametersHWInjections}.
Table~\ref{tab:StackSlideHWInj} shows the observed and injected SNR, and the 
square root of the observed and injected
StackSlide Power, $\sqrt{P}$.
The percent difference of the latter is given, since this compares
amplitudes, which are easier to compare with calibration errors.
The observed values were obtained by running the StackSlide code
using a template that exactly matches the
injection parameters, while the injected values were calculated using the 
parameters in Table~\ref{tab:ParametersHWInjections} and the equations
in Appendix \ref{sec:stackslidepowerandstats}. The SNR's of Pulsar0, Pulsar1, Pulsar5, and
Pulsar6 were too small to be detected, and Pulsar4 and Pulsar7 were out of
the frequency band of the all-sky search. Pulsar2, Pulsar3,
and Pulsar8 were detected as outliers with SNR~$> 7$ (as discussed in
Sec.~\ref{sec:results}) while
Pulsar9 was not loud enough to pass this requirement. In all cases the observed StackSlide Power agrees
well with that predicted, giving an end-to-end validation of the StackSlide code.

\begin{figure}
  \begin{center}
  \includegraphics[height=6.5cm]{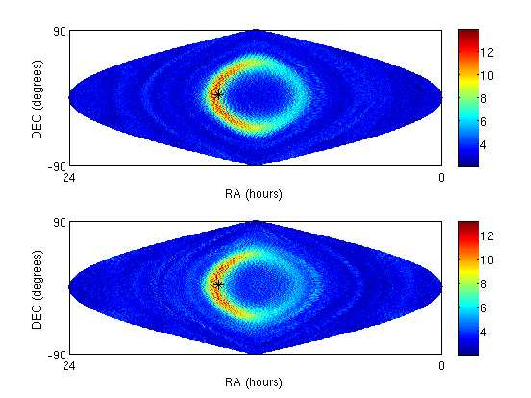}
  \caption{Detection of hardware injected Pulsar 2 by the StackSlide code in
  the H1 (top) and L1 (bottom) data.}
  \label{fig:S4H1L1StackSlideAllSkyPulsar2SNR}
  \end{center}
\end{figure}

As an example of an all-sky search for a band with an injection,
Fig.~\ref{fig:S4H1L1StackSlideAllSkyPulsar2SNR} shows the
detection of Pulsar2 for a search of the H1 (top) and L1 (bottom) data, and only during the
times the hardware injections were running. Later, when the entire S4 data set was analyzed Pulsar2
was still detected but with lower SNR, since this data includes times when the hardware
injections were absent. Also note that, as explained in section~\ref{subsec:powerflux},
because of strong correlations on the sky, a pulsar signal will be detected at many points that
lie in an annular region in the sky that surrounds the point corresponding to the average
orbital acceleration vector of the Earth, or its antipode. In fact, because of the large number of
templates searched, random noise usually causes the maximum detected SNR to occur in a template other than
the one which is closest to having the exact parameters of the signal.
For example, for the exact template and times matching the Pulsar2 
hardware injection, it was detected with 
SNR's of $8.92$ and $8.20$ in H1 and L1, respectively, as given in Table~\ref{tab:StackSlideHWInj},
while the largest SNR's shown in Fig.~\ref{fig:S4H1L1StackSlideAllSkyPulsar2SNR}
are $13.84$ and $13.29$. During the search of the full data set
(including times when Pulsar2 was off) it
was detected with SNR $11.09$ and $10.71$ in H1 and L1, respectively.

\subsection{Hough Validation}
\label{subsec:houghval}

\begin{figure}
  \begin{center}
  \includegraphics[height=7.4cm]{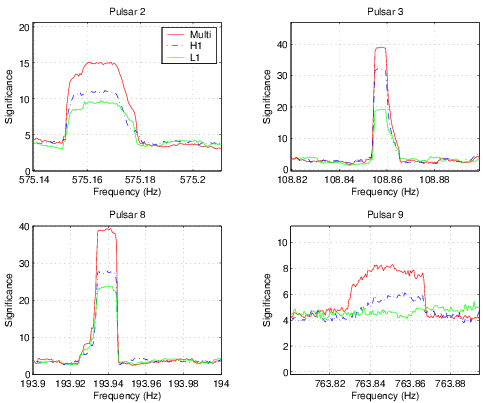}
  \caption{Maximum significance as a function of frequency
  corresponding to the 
  multi-interferometer search (using the data from the three detectors)
  and the  H1   and L1 alone.}
  \label{fig:InjectionsSignificance}
  \end{center}
\end{figure}

\begin{figure}
  \begin{center}
  \includegraphics[height=7cm]{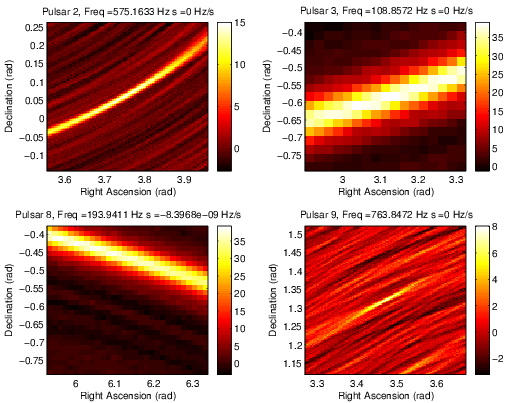}
  \caption{Maps of the Hough significance corresponding to the 
  multi-interferometer case for Pulsar2, Pulsar3, Pulsar8 and Pulsar9. 
  The location of the injected pulsars are the centers of
  the maps.
  For Pulsar2, Pulsar3 and Pulsar9, the maps correspond to the 
   frequency and spin-down values closest to the real injected ones. 
  For Pulsar8, we show the map containing 
  the maximum significance value.
  The discrepancy in sky location is due  to the mismatch in
  frequency and spin-down values between those used in the injections
  and those corresponding  to the Hough map.}
  \label{fig:MapSignificancePx}
  \end{center}
\end{figure}

\begin{table}
\begin{tabular}{cccccc}\hline
Pulsar  & $\,$ & Detector   &  $f_0$ range  & $f_0(max)$ & Significance \\
  & $\,$ &    &   (Hz) & (Hz) &  \\
\hline \hline
Pulsar2 & $\,$ & Multi-IFO & 575.15-575.18 & 575.1689	& 15.1195 \\
        & $\,$ &      H1   & 575.15-575.18 & 575.1667	& 11.1730  \\
        & $\,$ &      L1   & 575.15-575.18 & 575.1650  &  9.7635  \\
Pulsar3  & $\,$ & Multi-IFO & 108.855-108.86 & 108.8572 & 39.1000 \\
         & $\,$  &      H1   & 108.855-108.86 & 108.8572	& 32.2274 \\
	 & $\,$ &      L1   & 108.855-108.86 & 108.8589	& 19.2267 \\
Pulsar8  & $\,$ & Multi-IFO & 193.932-193.945 & 193.9411 & 39.2865\\
	 & $\,$ &      H1   & 193.932-193.945 & 193.9394 & 27.9008 \\
	 & $\,$ &      L1   & 193.932-193.945 & 193.9400 & 23.8270 \\
Pulsar9  & $\,$ & Multi-IFO & 763.83-763.87 & 763.8511 & 8.3159 \\
	 & $\,$ &      H1   & 763.83-763.87 & 763.8556 & 6.1268	\\
	 & $\,$ &      L1   &       -       &    -     & 5.4559 \\	
\hline
\end{tabular}
\caption{Results of the Hough search for the hardware injected signals for 
the multi-interferometer, H1 and L1 data.}
\label{tab:HoughHI}
\end{table}

Using the Hough search code,
four hardware-injected signals have been clearly detected by analyzing 
the data from the interval when the injections took place. These
correspond to Pulsar2, Pulsar3, Pulsar8 and Pulsar9. For each of these 
injected signals, a small-area search ($0.4$ rad$\times 0.4$ rad) was
performed, using a step size on the
spin-down parameter of  $-4.2 \times 10^{-10}~\mathrm{Hz}~\mathrm{s}^{-1}$. 
Given the large
spin-down value of Pulsar8 ($\sci{-8.65}{-9}~\mathrm{Hz}~\mathrm{s}^{-1}$),
we have used 23 values
of the spin-down spanning the range
[$\sci{-9.24}{-9}~\mathrm{Hz}~\mathrm{s}^{-1}$,
$0~\mathrm{Hz}~\mathrm{s}^{-1}$] to search for this pulsar. 
Because of its large amplitude, Pulsar8 can be detected even
with a large
mismatch in the spin-down value, although at the cost of lower SNR.

Figure~\ref{fig:InjectionsSignificance} shows the significance
maximized over different sky locations and spin-down values for the
different frequencies. These four hardware injected pulsars have been
clearly detected, with the exception of Pulsar9 in the L1 data.
Pulsar9 is marginally visible using the H1 data alone, with a
maximum significance of 6.13, but when we combine the data from the three
interferometers, the significance increases up to 8.32.  
Details are
given in Table~\ref{tab:HoughHI}, including the frequency range of the
detected signal, the frequency at which the maximum significance is
obtained and its significance value.

Figure~\ref{fig:MapSignificancePx} shows the Hough significance maps
for the multi-interferometer case.  The maps displayed correspond
either to the frequency and spin-down values nearest to the injected
ones, or to those in which the maximum significance was observed.
The location of the injected pulsars correspond to the center of each map.
Note that the true spin-down value of Pulsar8,
$\sci{-8.65}{-9}~\mathrm{Hz}~\mathrm{s}^{-1}$, lies between
the parameter values $\sci{-8.82}{-9}~\mathrm{Hz}~\mathrm{s}^{-1}$ and
$\sci{-8.40}{-9}~\mathrm{Hz}~\mathrm{s}^{-1}$ of the nearest templates used.

\subsection{PowerFlux validation}
\label{subsec:powerfluxmcvalidation}

\begin{table*}
\begin{tabular}{rrrrcccrrrr}\\
\hline\hline
 &  & \multicolumn{2}{c}{Detected $f_0$ (Hz)} &  & \multicolumn{2}{c}{$h_0$ upper limit} & \multicolumn{2}{c}{Det. polarization} & \multicolumn{2}{c}{Detected SNR} \\
Pulsar & $f_0$ (Hz) & H1 & L1 & True $h_0$ & H1 & L1 & H1 & L1 & H1 & L1 \\
\hline
Pulsar2 & 575.164 & 575.161 & 575.164 & $8.04\times10^{-24}$ & $3.18\times10^{-23}$ & $2.16\times10^{-23}$ & circular & circular & 16.59 & 15.33 \\
Pulsar3 & 108.857 & 108.858 & 108.858 & $3.26\times10^{-23}$ & $3.92\times10^{-23}$ & $3.36\times10^{-23}$ & circular & linear & 328.59 & 209.99 \\
Pulsar4 & 1402.110 & 1402.111 & 1402.113 & $4.56\times10^{-22}$ & $6.50\times10^{-22}$ & $5.32\times10^{-22}$ & linear & circular & 2765.71 & 1651.82 \\
Pulsar7 & 1220.933 & 1220.933 & -- & $1.32\times10^{-23}$ & $3.56\times10^{-23}$ & $2.88\times10^{-23}$ & circular & -- & 8.89 & -- \\
Pulsar8 & 193.950 & 193.951 & 193.948 & $3.18\times10^{-23}$ & $4.18\times10^{-23}$ & $3.52\times10^{-23}$ & linear & circular & 289.11 & 292.13 \\
Pulsar9 & 763.847 & 763.849 & -- & $8.13\times10^{-24}$ & $1.69\times10^{-23}$ & $1.97\times10^{-23}$ & circular & -- & 8.18 & -- \\
\hline
\end{tabular}

\caption{Results of PowerFlux analysis of the six S4 hardware pulsar injections 
for which there is detection (SNR$>$7). Shown are the true nominal pulsar frequency at
the start of the run (SSB frame), the frequency in each interferometer for
detected signals, the true $h_0$ value of the injection, the worst-case upper limit
from each interferometer, the polarization state for which the SNR is maximum in 
each interferometer, and the SNR of detected candidates.}
\label{tab:PowerFluxHWInj}
\end{table*}

Several cross checks have been performed to validate the PowerFlux search algorithm.
These validations range from simple and rapid Fourier-domain ``power injections'' to more 
precise time domain software simulations, to hardware signal injections carried out
during data taking. 

\begin{figure}
  \begin{center}
  \includegraphics[height=8cm]{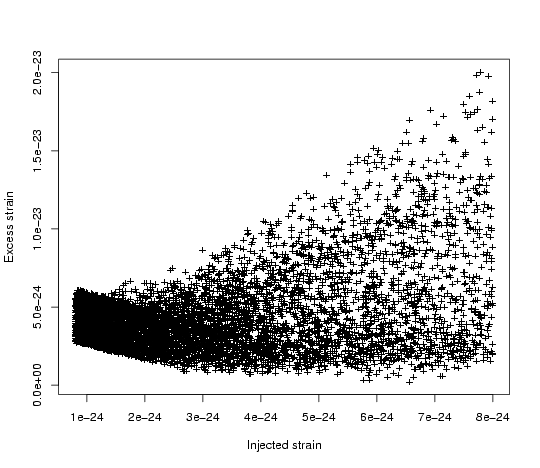}
  \caption{``Excess'' (upper limit minus injected) strain plotted vs
  injected signal strain for sample PowerFlux H1 elliptic-polarization near 140 Hz injections.}
  \label{fig:excessvsstrain} 
  \end{center}
\end{figure}

\begin{figure}
  \begin{center}
  \includegraphics[height=8cm]{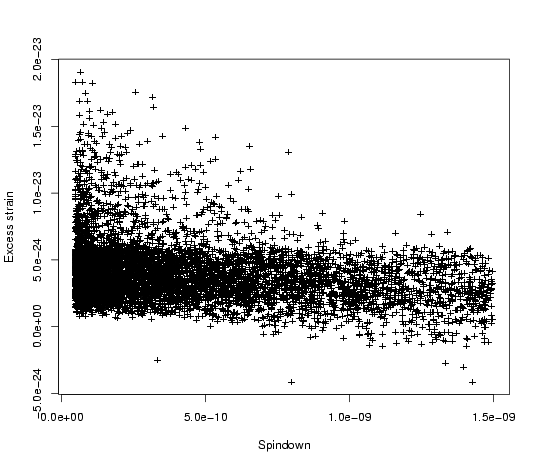}
  \caption{``Excess'' (upper limit minus injected) strain plotted vs
  injected signal spin-down for sample PowerFlux H1 elliptic-polarization near 140 Hz injections.}
  \label{fig:excessvsspindown} 
  \end{center}
\end{figure}

\begin{figure}
  \begin{center}
  \includegraphics[height=8cm]{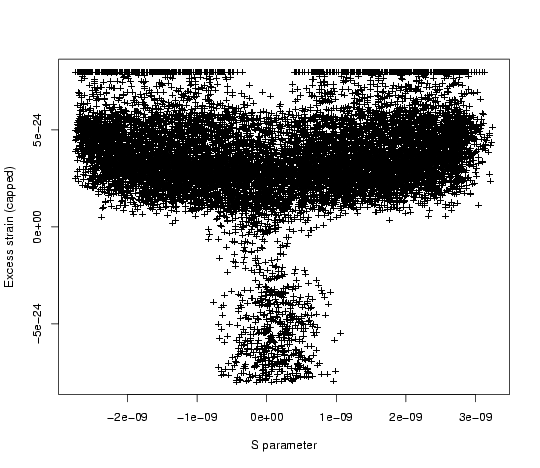}
  \caption{``Excess'' (upper limit minus injected) strain plotted vs
  S parameter defined in text, where values greater than $8\times10^{-24}$ have been ``capped'' at that ceiling value.}
  \label{fig:excessvss} 
  \end{center}
\end{figure}

Signal strain power injections have been carried out as part of PowerFlux algorithm development
and for parameter tuning. These software injections involve superimposing calculated powers for
assumed signals upon the LIGO power measurements and carrying out searches. For computational speed, when
testing signal detection efficiency, only a small region of the sky around the known
source direction is searched. A critical issue is whether the strict frequentist limits
set by the algorithm are sufficiently conservative to avoid undercoverage
of the intended frequentist confidence band. We present here
a set of figures that confirm overcoverage applies. Figure~\ref{fig:excessvsstrain} shows
the difference (``excess'') between the Feldman-Cousins 95\% confidence-level upper limit (conventional 97.5\% upper limit) on strain and
the injected strain for a sample of elliptic-polarization time-domain injections in the H1 
interferometer for the 140.50-140.75 Hz band. Injection amplitudes were distributed logarithmically,
while frequencies, spin-downs, sky locations, and orientations were distributed uniformly. One sees that
there is indeed no undercoverage (every excess strain value is above zero) over the range of
injection amplitudes. 
Figure~\ref{fig:excessvsspindown} shows the same ``excess'' plotted vs the injected spin-down
value, where the search assumes a spin-down value of zero, and
where the sample includes injections with actual spin-down values
more than a step size away from the the assumed value for the search
template. As one can see, in this frequency range,
a spin-down stepsize of $\sci{1.0}{-9}~\mathrm{Hz}~\mathrm{s}^{-1}$ 
 is safe (true spin-down no more $\sci{5.0}{-10}~\mathrm{Hz}~\mathrm{s}^{-1}$
away from the assumed search value). Figure~\ref{fig:excessvss} shows the ``excess'' plotted vs
the $S$ parameter that discriminates between sky regions of low and high Doppler stationarity.
As shown,  a value of $S_{\rm large}\sim\sci{1}{-9}~\mathrm{Hz}~\mathrm{s}^{-1}$ 
 is safe for these injections.
For this search we have chosen 51 spin-down steps of  $\sci{2}{-10}~\mathrm{Hz}~\mathrm{s}^{-1}$  
for 50-225 Hz and 11 steps of $\sci{1}{-9}~\mathrm{Hz}~\mathrm{s}^{-1}$ 
 for 200-1000 Hz. 

More computationally intensive full time-domain signal injections were also carried
out and the results found to be consistent with those from power injections,
within statistical errors. 

\begin{figure}
  \begin{center}
   \includegraphics[height=8.5cm,angle=270]{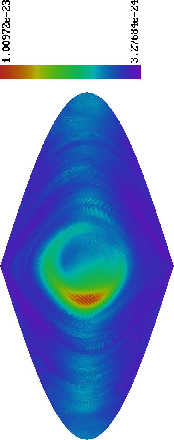}
  \caption{Sample sky map of Feldman Cousins upper limits on circularly polarized
  strain for a 0.25-Hz band containing hardware-injected Pulsar 2 at 575.16 Hz.
  Only the data (half the run) during which the pulsar injection was enabled has
  been analyzed for this plot. The injected pulsar ($h_0=\sci{8.0}{-24}$) stands out
  clearly above background.  (Right ascension increases positively toward the left
  and declination toward the top of the sky map.)}
  \label{fig:pulsar2skymap} 
  \end{center}
\end{figure}

In addition, the PowerFlux method was validated with the hardware signal injections
summarized in Table~\ref{tab:ParametersHWInjections}. 
The PowerFlux algorithm was run on all 10 isolated pulsars, including two outside the 50-1000 Hz search region,
and results found to agree well with expectation
for the strengths of the signals and the noise levels in their bands. 
Table~\ref{tab:PowerFluxHWInj} shows the results of the analysis for the six pulsars
for which a detection with SNR$>$7 is obtained by PowerFlux for
one or both of the 4-km interferometers.
Figure~\ref{fig:pulsar2skymap} shows a sky map of
PowerFlux $\psi=0$ polarization SNR for the 0.25 Hz band containing pulsar 2 (575.16 Hz).

\section{Results}
\label{sec:results}

All three methods described in Sections~\ref{sec:analysismethodoverview} 
and \ref{sec:analysismethoddetails} have been applied
in an all-sky search over a frequency range 50-1000 Hz. As described below, no evidence for a 
gravitational wave signal is observed in any of the searches, and upper limits
on sources are determined. For the StackSlide and Hough methods, 
95\% confidence-level frequentist upper limits are placed
on putative rotating neutron stars, assuming a uniform-sky and isotropic-orientation
parent sample. Depending on the source location and inclination, these limits
may overcover or undercover the true 95\% confidence-level band. 
For the PowerFlux method, strict frequentist upper limits are placed on linearly and circularly
polarized periodic gravitational wave sources, assuming {\it worst-case}
sky location, avoiding undercoverage. The limits on linear polarization are also
re-interpreted as limits on rotating neutron stars, assuming worst-case sky location
and worst-case star inclination. The following
subsections describe these results in detail.

\subsection{StackSlide Results}
\label{subsec:stackslideresults}

\subsubsection{Loudest powers and coincidence outliers}
\label{subsubsec:stackslidelpsandoutliers}

\begin{figure}
  \begin{center}
  \includegraphics[height=6.5cm]{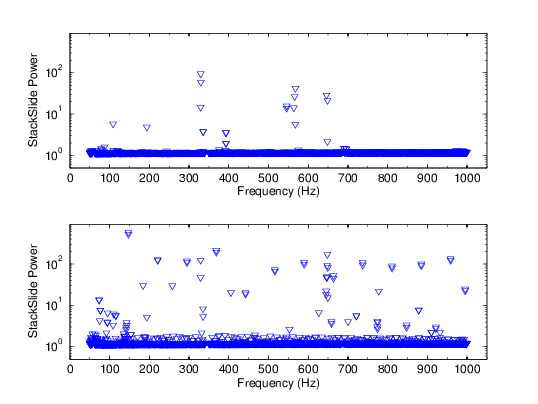}
  \caption{The loudest observed StackSlide Power for H1 (top) and L1 (bottom). Frequency bands 
   with the harmonics of 60 Hz and the violin modes have been removed.}
  \label{fig:S4H1L1StackSlideLEs}
  \end{center}
\end{figure}

The StackSlide method was applied to the S4 H1 and L1 data set,
as given in Sec.~\ref{subsec:stackslide}. As described in that section,
only the loudest StackSlide Power was returned from a search of the entire sky, the 
range of the frequency's time derivative, 
$[-1 \times 10^{-8}, 0]$~Hz~$\mathrm{s}^{-1}$, and for
each $0.25$~Hz band within $50-1000$~Hz. The results are shown
in Fig.~\ref{fig:S4H1L1StackSlideLEs}. 

\begin{figure}
  \begin{center}
  \includegraphics[height=6.5cm]{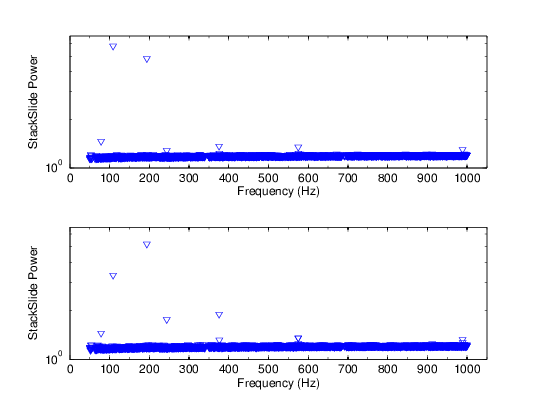}
  \caption{The loudest observed StackSlide Power for H1 (top) and L1 (bottom) with a simple veto
   applied: only outliers in each $0.25$~Hz band with SNR~$> 7$ in both interferometers that have
   a fractional
   frequency difference $\le 2.2 \times 10^{-4}$ are kept. These are shown against
   the background results that have SNR~$\le 7$ in both interferometers. Frequency bands
   with the harmonics of the 60 Hz and the violin modes have also been removed.}
  \label{fig:S4H1L1StackSlideLEsSimpleVetoes}
  \end{center}
\end{figure}

\begin{table}
\begin{tabular}{cccccc}\hline
     & $f_{H1}$~(Hz) & $f_{L1}$~(Hz) & H1 SNR   & L1 SNR  & Comment           \\
\hline \hline
1    &  78.618889    &  78.618889     &  14.82  &  13.58   & Inst. Lines      \\
2    & 108.856111    & 108.856111     & 152.11  &  69.79   & HW Inj. Pulsar3  \\
3    & 193.947778    & 193.949444     & 121.89  & 125.75   & HW Inj. Pulsar8  \\
4    & 244.148889    & 244.157778     &   9.00  &  22.89   & Inst. Lines      \\
5    & 375.793889    & 375.806667     &  11.68  &  27.09   & HW Inj. Pulsar11 \\
6    & 376.271111    & 376.281667     &   7.47  &  9.46    & HW Inj. Pulsar11 \\
7    & 575.162778    & 575.153333     &  11.09  & 10.71    & HW Inj. Pulsar2  \\
8    & 575.250000    & 575.371667     &   7.49  &  7.51    & Inst. Lines      \\
9    & 575.250000    & 575.153333     &   7.49  & 10.71    & Inst. \& Pulsar2 \\
10   & 580.682778    & 580.734444     &   7.02  &  7.19    & Inst. Lines      \\
11   & 912.307778    & 912.271111     &   7.02  &  7.37    & Inst. Lines      \\
12   & 988.919444    & 988.960556     &   9.56  &  9.75    & Inst. Lines      \\
13   & 988.919444    & 989.000000     &   9.56  &  8.12    & Inst. Lines      \\
14   & 993.356111    & 993.523333     &   7.08  &  7.12    & Inst. Lines      \\
\hline
\end{tabular}
\caption{StackSlide outliers with SNR~$> 7$ in both interferometers, with fraction
difference in frequency less than or equal to $2.2 \times 10^{-4}$, 
and after removal of the bands with $60$~Hz harmonics
and the violin modes.}
\label{tab:StackSlideOutliers}
\end{table}

Many of the StackSlide results have power
greater than expected due to random chance alone (for Gaussian noise). 
To identify the most interesting subset
of these cases, a simple coincidence test was applied: only results with an SNR greater than $7$
in both H1 and L1 and with a fractional difference in frequency, measured in the SSB, less than
or equal to $2.2 \times 10^{-4}$ were identified as outliers for further follow-up. The requirement on
frequency agreement comes from the worst-case scenario where a signal is detected on opposite
sides of the sky with opposite Doppler shifts of $1 + v/c$ and $1 - v/c$, giving a maximum
fraction difference in the detected frequency at the SSB of $2 v/c \le 2.2 \times 10^{-4}$.
The results after applying this simple coincidence test are shown
in Fig.~\ref{fig:S4H1L1StackSlideLEsSimpleVetoes}.  The outliers that passed the test are
shown in Table~\ref{tab:StackSlideOutliers}.

Note that the coincidence test used on the StackSlide results
is very conservative in that it only covers the worst-case
frequency difference, and makes no requirement on consistency in
sky position or the frequency's time derivative. 
However it is meant to find only the most prominent outliers. Since an automated follow-up
of possible candidates is not yet in place, the follow-up is carried out manually. This dictated
using a large threshold on SNR.  Also, since the false dismissal rate of the coincidence test used
was not determined (though it is assumed to be essentially zero) it is not used in this paper when
setting upper limits. Monte Carlo studies will be needed to find appropriate thresholds on SNR and the
size of coincidence windows, so that proper false alarm and false dismissal rates can be
determined; such studies will be carried out when analyzing future data sets.

\begin{figure}
  \begin{center}
  \includegraphics[height=6.5cm]{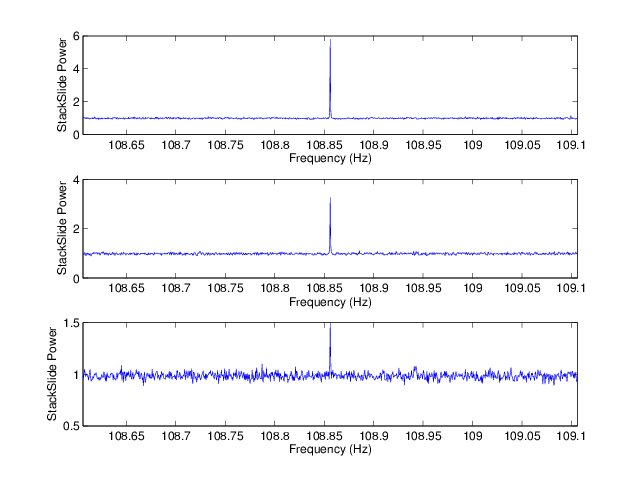}
  \caption{The StackSlide Power vs. frequency for H1 (top), L1 (middle) and H2 (bottom)
   using the sky position and the $\fdot$ value of the template that gives the outlier
   in H1, for outlier number~$2$ given in Table~\ref{tab:StackSlideOutliers}. 
   Comparing with Tables~\ref{tab:ParametersHWInjections} and \ref{tab:StackSlideHWInj}   
   this outlier is identified as due to hardware injection Pulsar3.}
  \label{fig:S4H1L1H2StackSlideOutliersWithHWInjs}
  \end{center}
\end{figure}

\begin{figure}
  \begin{center}
  \includegraphics[height=6.5cm]{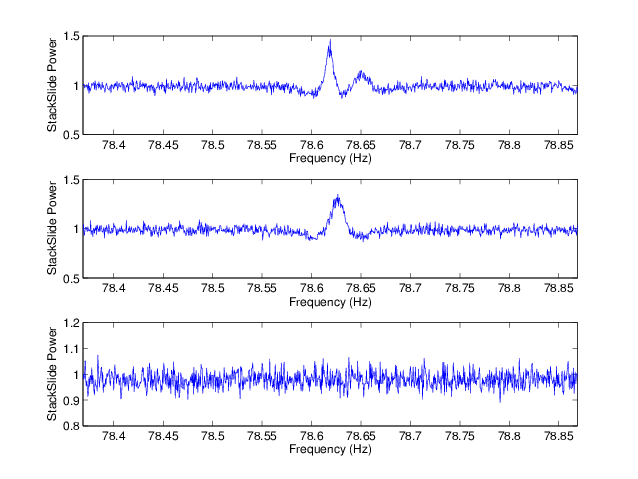}
  \caption{The StackSlide Power vs. frequency for H1 (top), L1 (middle) and H2 (bottom)
   using the sky position and the $\fdot$ value of the template that gives the outlier
   in H1 for outlier number~$1$ given in Table~\ref{tab:StackSlideOutliers}.}
  \label{fig:S4H1L1H2StackSlideOutliersOtherOutliersFirstSet}
  \end{center}
\end{figure}

Three types of qualitative follow-up tests were performed on each of the outliers in
Table~\ref{tab:StackSlideOutliers}. First, using the sky position and the
$\fdot$ value of the template that gives the outlier in H1, the StackSlide Power
was found using the same values for these in L1 and H2 for a frequency band around
that of the outlier in H1. For a fixed sky position and $\fdot$, a true
gravitational-wave signal should show up in all three detectors as a narrow line
at nearly the same frequency (though with an SNR corresponding to half the
length displacement in H2 compared with that in H1 and L1). Second, the StackSlide Power was computed
for the frequency bands containing the outliers, with sliding turned off. If an instrumental line
is the underlying cause of the outlier, a stronger and narrower peak will tend to
show up in this case. Third, the StackSlide Power was computed for each H1 outlier template,
using half (and some other fractions) of the data. This should reduce the SNR of a 
true signal by roughly the square root of the fractional reduction of the data, but
identify transient signals, which would fail this test by showing up in certain stretches of the
data with more SNR while dissappearing in other stretches. This would be true of the hardware
injections which were not always on during the run, or temporary disturbances of the instrument
which appear to look like signals only for limited periods of time. (The search described here was
not designed to find truely transient gravitational-wave signals.)

The follow-up tests on the outliers given in Table~\ref{tab:StackSlideOutliers} found that none
is qualitatively consistent with a true gravitational-wave signal. The three loudest hardware
injections of periodic gravitational waves from fake isolated sources were found (indicated as
Pulsar3, Pulsar8, and Pulsar2), as well as interference from a fake source in a binary
system (Pulsar11). All of the outliers due to the hardware injections show up in the H1 template as
relatively narrow lines in all three detectors, for example
as shown in Fig.~\ref{fig:S4H1L1H2StackSlideOutliersWithHWInjs}.
These outliers, on the other hand, fail the third test when looking at times the hardware injections
were turned off. In particular, this test, along with the frequencies in
Table~\ref{tab:ParametersHWInjections}, confirms
the identification of outliers~$5$ and~$6$ as due to Pulsar11.
The other hardware injections also are identified as such via their detected
frequencies in Table~\ref{tab:ParametersHWInjections} and SNRs in Table~\ref{tab:StackSlideHWInj}. 
In comparison, none of the other outliers qualitatively passes the first test, for example
as shown in Fig.~\ref{fig:S4H1L1H2StackSlideOutliersOtherOutliersFirstSet}.
The second test was less conclusive, since
some of the outliers lie at points on the sky that receive little Doppler modulation,
but based on the first test we conclude that the remaining outliers are only
consistent with instrumental line artifacts. 
These results are summarized in column six of Table~\ref{tab:StackSlideOutliers}.
In future searches, tests of the type used here should be studied using Monte Carlo
simulations, to make them more quantitative.

\subsubsection{StackSlide upper limits}
\label{subsubsec:stackslideupperlimits}

\begin{figure*}
  \begin{center}
  \includegraphics[height=13cm]{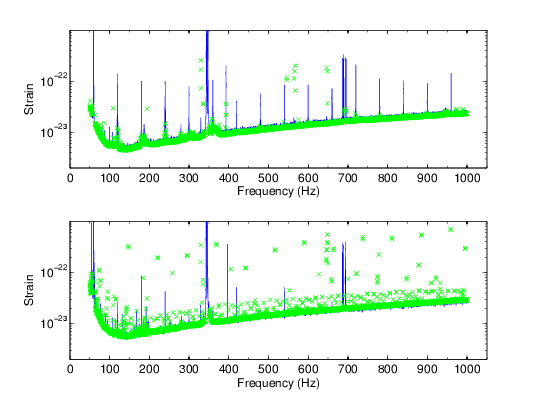}
  \caption{The solid curve shows the characteristic amplitude given by 
  Eq.~(\ref{eq:stackslidecharamplitudeallsky}) and crosses show the measured upper limits
  on $h_0$ for the StackSlide search of the H1 (top) and L1 (bottom) data.}
  \label{fig:S4H1StackSlideCharAmpvsMeasuredULs}
  \end{center}
\end{figure*}

\begin{table}
\begin{tabular}{ccc}\hline
Detector &  Band (Hz)   & $h_0^{95\%}$ \\
\hline \hline
H1 & 139.50-139.75 & $4.39\times 10^{-24}$ \\
L1 & 140.75-141.00 & $5.36\times 10^{-24}$ \\
\hline
\end{tabular}
\caption{Best StackSlide all-sky $h_0$ upper limits obtained on the strength of
gravitational waves from isolated neutron stars.}
\label{tab:StackSlideBestULs}
\end{table}

The StackSlide $95\%$ confidence upper limits on $h_0$ are shown as crosses for H1 (top) and L1 (bottom)
respectively in 
Fig.~\ref{fig:S4H1StackSlideCharAmpvsMeasuredULs}, 
while the solid curves in this figure
show the corresponding characteristic amplitudes given by
Eq.~(\ref{eq:stackslidecharamplitudeallsky})
in Appendix~\ref{sec:stackslidepowerandstats}.
The characteristic amplitudes were calculated
using an estimate of the noise from a typical time during
the run, but include bands with the power line and violin line harmonics which were excluded from 
the StackSlide search. The best upper limits over the entire search band 
are given in Table~\ref{tab:StackSlideBestULs}.  The uncertainties
in the upper limits and confidence due to the method used are less than or equal to $3\%$
and $5.3\%$ respectively; random and systematic errors from the calibration increase
these uncertainties to about $10\%$.

\subsection{Hough results}
\label{subsec:houghresults} 

\subsubsection{Number Counts}
\label{subsubsec:numbercounts}

For the S4 data set, there are a total of $N=2966$ SFTs from the three
interferometers, giving an expected average number count for pure
noise of $\bar{n} = Np \sim 593$.  The standard deviation $\sigma$ now
depends on the sky-patch according to \eqref{eq:4}.  For reference, if
we had chosen unit weights, the standard deviation assuming pure
Gaussian noise would have been $\sim 22$ for the multi-interferometer
search.  To compare number counts directly across
different sky-patches, we employ the \emph{significance} $s$ of a number
count defined in Eq.~(\ref{eq:5}).

\begin{figure}
  \begin{center}
  \includegraphics[height=6.5cm]{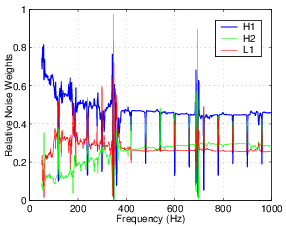}
  \caption{Relative contributions of the three interferometers in the
    Hough multi-interferometer search. The noise weights are calculated in
    $1\,$Hz bands.}
  \label{fig:HoughWeights}
  \end{center}
\end{figure}

Since the three interferometers have different noise floors and duty
factors, we would like to know their relative contributions to the
total Hough number count, and whether any of the interferometers
should be excluded from the search, or if all of them should be
included.  For this purpose, for the moment let us ignore the beam
pattern functions and consider just the noise weighting: $w_i \propto
1/S^\srchTemplateInd_{\iSubSupInd}$.  
The relative contribution of a particular interferometer, say $I$, is 
given by the ratio
\begin{equation}
  \label{eq:6}
  r_I = \frac{\sum_{i\in I} w_i}{\sum_{i=1}^N w_i}\,, \qquad I =
  \textrm{H1, L1, H2}\,.
\end{equation}
The numerator is a sum of the weights for the $I^{th}$ interferometer
while the denominator is the sum of all the weights.  This
figure-of-merit incorporates both the noise level of data from an
interferometer, and also its duty cycle as determined by the number of
SFTs available for that interferometer.  Figure~\ref{fig:HoughWeights}
shows the relative contributions from H1, L1, and H2 for the duration
of the S4 run.  From the plot, we see that H1 clearly contributes the
most.  H2 contributes least at low frequencies while L1 contributes
least at higher frequencies.  Hence all three LIGO interferometers are
included in this search.  For comparison purposes and for coincidence
analysis, we have also analyzed the data from H1 and L1 separately.

\begin{figure}
  \begin{center}
  \includegraphics[height=9.5cm, width=9.5cm]{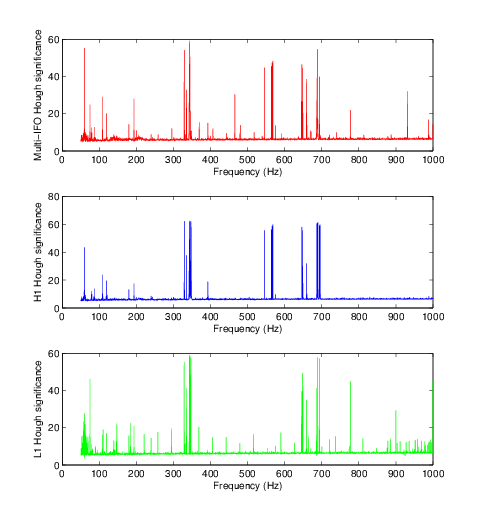}
  \caption{The measured loudest significance in each 0.25 Hz from the Hough
  search of the multi-interferometer (top), H1 (middle) and L1(bottom) data.}
  \label{fig:HoughSignificance}
  \end{center}
\end{figure}
Figure \ref{fig:HoughSignificance} shows the result of the Hough
search using data from all three LIGO interferometers, either combined
in a multi-interferometer search, or just for H1 and L1 data.  This
figure shows the loudest significance in every $0.25\,$Hz band,
maximized over all sky-positions, frequencies and spin-downs for the
three searches.  Line cleaning was used as described before.
 In the bands in which there are no spectral disturbances the significance distribution
 agrees very well with the theoretical expected distribution as was shown 
 in Fig.~\ref{fig:H1histo150}.

\subsubsection{Study of coincidence outliers}
\label{subsubsec:houghout}

There are many outliers from the Hough search with significance values
higher than expected for Gaussian noise, as shown in
Fig.~\ref{fig:HoughSignificance}.  
Many of the large outliers correspond to well known instrumental
artifacts described earlier, such as the power mains harmonics or the
violin modes. 

 Note the relation between significance and false alarm
which can be derived from equations~(\ref{eq:nth}) and (\ref{eq:5})
for Gaussian noise:
\begin{equation}
  \label{eq:sa}
  \alpha_\H= 0.5\,\textrm{erfc} (s/\sqrt{2}) \,.
\end{equation}
To identify interesting candidates, we consider only those that have a
significance greater than 7 in the multi-interferometer search (the
most sensitive one). This is the same threshold considered by the 
StackSlide and PowerFlux searches.  For the Hough search, this threshold
corresponds to a false alarm rate of $1.3\times
10^{-12}$. With this threshold, 
we would expect about 6 candidates
in a 100 Hz band around 1~kHz for Gaussian noise, since the number of 
 templates analyzed in a 1~Hz band around 1~kHz is about
$n=4.4\times 10^{10}$. If we would like to set a different threshold in order to
select, say one event in a 1 Hz band, then we should increase the false alarm to 
$\alpha_\H =1/n =2.2\times 10^{-11}$.

In order to exclude spurious events due to instrumental noise in just
one detector, we pass these candidates through a simple coincidence
test in both the H1 and the L1 data. Since the single detector search
is less sensitive than the multi-interferometer one, we consider
events from H1 and L1 with a significance greater than 6.6,
corresponding to a false alarm rate of $2.0\times 10^{-11}$.  The numbers of
templates analyzed using the H1 or L1 data are the same as for the
multi-interferometer search.

\begin{table}
\begin{tabular}{cccccc}\hline
     &           &  \multicolumn{3}{c}{Hough significance}\\
     & Band~(Hz) & Multi-IFO & H1    & L1   & Comment           \\
\hline \hline
1    &  78.602-78.631  & 12.466   &  12.023 & 10.953   & Inst. Lines      \\
2    & 108.850-108.875 & 29.006   &  23.528 & 16.090   & Inj. Pulsar3  \\
3    & 130.402-130.407 &  7.146   &   6.637 &  6.989   & ?  \\
4    & 193.92-193.96   & 27.911   &  17.327 & 20.890   & Inj. Pulsar8  \\
5    & 575.15-575.23   & 13.584   &   9.620 & 10.097   & Inj. Pulsar2  \\
6    & 721.45-721.50   &  8.560   &   6.821 & 13.647   & L1 Inst. Lines   \\
7    & 988.80-988.95   &  7.873   &   8.322 &  7.475   & Inst. Lines      \\
\hline
\end{tabular}
\caption{Hough outliers that have survived the coincidence analysis
in frequency, excluding those related to $60$~Hz harmonics
and the violin modes.}
\label{tab:HoughOutliers}
\end{table}

\begin{figure}
  \begin{center}
  \includegraphics[height=6.8cm]{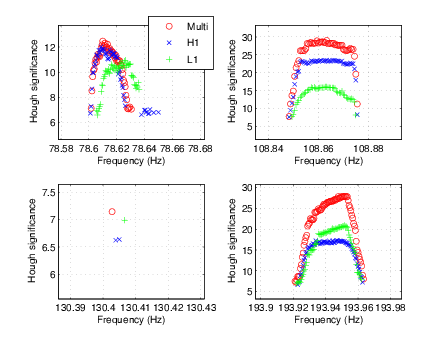}
  \includegraphics[height=6.8cm]{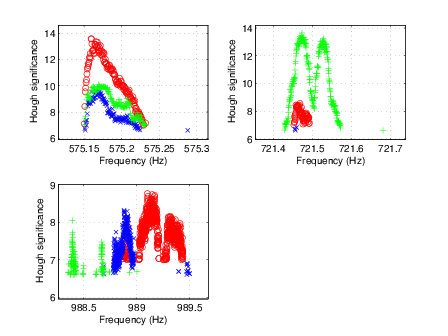}
  \caption{Hough  significance of the outliers that have survived 
  the coincidence analysis
without considering the bands contaminated with  $60$~Hz harmonics
or the violin modes.
Points are plotted only for multi-interferometer templates with
significance greater than $7$ and for single-interferometer
templates with significance greater than $6.6$.
}
  \label{fig:HoughOutliers}
  \end{center}
\end{figure}

\begin{table}
\begin{tabular}{cccccc}\hline
Detector &   $s$     & $f_0$ (Hz) & $df/dt$ ($\mathrm{Hz}~\mathrm{s}^{-1}$) &   $\alpha$ (rad)& $\delta$ (rad)\\
\hline \hline
Multi-IFO & 7.146 & 130.4028 & $\sci{-1.745}{-9}$  & 0.8798 & -1.2385 \\
H1 	  & 6.622 & 130.4039 & $\sci{-1.334}{-9}$  & 2.1889 &  0.7797\\
H1 	  & 6.637 & 130.4050 & $\sci{-1.334}{-9}$  & 2.0556 &  0.6115 \\
L1 	  & 6.989 & 130.4067 & $\sci{-1.963}{-9}$  & 1.1690 & -1.0104 \\
\hline
\end{tabular}
\caption{Parameters of the candidate events with a significance greater than
6.6 in the multi-interferometer, H1 and L1 data searches around the 
Hough outlier number 3. The parameters correspond to the significance,
 frequency and spin-down for the reference time  of the beginning of S4,
 and sky locations.}
\label{tab:HoughOutliers130}
\end{table}
  
The coincidence test applied first in frequency is similar to the one
described for the StackSlide search, using a coincidence frequency
window as broad as the size of the maximum Doppler shift expected at a
given frequency.  Of the initial 3800 0.25-Hz bands investigated, 276
yielded outliers in the multi-interferometer search with a
significance higher than 7.  Requiring those bands (or 
neighboring bands) to have outliers in H1 higher than 6.6, reduced by half
the number of surviving bands. These remaining bands were studied in
detail and, after eliminating power line harmonics and the violin modes,
27 candidates remained.  Applying again the same coincidence test with
the L1 data, we are left with only 7 coincidence outliers that are
listed on Table~\ref{tab:HoughOutliers} and displayed in
Fig.~\ref{fig:HoughOutliers}.
 
Except for the third outlier, the coincidence can be attributed to 
instrumental lines in the detectors or to the hardware pulsar injections.
Table~\ref{tab:HoughOutliers130} summarizes the parameters of the
third coincidence candidate in the 130.40-130.41~Hz frequency band,
including all the events that in any of the searches had a
significance larger than 6.6.  As can be seen from the Table, the
events from the different data sets correspond to widely separated 
sky locations. Hence no 
detections were made in the Hough search of the S4 data.
 
In future searches we plan to use lower thresholds in the
semi-coherent step in order to point to interesting areas in parameter
space to be followed up, using a hierarchical scheme with alternating
coherent and semicoherent steps. In what follows we will
concentrate on setting upper limits on the amplitude $h_0$ in each of
the 0.25 Hz bands.  

\subsubsection{Upper limits}
\label{subsubsec:houghul}

\begin{figure*}
  \begin{center}
  \includegraphics[height=13cm, width=14cm]{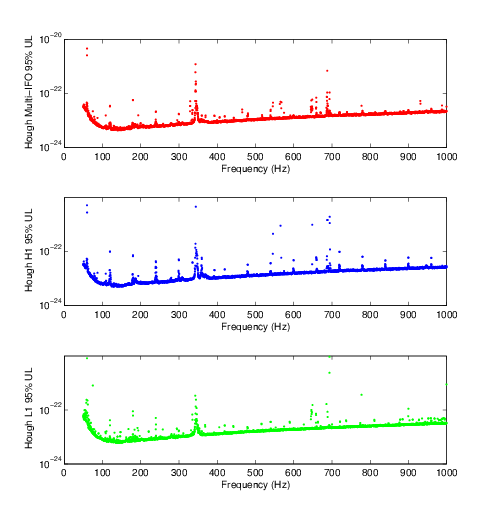}
  \caption{ The 95$\%$ confidence all-sky upper limits on $h_0$ from the
  Hough search of the multi-interferometer (top), 
  H1 (middle) and L1 (bottom) data. These upper limits have been obtained 
  by means of Monte-Carlo injections in each  0.25 Hz band.
  }
  \label{fig:HoughUL}
  \end{center}
\end{figure*}

\begin{figure}
  \begin{center}
  \includegraphics[height=6.5cm]{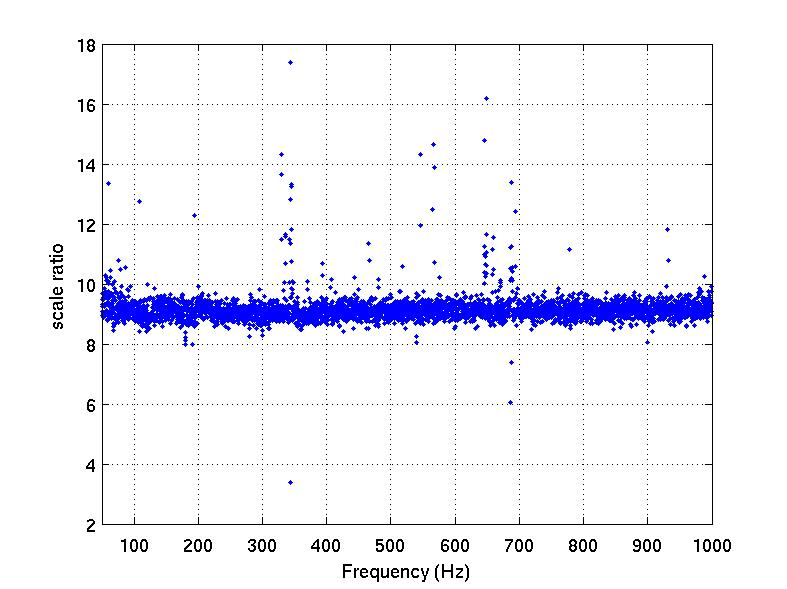}
  \caption{ Ratio of the upper limits measured by means of Monte-Carlo
  injections in the multi-interferometer Hough search to the quantity
  $h_0^{95\%}/C$ as defined in Equation~(\ref{eq:8}).
  The value of $\S$ in equation~(\ref{eq:8}) is computed 
  using the false alarm $\alpha_\H$ corresponding to the observed loudest event, 
  in a given frequency band, and for a false dismissal rate $\beta_\H=0.05$, 
  in correspondence to the desired confidence level of the upper limit. 
The comparison is performed in each 0.25 Hz band.
Analysis of the full bandwidth, and also  in 
  different 100 Hz bands, yield a scale factor $C$ to be
   $9.2\pm0.5$.}
  \label{fig:HoughScaleFit}
  \end{center}
\end{figure}

\begin{figure}
 \begin{center}
  \includegraphics[width=9.5cm]{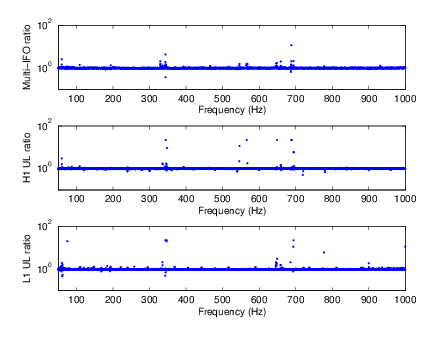}
  \caption{Ratio of the 95$\%$ confidence all-sky upper limits on $h_0$ 
  obtained from the Hough search by means of Monte Carlo injections 
  to those predicted by Eq.~(\ref{eq:8})
  of the multi-interferometer (top), 
  H1 (middle) and L1(bottom) data.
  The comparison is performed in  0.25 Hz bands. The scale factors $C$ used
  are 9.2 for the multi-interferometer search, 9.7 for H1 and 9.3 for L1.
  }
  \label{fig:HoughMultiULPredictionComparison}
  \end{center}
\end{figure}

\begin{table}
\begin{tabular}{cccc}\hline
Detector &  $\quad$ &  Band (Hz)   & $h_0^{95\%}$  \\
\hline \hline
H1+H2+L1 &  $\quad$ & 140.00-140.25 & $4.28\times 10^{-24}$  \\
H1 &  $\quad$ & 129.00-129.25 & $5.02\times 10^{-24}$ \\
L1 &  $\quad$ & 140.25-140.50 & $5.89\times 10^{-24}$ \\
\hline
\end{tabular}
\caption{Best Hough all-sky  upper limits obtained on the strength of
gravitational waves from isolated neutron stars.}
\label{tab:HoughBestULs}
\end{table}

As in the previous S2 Hough search~\cite{S2HoughPaper}, we set a
population based frequentist upper limit using Monte Carlo signal
software injections.  We draw attention to two important differences
from that analysis:
\begin{itemize}

\item In \cite{S2HoughPaper}, known spectral disturbances were handled
  by simply avoiding all the frequency bins which could have been
  affected by Doppler broadening.  Thus, the loudest event was
  obtained by excluding such frequency bins, and the subsequent Monte
  Carlo simulations also did not perform any signal injections in
  these bins.  Here we follow the same approach as used in the
  StackSlide search; we use the spectral line removal procedure
  described in section~\ref{subsubsec:stackslideimpl}.  For
  consistency, the same line removal procedure is followed in the
  Monte Carlo simulation after every software injection.

\item Recall that the calculation of the weights depends on the
  sky-patch, and the search has been carried out by breaking up the
  sky in 92 patches.  Thus, for every randomly injected signal, we
  calculate the weights corresponding to the center of the
  corresponding sky patch.  The analysis of \cite{S2HoughPaper} did not
  use any weights and this extra step was not required.
\end{itemize}
The 95$\%$ confidence
all-sky upper limit results on $h_0$ from the Hough search for the
multi-interferometer, H1 and L1 data are shown in Fig.~\ref{fig:HoughUL}.
These upper limits have been obtained by means of Monte-Carlo
injections in each 0.25 Hz band in the same way as described in
\cite{S2HoughPaper}.  The best upper limit over the entire search band
corresponds to $4.28\times
10^{-24}$ for the multi-interferometer case in the $140.00-140.25\,$Hz
band.
The results are summarized in Table~\ref{tab:HoughBestULs}.
  
Let us now understand some features of the upper-limit results.
First, it turns out that it is possible to accurately estimate the
upper limits without extensive Monte Carlo simulations. 
From~\eqref{eq:hough_h0}, and setting $w_i \propto X_i$, we expect
that the upper limits are:
\begin{equation}
  \label{eq:7}
  h_0^{95\%} \propto \left(
    \frac{1}{||\vec{X}||}\right)^{1/2}\sqrt{\frac{ \S}{\Tcoh}}  \,.
\end{equation}
Recall that $X_i$ 
contains contributions both from the sky-location-dependent
antenna pattern functions and from the sky-location-independent noise
floor estimates.  However, since we are setting upper limits for a population
uniformly distributed in the sky, we might expect that the $S^\srchTemplateInd_{\iSubSupInd}$
are more important for estimating the value of $h_0^{95\%}$. From 
Eq.~(\ref{eq:optimalweights}) and averaging over the sky we get
\begin{equation}
  ||\vec{X}|| \propto \sqrt{\sum_{i=0}^{N-1} \left(\frac{1}{S^\srchTemplateInd_{\iSubSupInd}}\right)^2}
    \,,
\end{equation}
 and thus,
up to a constant factor $C$, the estimated upper limits are given by
\begin{equation}
  \label{eq:8}
  h_0^{95\%} = C \left( \frac{1}{
      \sum_{i=0}^{N-1}(S^\srchTemplateInd_{\iSubSupInd})^{-2}}\right)^{1/4}\sqrt{\frac{
      \S}{\Tcoh}}  \,. 
\end{equation}
The value of $\S$ is calculated from Eq.~(\ref{eq:sdef})
using the false alarm $\alpha_\H$
corresponding to the significance of the observed loudest event in a
particular frequency band. The value of the false dismissal rate
$\beta_\H$ corresponds to the desired confidence level of the upper limit
(in this case $95\%$).  To show that such a fit is viable,
Fig.~\ref{fig:HoughScaleFit} plots the value of the constant $C$
appearing in the above equation for every $0.25\,$Hz frequency band,
using the measured upper limits.  It turns out that $C=9.2\pm 0.5$.
The exact value of $C$ depends on the interferometer and the search
performed, but it is still found to lie within this range.  This scale
factor $C=9.2\pm 0.5$ is about two times worse than we would expect
if we were performing a targeted (multi-interferometer with weights)
search with no mismatch. This factor of two is also in very good
agreement with what was reported in the S2 search \cite{S2HoughPaper}.

The utility of this fit is that having determined the value of $C$ in
a small frequency range, it can be extrapolated to cover the full
bandwidth without performing any further Monte Carlo simulations.
Figure~\ref{fig:HoughMultiULPredictionComparison} plots the ratio of
the measured
upper limits to the estimated values showing the
accuracy of the fit. 
The scale factors $C$ used are 9.2 for the
multi-interferometer search, 9.7 for H1 and 9.3 for L1.  The scale
factors have been obtained in all cases by comparing the measured upper
limits by means of Monte Carlo injections to the quantity
$h_0^{95\%}/C$ as defined in Equation~(\ref{eq:8}), using the full
bandwidth of the search.  
These estimated upper limits have an error
smaller than $5\%$ for bands free of large instrumental
disturbances. 

\begin{figure}
  \begin{center}
  \includegraphics[height=6.5cm]{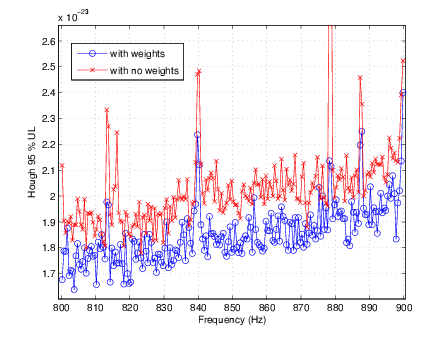}
  \caption{Comparison of the upper limits obtained using 500 Monte
    Carlo injections with and without weights in 0.5 Hz bands for the
    Hough multi-interferometer search. The use of the weights improves
    the upper limits by a $\sim$9$\%$ factor. 
    }
  \label{fig:HoughULweightsComparison}
  \end{center}
\end{figure}

We conclude this section by quantifying the improvement in sensitivity
caused by using the weights.
Figure~\ref{fig:HoughULweightsComparison} shows the comparison between
the weighted and un-weighted results in the $800$-$900\,$Hz frequency
range.  The average improvement is $\sim$$9\%$ in this band.
It is easy to see that the improvement as compared to the
unweighted Hough search will be larger if the variation of
$S^\srchTemplateInd_{\iSubSupInd}$ and the beam pattern functions is large across the SFTs.
Since the variation in $S^\srchTemplateInd_{\iSubSupInd}$ is larger in a multi-interferometer
search, we expect this improvement to be much more significant in a
multi-interferometer search.  For the case of analyzing
data from a single interferometer, for example H1, the improvement in
the upper limits due to the weights turns out to be only $\sim 6\%$.
Also, the
improvement can be increased by choosing smaller sky-patches so that
the weight calculation is more optimal. 
In particular, if there would not be any sky mismatch in computing the weights, only
due to the amplitude modulation, i.e., in the presence of Gaussian and stationary noise, 
we would expect an average increase of sensitivity of $\sim$$10\%$, and it could be up
to $\sim$$12\%$ for optimally oriented pulsars. These results have been
verified experimentally by means of a set of Monte-Carlo tests \cite{badrisintes}.


\subsection{PowerFlux results}
\label{subsec:powerfluxresults}

\subsubsection{Single-interferometer results}

\begin{figure}
  \begin{center}
  \includegraphics[height=6.5cm]{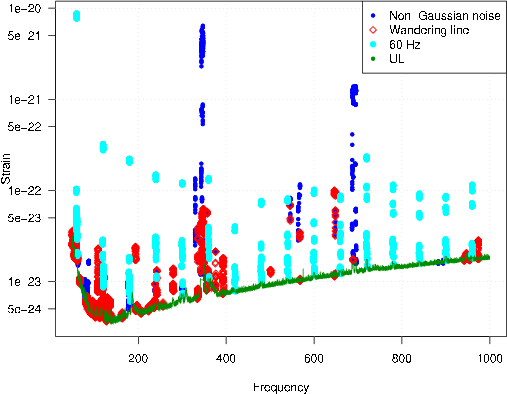}
  \caption{PowerFlux limits on linearly polarized CW radiation amplitude for
  the H1 data from the S4 run. Bands flagged as non-Gaussian (instrumental artifacts)
  or near 60-Hz harmonics, and for which derived upper limits are unreliable,
  are indicated by color. Diamonds indicate
  bands for which wandering instrumental lines (or very strong
injected signals) lead to degraded upper limits.}
  \label{fig:PowerFluxLinLimitsH1} 
  \end{center}
\end{figure}

\begin{figure}
  \begin{center}
  \includegraphics[height=6.5cm]{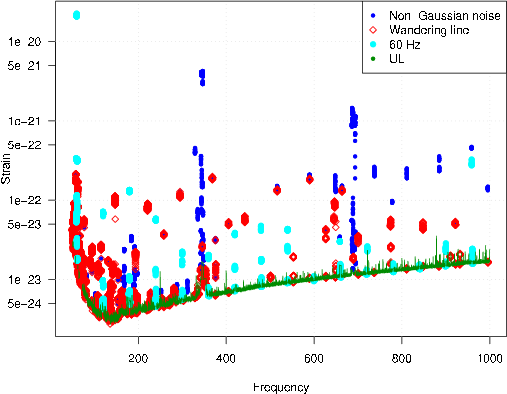}
  \caption{PowerFlux limits on linearly polarized CW radiation amplitude for
  the L1 data from the S4 run, with the same color coding as in the
  preceding figure.}
  \label{fig:PowerFluxLinLimitsL1} 
  \end{center}
\end{figure}

\def\psiproj{\psi_{\rm PROJ}}
The PowerFlux method has been applied to the S4 data sample in the range 50-1000 Hz.
Five polarization projections are sampled for each grid point: four linear polarizations with
$\psi$ = 0, $\pi/8$, $\pi/4$, $3\,\pi/8$; and circular polarization. For each sky grid point in
the ``good sky'' defined above
and each of the 501 frequency bins (there is slight overlap of 0.25 Hz bands), the Feldman-Cousins~\cite{FeldmanCousins} 95\%\ CL
upper limit is computed, as described in section~\ref{subsubsec:powerfluxupperlimitsetting}, 
for each polarization projection.
Worst-case upper limits on linear polarization for each grid point and frequency are taken to be the highest
linear-polarization-projection strain limit divided by $\cos(\pi/8)$ to correct for worst-case
polarization mismatch. The highest limit for all frequency bins in the 0.25 Hz band and over all 
sampled sky
points is taken to be the broad-sky limit for that 0.25 Hz band. 
Figures~\ref{fig:PowerFluxLinLimitsH1}-\ref{fig:PowerFluxLinLimitsL1} show the resulting broad-sky limits on linearly polarized
periodic sources from H1 and L1.
Bands flagged as non-Gaussian (instrumental artifacts leading to failure of the KS test)
or near 60-Hz harmonics are indicated by color. The derived upper limits for these bands
are considered unreliable. Diamonds indicate
bands for which wandering instrumental lines (or very strong
injected signals) lead to degraded upper limits.
An exceedingly strong pulsar can be identified as a wandering line, and several strong
hardware-injected pulsars are marked in the figures as such.

These limits on linearly polarized radiation and the corresponding limits on circularly
polarized radiation can be interpreted as 
worst-case and best-case limits on a triaxial-ellipsoid,
non-precessing neutron star, respectively, as discussed in Appendix~\ref{sec:polarization}.
Multiplying the linear-polarization
limits by a factor of two leads to the {\it worst-case} H1 limits on $h_0$ shown in 
Figs.~\ref{fig:H1PSH_limits}--\ref{fig:L1PSH_limits}. The circular-polarization limits 
require no scale correction.
Note that the StackSlide and Hough H1 limits
shown on the same figure apply to a uniform-sky, uniform-orientation 
population of pulsars. 

\subsubsection{Coincidence followup of loud candidates}
All outliers (SNR$>$7, diamonds, and non-Gaussian bands)
in the single-interferometer
analysis are checked for coincidence between H1 and L1.
In this followup, agreement is required in frequency
to within 10 mHz, in spin-down to within $\sci{1}{-10}~\mathrm{Hz}~\mathrm{s}^{-1}$, and in both right ascension and declination 
to within 0.5 radians.  The only surviving candidates are 
associated with hardware-injected pulsars 2, 3, 4, and 8 (see Table~\ref{tab:PowerFluxHWInj}),
1-Hz harmonics, violin modes, and instrumental lines in both detectors near 78.6 Hz (also seen in
the StackSlide and Hough searches). The source of these lines remains unknown, but followup consistency
checks described in section~\ref{subsec:stackslideresults} rule out an astrophysical explanation.

From this coincidence analysis, we see no evidence of a strong pulsar signal in the S4 data.
It should be noted, however, that the SNR threshold of 7 is relatively high. A lower threshold
and a more refined algorithm for location and frequency coincidence is under development 
for future searches.

\section{Comparison of the Three Methods}
\label{sec:comparisonresults}

\begin{figure*}
  \begin{center}
  \includegraphics[height=12cm]{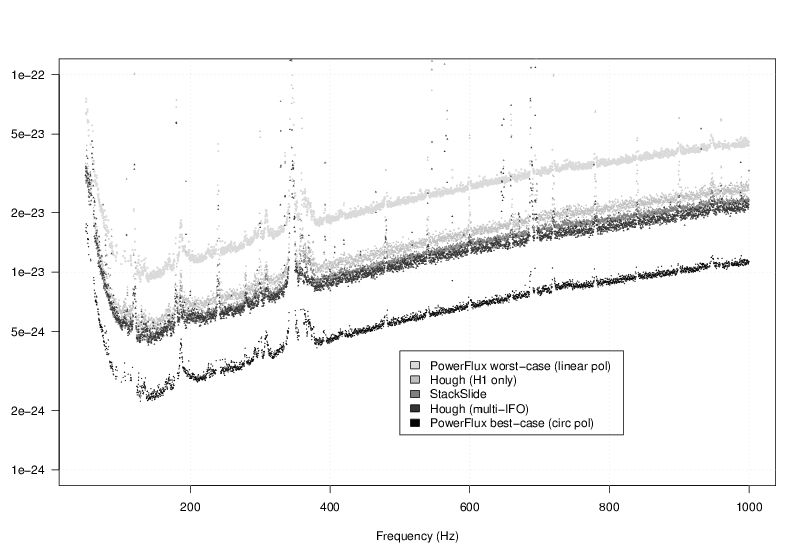}
  \caption{H1 Upper limits (95\% CL) on $h_0$ from the three methods. The StackSlide and
   Hough limits are population-based, while those from PowerFlux are strict and apply,
   respectively, to the most favorable and least favorable pulsar inclinations. Also
   shown are the multi-interferometer limits from the Hough search.}
  \label{fig:H1PSH_limits} 
  \end{center}
\end{figure*}

\begin{figure*}
  \begin{center}
  \includegraphics[height=12cm]{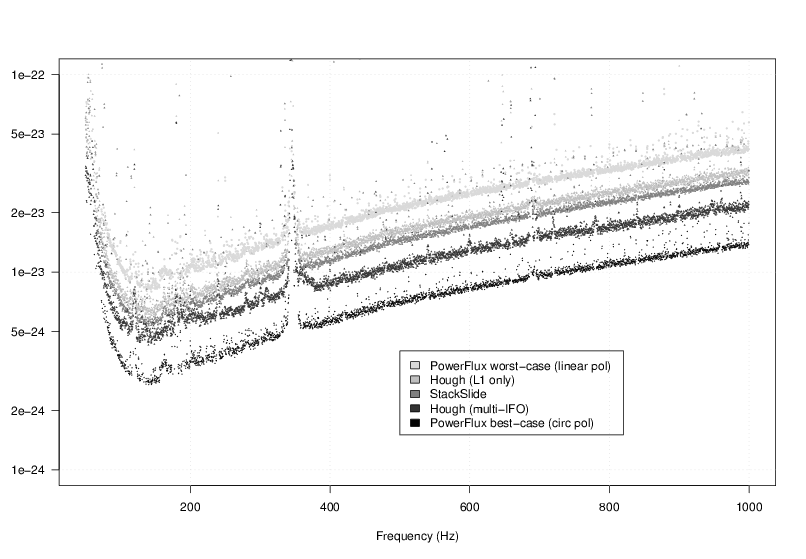}
  \caption{L1 Upper limits (95\% CL) on $h_0$ from the three methods. The StackSlide and
   Hough limits are population-based, while those from PowerFlux are strict and apply,
   respectively, to the most favorable and least favorable pulsar inclinations. Also
   shown are the multi-interferometer limits from the Hough search.}
  \label{fig:L1PSH_limits} 
  \end{center}
\end{figure*}

Figures~\ref{fig:H1PSH_limits} and~\ref{fig:L1PSH_limits} show superimposed the
final upper limits on $h_0$ from the StackSlide, Hough, and PowerFlux methods when applied
to the S4 single-interferometer H1 and L1 data, respectively. As one might have expected, we
see that the StackSlide and Hough population-based limits lie between the 
best-case and worst-case $h_0$ strict limits from PowerFlux. 
As indicated in Figs.~\ref{fig:H1PSH_limits}--\ref{fig:L1PSH_limits}, the Hough search
sensitivity improves with the summing of powers
from two or more interferometers. 

To be more precise as to expectations, we have directly 
compared detection efficiencies of the three methods
in frequency bands with different noise characteristics.  As discussed
above, we expect overall improved performance of Powerflux with respect to
StackSlide and Hough, except possibly for frequency bands marked by
extreme non-Gaussianity or non-stationarity, where the Hough integer
truncation of extreme power outliers can provide more robustness.  We
do not consider computational efficiency, which could play an
important role in deciding which algorithm to use in
computationally limited hierarchical searches.

\begin{figure}
  \begin{center}
  \includegraphics[height=5.8cm]{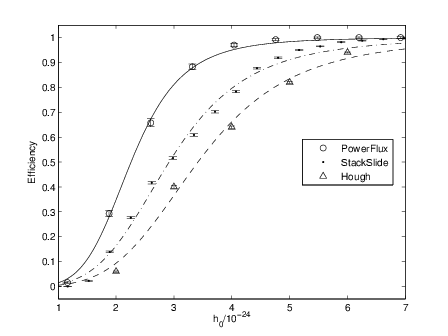}
  \caption{Comparison of StackSlide, Hough, and PowerFlux efficiencies
    (SNR $>$ 7) {\it vs} injected strain amplitude $h_0$ for the band
    140.50-140.75 Hz for H1.  From left to right, the curves
    correspond to PowerFlux, StackSlide, and Hough.  
    This band is typical of those without large outliers. }
  \label{fig:NewEffic3MethodsBand1} 
  \end{center}
\end{figure}

\begin{figure}
  \begin{center}
  \includegraphics[height=5.8cm]{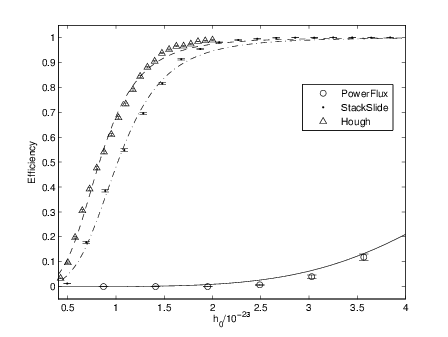}
  \caption{Detection efficiency curves for the frequency band 357-357.25 Hz, for
    H1. This band has a transient spectral disturbance affecting some
    of the SFTs. The Hough transform method proves to be robust
    against such non-stationarities and is more sensitive than
    StackSlide or PowerFlux in this band. The SNR thresholds used
    to generate these curves were 6.3, 5.2, and 30, respectively,
    for the StackSlide, Hough, and PowerFlux methods, where the StackSlide
    and PowerFlux thresholds correspond to the loudest candidates
    in that band in the data.}
  \label{fig:NewEffic3MethodsBand3} 
  \end{center}
\end{figure}

A comparison is shown in
Figs.~\ref{fig:NewEffic3MethodsBand1} and \ref{fig:NewEffic3MethodsBand3}
among the efficiencies of the three methods for two particular 0.25
Hz bands for H1: 140.5--140.75 Hz and 357--357.25 Hz.
The horizontal axis in each case is the $h_0$ of Monte Carlo software
injections with random sky-locations, spin-downs and orientations.  The
noise in the two bands have qualitatively different features.  The
140.5-140.75 Hz band is a typical ``clean'' band with Gaussian noise
and no observable spectral features.  As expected,
Fig.~\ref{fig:NewEffic3MethodsBand1} shows that the efficiency for
the PowerFlux method is higher than that for StackSlide, 
while that of StackSlide is better than that for Hough.
In other bands, where there are stationary spectral disturbances,
we find that PowerFlux remains the most efficient method. 

The noise in the band 357-357.25 Hz is non-Gaussian and displays a large transient
spectral disturbance, in addition to stationary line noise at 357 Hz itself.
The stationary 357 Hz line was removed during the StackSlide and Hough searches, avoided
during the PowerFlux search, and handled self-consistently during Monte Carlo
software injections. In this band,
the Hough transform method proves to be robust against transient noise,
and more sensitive than the StackSlide or PowerFlux implementations 
(see Fig.~\ref{fig:NewEffic3MethodsBand1}).
In fact, no PowerFlux upper limit is quoted for this band because of the
large non-Gaussianity detected during noise decomposition. Note that 
the SNR thresholds used for Stackslide, Hough and PowerFlux in Fig.~\ref{fig:NewEffic3MethodsBand3}
are set to 6.3, 5.2 and 30, respectively, to match their loudest events in
this band of the data.

\section{Summary, Astrophysical Reach, and Outlook}
\label{sec:summary}

In summary, we have set upper limits on the strength of continuous-wave
gravitational radiation over a range in frequencies from 50 Hz to 1000 Hz,
using three different semi-coherent methods for summing of strain power from
the LIGO interferometers. Upper limits have been derived using both a 
population-based method applicable to  the entire sky and a strict method
applicable to regions of the sky for which received frequencies were not
stationary during the S4 data run.

The limits have been
interpreted in terms of amplitudes $h_0$ for pulsars and in terms of 
linear and circular polarization amplitudes, corresponding to least favorable
and most favorable pulsar inclinations, respectively. As a reminder,
sets of known instrumental spectral lines have been cleaned
from the data prior to setting the population-based StackSlide and Hough
upper limits (Tables~\ref{tab:StackSlideCleanedLines}, 
\ref{tab:StackSlideExcludedBands}, and \ref{tab:HoughCleanedLines}), 
while regions of the sky (defined by cutoff values on the $S$ parameter
(Equations~\ref{eq:sparam} and \ref{eq:slarge})
have been excluded in the strict PowerFlux upper limits. The numerical
values of the upper limits can be obtained separately\cite{epaps}.

We have reached an important milestone on the road to astrophysically
interesting all-sky results:
Our best upper limits on $h_0$ are comparable to the value of a few times
$10^{-24}$ at which one might optimistically expect to see the strongest
signal from a previously unknown neutron star according to a generic
argument originally made by Blandford (unpublished) and extended in our
previous search for such objects in S2 data \cite{S2FstatPaper}.
The value from Blandford's argument does not depend on the distance to the
star or its ellipticity, both of which are highly uncertain.

\begin{figure*}
\includegraphics[width=6.5in]{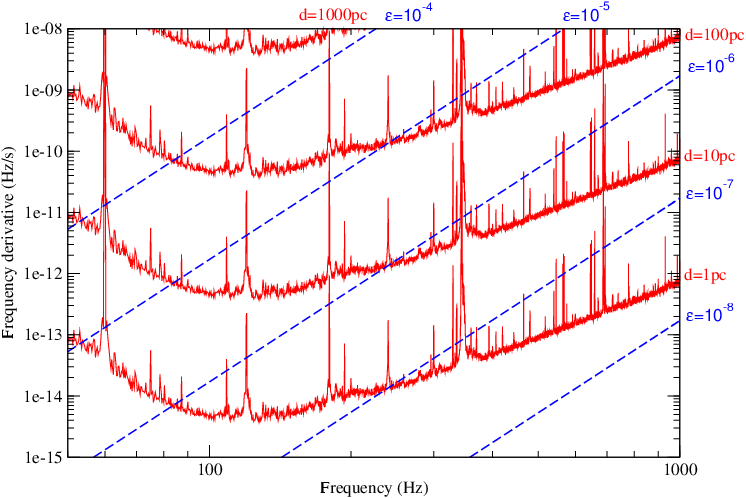}
\caption{
\label{range-hough}
Range of the multi-interferometer Hough transform search for neutron stars
spinning down solely due to gravitational radiation.  This is a
superposition of two contour plots.  The solid lines (red in the color
version) are contours of the maximum distance $d$ at which a neutron
star could be detected as a function of gravitational-wave frequency
$f$ and its derivative $\dot{f}$.  The dashed lines 
are contours of the corresponding ellipticity
$\epsilon(f,\dot{f})$.  In concert these quantities tell us the
maximum range of the search in terms of various populations (see text
for details).  }
\end{figure*}

We find the next milestone by considering the maximum distance to which a
signal could be detected and the ellipticity needed to generate a signal
of the required strength at that distance.
Both quantities are of interest since there are theoretical limits on the
ellipticity, and both quantities are functions of the gravitational-wave
frequency $f$ and its derivative $\dot{f}$.
Figure~\ref{range-hough} is a contour plot of both quantities
simultaneously, which we explain here in more detail.
The Hough transform multi-interferometer upper limits on $h_0$ are used
for illustration because they fall in the middle of the range of values for
the different searches (see Fig.~\ref{fig:H1PSH_limits}).
The maximum distance $d(f,\dot{f})$ is obtained by equating the 95\%
confidence upper limits on $h_0$ for the multiple-interferometer plot in
Fig.~\ref{fig:HoughUL} to the spin-down limit given in Eq.~(\ref{hsd}).
This tacitly assumes that $\dot{f}$ is entirely due to emission of
gravitational radiation, which implies the ellipticity given in
Eq.~(\ref{esd}) regardless of the data and the distance to the source.
If we relaxed this assumption, knowing that neutron stars spin down due to
electromagnetic wave emission, relativistic particle winds, and other
factors as well, the maximum distance and required ellipticity for a given
$f$ and $\dot{f}$ would both be reduced.
The degree of reduction would, however, be highly uncertain.

We can use the combined contour plot in  Fig.~\ref{range-hough} to answer
questions about the astrophysical significance of our results.
Here we ask and answer several salient questions.
First, what is the maximum range of the Hough transform search?
The answer is obtained from looking at the top of Fig.~\ref{range-hough}:
We could detect isolated pulsars to about 1~kpc, but only for a star 
radiating at a frequency near 100~Hz and then only if
that star has an ellipticity somewhat more than $10^{-4}$, which is allowed
only in the most extreme equations of state \cite{Owen:2005fn, Xu:2003xe,
Mannarelli:2007bs}.
Second, what is the maximum range of detection for a normal neutron star?
Normal neutron stars are expected to have $\epsilon < 10^{-6}$ based on theoretical predictions
\cite{Ushomirsky:2000ax}.
By tracing the $\epsilon=10^{-6}$ contour, we find that the maximum range is about 50~pc
at the highest frequencies (1~kHz), falling with frequency to less than
2~pc below 100~Hz.
Third, what is the maximum range for a recycled millisecond pulsar?
Based on the observed sample \cite{ATNF}, recycled pulsars usually
have small $|\dot{f}|$ values, corresponding to $\epsilon_\mathrm{sd}$ usually less than $10^{-8}$.
Unfortunately the $\epsilon = 10^{-8}$ contour corresponds to
$d<1$~pc at all frequencies in the LIGO band.

Figure~\ref{range-hough} then demonstrates that we have reached a second
milestone not achieved in our previous all-sky searches \cite{S2HoughPaper,
S2FstatPaper}:
The multi-interferometer Hough transform search could have detected an
object at the distance of the nearest known neutron star RX~J1856.5$-$3754,
which is about 110--170~pc from Earth \cite{Walter:2002uq,
vanKerkwijk:2006nr}.
We could not have detected that particular star, since the recently
observed 7~s rotation period \cite{Tiengo:2006eb} puts the gravitational
wave frequency well out of the LIGO band.
But the top of Fig.~\ref{range-hough} shows that we could have detected a
Crab-like pulsar ($f \approx 100$~Hz, $\dot{f} \approx 10^{-10}~\mathrm{Hz}~\mathrm{s}^{-1}$) at
that distance if gravitational radiation dominated its spin-down.

For the ongoing S5 data run, expected to finish data collection in late 2007,
several refinements of these
methods are under development. 
The StackSlide and Hough methods
can be made more sensitive than PowerFlux by starting with the maximum likelihood statistic
(known as the $\cal F$-statistic \cite{jks,hough04,S2FstatPaper}) rather than SFT power.
This increases the time-baseline of the coherent step in a hierarchical search, 
though at increased computational cost. The lower computational cost of
the Hough search would be an advantage in this case. Multi-interferometer searches also increase
the sensitivity, while reducing outliers (false-alarms), without having to increase greatly
the size of the parameter space used, as illustrated by the Hough search in this paper.
A multi-interferometer version of PowerFlux is under development, as well as hierarchical 
multi-interferometer searches that use the Hough and StackSlide method on the $\cal F$-statistic. 

Thus, PowerFlux will be the primary tool used for semi-coherent searches using SFTs, while the Hough and
StackSlide methods will be used in multi-interferometer hierarchical searches.
Strong candidates from the PowerFlux search will be fed into the latter type of search as well.
The parameter space searches described here do not take into account the correlations that
exist between points in the four or five dimensional parameter space (including those on the sky). 
A map of the mismatch between a signal and the parameter-space
templates can be used to generate a parameter-space metric to reduce further the number of points needed
to conduct a search, a method under development for the hierarchical searches.
Finally, the strain noise of the S5 data is lower by about a factor of 2, and the run will
accumulate at least $1$~year of science mode data.

\begin{appendix}

\section{PowerFlux polarization projection relations}
\label{sec:polarization}

The PowerFlux method uses circular and four linear polarization ``projections'' to
increase sensitivity to different source polarizations~\cite{PowerFluxPolarizationNote}. 
The projections are 
necessarily imperfect because the interferometer itself is a polarimeter continually
changing its orientation with respect to a source on the sky. There is ``leakage'' of one
polarization into another's projection.
In this appendix we present the formulae used by PowerFlux to define these 
imperfect projections and discuss corrections one can make for leakage in
followup studies of candidates. 

As described in section~\ref{subsubsec:signalestimator}, the signal estimator used by
PowerFlux for frequency bin $k$ and projection polarization angle $\psi'$ is
\begin{equation}
R \quad = \quad {2 \over \Tcoh} \sum_i W_{\iSubSupInd} {P_{\iSubSupInd}\over|F_{\psi'(+)}^\iSubSupInd|^2}\>/\>\sum_i W_{\iSubSupInd},
\end{equation}
where $W_{\iSubSupInd} \equiv |F_{\psi'(+)}^\iSubSupInd|^4/(\bar P_{\iSubSupInd})^2$ is the weight for
SFT $i$ and $F_{\psi'([+/\times])}^\iSubSupInd$ is the antenna pattern factor for a source 
with $[+,\times]$ polarization with respect to a major axis of polarization angle $\psi'$.

For a source of true polarization angle $\psi$ and plus / cross amplitudes $A_+$ and $A_\times$,
where $h_+'(t) = A_+ \cos(\omega t+\Phi)$ and $h_\times'(t) = A_\times\sin(\omega t+\Phi)$,
the strain amplitudes projected onto the $+$ and $\times$ axes for a polarization angle $\psi'$
are 
\begin{eqnarray}
h_+ & = & A_+ \cos(\omega t)\cos(\Delta\psi) \nonumber\\
    &   & -A_\times\sin(\omega t)\sin(\Delta\psi), \\
h_\times & = & A_+ \cos(\omega t)\sin(\Delta\psi) \nonumber\\
    &   & +A_\times\sin(\omega t)\cos(\Delta\psi), 
\end{eqnarray}
where $\Delta\psi \equiv 2(\psi-\psi')$, where the SFT-dependent phase constant $\Phi_0$ has
been taken to be zero, for convenience, and where frequency variation of the source
during each 30-minute SFT interval has been neglected. 
Averaging the detectable signal power
$(F_{\psi'(+)}h_++F_{\psi'(\times)} h_\times)^2$ over one SFT interval $i$,
one obtains approximately (neglecting antenna rotation during the half-hour interval):
\begin{eqnarray}
\langle P_{\rm signal}\rangle & & =  {1\over4}\bigl[(F_+^2+F_\times^2)(A_+^2+A_\times^2) \nonumber \\
 & & + (F_+^2-F_\times^2)(A_+^2-A_\times^2)\cos(2\,\Delta\psi) \nonumber \\
 & & + 2\,F_+F_\times(A_+^2-A_\times^2)\sin(2\,\Delta\psi)\bigr].
\end{eqnarray}

Note that for a linearly polarized source with polarization angle $\psi=\psi'$
(so that $\Delta\psi = 0$) and amplitude $A_+ = h_0^{\rm Lin}$, $A_\times = 0$, one obtains
\begin{equation}
\langle P_{\rm signal}\rangle\quad =\quad {1\over2}F_+^2(h_0^{\rm Lin})^2,
\end{equation} 
and that for a circularly polarized source of amplitude $A_+ = A_\times = h_0^{\rm Circ}$,
\begin{equation}
\langle P_{\rm signal}\rangle \quad = \quad {1\over2}(F_+^2+F_\times^2)(h_0^{\rm Circ})^2,
\end{equation}
as expected. 

For an average of powers from many SFT's, weighted
according to detector noise and antenna pattern via $W_i$, the 
expectation value of the signal estimator depends on 
\begin{eqnarray}
\langle P^{\rm Det}\rangle \quad & = & \quad \langle P_{\rm signal}\rangle + \langle n(\psi')^2\rangle \nonumber \\
& & + 2\langle P_{\rm signal}n(\psi')\rangle,
\end{eqnarray}
where $n_{\iSubSupInd}$ is the expected power from noise alone, where $\langle P_{\rm signal}n\rangle$ is
assumed to vanish (signal uncorrelated with noise), and where the 
frequency bin index $k$ is omitted
for simplicity. 

For a true source with parameters $\psi$, $A_+$, and $A_\times$, this expectation
value can be written:
\begin{eqnarray}
\langle P^{\rm Det}\rangle\quad & = & \langle n(\psi')^2\rangle \nonumber \\
& & + {1\over4}\bigl[(1+\beta_2)(A_+^2+A_\times^2) \nonumber \\
& & +(1-\beta_2)(A_+^2-A_\times^2)\cos(2\Delta\psi) \nonumber \\
& & +2\,\beta_1\,(A_+^2-A_\times^2)
\sin(2\Delta\psi)\bigr],
\end{eqnarray}
where the correction coefficients
\begin{eqnarray}
\beta_1 & = & {\sum_i W_{\iSubSupInd}\>F_\times/F_+\over\sum_i W_{\iSubSupInd}}, \\
\beta_2 & = & {\sum_i W_{\iSubSupInd}\>F_\times^2/F_+^2\over\sum_i W_{\iSubSupInd}}, 
\end{eqnarray}
depend implicitly on $\psi'$ through $F_+$ and $F_\times$.

For a linearly polarized source with polarization angle $\psi=\psi'$, one obtains
\begin{equation}
\langle P^{\rm Det}\rangle \quad  = \quad \langle n(\psi')^2\rangle 
+ {1\over2}(h_0^{\rm Lin})^2
\end{equation}
and for a circularly polarized source one obtains:
\begin{equation}
\langle P_{\rm Det}\rangle \quad = \quad \langle n(\psi')^2\rangle + {1\over2}(h_0^{\rm Circ})^2(1+\beta_2).
\end{equation}

These formulae permit corrections for polarization leakage to be applied for
a hypothetical source, allowing for estimation of $\psi$, $A_+$, and $A_\times$
from a sampling of polarization projection measurements. In practice, however, the
calculation of the $\beta$ coefficients is computationally costly in an
all-sky search and is disabled by default. Instead, upper limits on linearly
polarized sources (worst-case pulsar inclination) are derived from the maximum
limit over all four linear polarization projections, as described in 
section~\ref{subsubsec:signalestimator}. In followup investigations of outliers,
however, these formulae permit greater discrimination of candidates, now in
use for PowerFlux searches of the data from the ongoing S5 data run.

\section{StackSlide Power And Statistics}
\label{sec:stackslidepowerandstats}

\subsubsection{Approximate Form For The StackSlide Power}
\label{sec:stackslideapproxP}

It is useful to have an analytic approximation for the StackSlide Power $P$.
For a single SFT (dropping the SFT index $i$)
expressing the phase in a first-order Taylor expansion about the midpoint time,
$t_{1/2}$, of the interval used to generate an SFT, we can write
\begin{equation} \label{eq:Phioft}
\phi(t) \cong \phi_{1/2} + 2\pi f_{1/2}(t - t_{1/2}) \,,
\end{equation}
where $\phi_{1/2}$ and $f_{1/2}$ are the phase and frequency at time $t_{1/2}$.
Treating the values of $F_+$ and $F_\times$ as constants equal to their
values at time $t_{1/2}$,
the signal strain at discrete time $t_j$ is approximately,
\begin{eqnarray}
h_j \cong F_{+} \hpluszero {\rm cos} (\phi_0 + 2\pi f_{1/2}t_j) \nonumber \\
\qquad + F_{\times} \hcrosszero {\rm sin} (\phi_0 + 2\pi f_{1/2}t_j) \,, 
\end{eqnarray}
where $j=0$ gives the start time of the SFT,
and $\phi_0$ is the approximate phase at the start of the SFT
(not the initial phase at the start of the observation), i.e.,
\begin{equation} \label{eq:phi0sft}
\phi_0 \equiv \phi_{1/2} - 2 \pi f_{1/2} (\Tcoh / 2) \,.
\end{equation}
Using these approximations, the Discrete Fourier Transform, given by Eq.~(\ref{eq:DFT}), of $h_j$ is 
\begin{eqnarray} \label{eq:hk}
{\tilde{h}_k \over \Tcoh} \cong 
 e^{{\mathrm i}\phi_0} \Biggl [ { F_{+} \hpluszero \over 2 }   
- {\mathrm i} { F_{\times} \hcrosszero \over 2 } \Biggl ]
\Biggl [ { {\rm sin} (2\pi\Delta\kappa) \over 2 \pi \Delta\kappa }  \nonumber \\
+ {\mathrm i} { 1 - {\rm cos} (2\pi\Delta\kappa) \over 2 \pi \Delta\kappa } \Biggr ]\,, \qquad \qquad \qquad
\end{eqnarray}
where $\Delta \kappa \equiv \kappa - k$ and 
$\kappa \equiv f_{1/2} T_{\rm coh} $ is usually not an integer. 
Equation~(\ref{eq:hk}) holds for $0 < \kappa < M/2$ and
$|\kappa - k| << M$, which is true for all of the frequencies 
over which we search.

If the discrete time samples of the data from the detector consist of a signal plus noise
the expected value of $P$ is approximated by
\begin{equation} \label{eq:eststackslidepwr}
P \cong P_0 + {1 \over 2} \langle d^2 \rangle \,,
\end{equation}
where the mean value of $P_0$ is $1$ and its standard deviation is $1/\sqrt{N}$ due to the 
normalization used, and
\begin{eqnarray} \label{eq:aveoptimalsnr}
\langle d^2 \rangle \cong \Biggl [ \hpluszero^2 \left \langle {F_{+}^2 \over S_k} {\sin^2(\pi \Delta \kappa) 
\over \pi^2 \Delta \kappa^2}  \right \rangle \qquad \qquad \qquad \nonumber \\
\qquad \qquad \qquad + \hcrosszero^2 \left \langle {F_{\times}^2 \over S_k} {\sin^2(\pi \Delta \kappa) 
\over \pi^2 \Delta \kappa^2} \right \rangle \Biggr ] \Tcoh \,, \qquad
\end{eqnarray}
is an approximate form for the square of the optimal SNR defined in Eq.~(71) in reference \cite{jks}
averaged over SFTs  (i.e., the angle brackets on $\langle d^2 \rangle$ represent an average over SFTs)
and where for each SFT the index $k$ is the nearest integer value to $\kappa$.
Thus, the relevant range for $\Delta \kappa$ is $0$ to $0.5$, corresponding to a frequency mismatch of
$0$ to $1/2$ of an SFT frequency bin.

\subsubsection{StackSlide Statistics}
\label{sec:stackslidestats}

It can be seen from Eq.~(\ref{eq:stackslidepower}) that, for Gaussian
noise in the absence of a signal,  $2NP$ is a $\chi^2$ variable 
with $2N$ degrees of freedom \cite{StackSlideTechNote}. Thus, the quantity
\begin{equation} \label{eq:stackslidechi2rho}
\varrho \equiv 2NP
\end{equation}
follows the
$\chi^2$ distribution:
\begin{equation} \label{eq:chi2dist2N}
{\cal P}(\varrho; N)d\varrho = {1 \over 2^N \Gamma(N)} \varrho^{N-1}e^{-\varrho/2} d\varrho \,.
\end{equation}
When a signal is present, $\varrho$ follows a non-central $\chi^2$ distribution
with $2N$ degrees of freedom and a
non-centrality parameter $N \langle d^2 \rangle$ such that
\begin{eqnarray} \label{eq:noncentralchi2dist2N}
{\cal P}(\varrho; N \langle d^2 \rangle) d\varrho = \qquad \qquad \qquad \qquad \qquad \qquad \nonumber \\
{ I_{N-1} \biggl ( \sqrt{\varrho N \langle d^2 \rangle} \biggr )
\over (N \langle d^2 \rangle)^{{N - 1}} }
\varrho^{{N - 1 \over 2}} e^{-( \varrho + N \langle d^2 \rangle )/2} d\varrho \, ,
\end{eqnarray}
where the form given here is based on that given in \cite{jksIII}, and
$I_{N-1}$ is the modified Bessel function of the first kind and order $N-1$.   

The distribution described by Eqs.~(\ref{eq:chi2dist2N}) and (\ref{eq:noncentralchi2dist2N}) can be
used to find the minimum optimal signal-to-noise ratio that can be
detected using the StackSlide search for fixed false alarm and false dismissal rates, for a targeted
search. For a $1\%$ false alarm rate, a $10\%$ false dismissal rate, and large $N$
Eqs.~(\ref{eq:stackslidechi2rho}) and (\ref{eq:noncentralchi2dist2N}) give 
$\langle d^2 \rangle = 7.385/\sqrt{N}$ (See also \cite{StackSlideTechNote}),
while averaging Eq.~(\ref{eq:aveoptimalsnr}) independently over the source
sky position, inclination angle, polarization angle, and mismatch in frequency
gives $\langle d^2 \rangle = 0.7737 (4 / 25)(h_0^2 T_{\rm coh} / S)$ (see also Eq.~5.35 in \cite{hough04} ).
Equating these and solving for $h_0$, the characteristic amplitude for a targeted StackSlide search with
a $1\%$ false-alarm rate, $10\%$ false-dismissal rate is:
\begin{equation} \label{eq:stackslidecharamplitude}
\langle h_0 \rangle_{\rm targeted} = 7.7 \sqrt{S} / ( T_{\rm coh} T_{\rm obs}^*)^{1/4}  \, ,
\end{equation}
where $T_{\rm obs}^* = N\Tcoh$ is the actual duration of the data,
which is shorter than the total observation time, $T_{\rm obs}$, because gaps exist in the data for times
when the detectors were not operating in science mode. Comparing this expression with Eq.~5.35 in \cite{hough04}
the StackSlide characteristic amplitude given in Eq.~(\ref{eq:stackslidecharamplitude})
is found to be about $10\%$ lower than a similar estimate for the standard Hough search.
Note that in this paper an improved version of the Hough method is presented. 
Also, in this paper an all-sky search for the loudest StackSlide Power is carried out,
covering up to $1.88 \times 10^{9}$ templates, and only the loudest StackSlide Power
is returned from the search, corresponding to a false alarm rate of $5.32 \times 10^{-10}$.  
Furthermore, the upper limits are found by injecting a family of signals, each of which has a StackSlide
Power drawn from a different noncentral chi-squared distribution. Using the
results from Sec.~\ref{sec:results},
for an all-sky StackSlide search the $95\%$ confidence all-sky upper limits are found
empirically to be approximately given by:
\begin{equation} \label{eq:stackslidecharamplitudeallsky}
\langle h_0 \rangle_{\rm all-sky} = 23 \sqrt{S} / ( T_{\rm coh} T_{\rm obs}^*)^{1/4}  \, .
\end{equation}

\end{appendix}

\section{Acknowledgments}
The authors gratefully acknowledge the support of the United States
National Science Foundation for the construction and operation of
the LIGO Laboratory and the Particle Physics and Astronomy Research
Council of the United Kingdom, the Max-Planck-Society and the State
of Niedersachsen/Germany for support of the construction and
operation of the GEO600 detector. The authors also gratefully
acknowledge the support of the research by these agencies and by the
Australian Research Council, the Natural Sciences and Engineering
Research Council of Canada, the Council of Scientific and Industrial
Research of India, the Department of Science and Technology of
India, the Spanish Ministerio de Educacion y Ciencia, The National
Aeronautics and Space Administration, the John Simon Guggenheim
Foundation, the Alexander von Humboldt Foundation, the Leverhulme
Trust, the David and Lucile Packard Foundation, the Research
Corporation, and the Alfred P. Sloan Foundation.

This document  has been assigned LIGO Laboratory document number
LIGO-P060010-06-Z.


%

\end{document}